%% file: thesis.tex
\documentclass[12pt]{my_report}
\usepackage{natbib,graphicx,amsmath,amssymb,bm}
\usepackage[dvipdfm]{color}
\usepackage{my_thesis,epsfig}
\usepackage[dvipdfm]{hyperref}
\newcommand{\mev}{\, \text{MeV}}
\newcommand{\gev}{\, \text{GeV}}
\newcommand{\sumint}{\int  \!\!\!\!\!\!\!
\sum}
\newcommand{\bra}{\big\langle}
\newcommand{\ket}{\big\rangle}
\newcommand{\threej}[6]{
     \left( \begin{array}{ccc}
              #1 & #3 & #5 \\
              #2 & #4 & #6
            \end{array}  \right) }
\newcommand{\sixj}[6]{
     \left\{ \begin{array}{ccc}
              #1 & #3 & #5 \\
              #2 & #4 & #6
            \end{array}  \right\} }
\newcommand{\ninej}[9]{
     \left\{ \begin{array}{ccc}
              #1 & #2 & #3 \\
              #4 & #5 & #6 \\
              #7 & #8 & #9
            \end{array}  \right\} }
\newcommand{\quark}[2]{
     \left( \begin{array}{c}
              #1 \\
              #2
            \end{array}  \right) }
\newcommand{\gcq}{\, \text{g}/\text{cm}^{3}}

\newcommand{\tnu}{{T_\nu}}
\begin{document}

    \title{Electro-Weak Interactions\\ \vskip 0.3 in  in Light Nuclei}
    \author{Doron Gazit}
    \principaladviser{Nir Barnea}
    \beforepreface
    \include{dedication}
    \include{Abstract}
    \afterpreface

    \include{Introduction}
    \include{ew_chapter}
    \include{eft_chapter}
    \include{lit_chapter}
    \include{HH_chapter}
    \include{Photo_chapter}
    \include{nu_scattering}
    \include{summary}
    \appendix
    \include{app_SM_current}
    \include{app_lepton_current}
    \include{app_spinors}
    \include{app_feynman}
    \include{app_2BME}
    \bibliographystyle{astron}
    \bibliography{thesis}
    \include{Acknowledgments}
\end{document}

%% file: dedication.tex
\vspace{14 cm}\hspace{2 cm}\textbf{To my beloved wife Shani,}
\\ \vspace{14 cm} \hspace{2 cm} \textbf{my strength and guide in life. }

%% file: Abstract.tex
\prefacesection{Abstract}
In this thesis, conducted in the field of few--body nuclear physics,
I studied low energy inelastic reactions of electro--weak probes on
light nuclei.

Such interactions hold experimental and theoretical attention in
recent years because of the role they play in various fields of
physics, stretching from checking the standard model limits; through
nuclear structure and dynamics research; up to microscopic
interaction in astrophysical phenomena, such as supernova explosion
and the nucleosynthesis of the elements.

The reactions considered in this work are:
\begin{itemize}
\item[(i)] Neutrino interaction with $A=3$ and $^4$He nuclei,
 in energies typical to core-collapse supernova. For $^4$He both the
 neutral and charged current reactions were calculated, whereas only
 neutral current reactions are calculated for $A=3$ nuclei.
\item[(ii)] Photoabsorption on $^4$He. This long standing problem has been
the attraction of theoretical and experimental efforts in the last
three decades.
\end{itemize}
A complete theoretical description of these reactions holds many
obstacles. The most significant one arises from the fact that light
nuclei are weakly bound and as such they have few, if any, excited
states. Thus, evaluation of the nuclear dynamics usually includes a
transition between the nucleus ground state and a resonance in the
continuum, namely fragments of the initial nucleus interacting with
each other in what is commonly known as final state interaction
(FSI). As a result, a theoretical study of the reaction should
contain both the ground state and the final state interaction. In
order to avoid a calculation of a continuum final state, which is
currently out of reach already for $A=4$, I used an integral
transform with a lorentzian kernel. This novel method, the Lorentz
integral transform (LIT) method allows a reduction of the full
scattering problem into a Schr{\"{o}}dinger like equation with
boundary conditions of a bound state. This equation is solved using
the recently developed method of effective interaction
hyperspherical harmonics (EIHH). The combination of these two
methods has been applied in the literature for electro--magnetic and
strong reactions. In my thesis, the approach was applied for the
first time for weak reactions, and
results in a percentage level numerical accuracy.\\

Photon reactions on nuclear targets are a common experimental method
for investigating the nuclear structure and dynamics. In
electromagnetic processes, the nuclear forces manifest themselves
also as exchange currents, due to gauge invariance. However, one
does not have to explicitly calculate their contribution. The
conservation of the electromagnetic current allows an application of
the Siegert theorem, which in low energy suggests that these
currents are included in the single nucleon charge operator. Thus,
the scattering operators are model independent. This makes
photonuclear processes ideal for experimental study of the nuclear
force and structure.

A fundamental example of electromagnetic response which drew
continuous interest in the last three decades, both in theory and in
experiment, is the $^4$He photodisintegration process. The
$\alpha$-particle is drawing such a great attention because it has
some typical features of heavier systems (e.g. binding energy per
nucleon), which make it an important link between the classical
few-body systems, viz deuteron, triton  and $^3$He, and more complex
nuclei. For example in $^4$He one can study the possible emergence
of collective phenomena typical of complex nuclei like the giant
dipole resonance. Furthermore, $^4$He is the ideal testing ground
for microscopic two- and three-body forces, which are fitted in the
two- and three-body systems. One expects that the 3NF is of
considerably greater relevance in the four-body system, as the
number of triplets is bigger, and due to its higher density.

The microscopic calculation of the $^4$He photoabsorption process
presented here is the first evaluation of the cross--section which
includes realistic NN potential, Argonne $v_{18}$, and 3NF, Urbana
IX, and full final state interaction, taken into account via the LIT
method. The resulting cross--section is characterized by a
pronounced peak, due to a dipole excitation. We have found that the
3NF only slightly lowers the peak (by about 6\%). This effect is
much smaller than the 20\% effect on the binding energy, and than
the influence expected when compared to the
$^3\text{H}/^3\text{He}(\gamma)$ processes. The effect of the 3NF
grows at higher energy transfer to about 35\%.

The comparison to experiments is unclear, mainly due to the poor
status of the experimental data, in which the differences reach a
factor of 2. However, close to threshold the theoretical cross
section agrees quite well with all experiments. In the giant
resonance region of the cross--section, where there are several
experimental evaluations of the cross--sections which in some cases
do not agree, the theoretical results are in good agreement with the
majority of the available data.

The photoabsorption process can be used to infer properties of the
ground state. In fact, moments of the total photoabsorption
cross--section lead to known sum--rules, which are indicative for
different characteristics of the ground state, such as the charge
radius and the nucleon-nucleon distance. In particular, the
configuration tetrahedral symmetry of $^4$He is tested, and this
symmetry is found to be slightly broken. This observation, which can
be tested in experiments, suggests that the mean distance between
identical nucleons is slightly (6\%)
larger than the mean distance between proton and neutron.\\

The second branch of the work investigates inelastic neutrino
scattering on $A=3$ and $^4$He nuclei in core--collapse supernova
scenario. Inelastic neutrino interaction with nuclei can potentially
influence several properties of the physics in the supernova.
Firstly, they may change the structure of the neutrino signals.
Secondly, they can deposit energy in the matter behind the stalled
shock, thus might change its temperature and even revive the shock.
Finally, they change the chemical composition of the star by
breaking nuclei to fragments. The high abundance of $^4$He nuclei
both in the shocked area and in the outer layers of the star, may
result in an effect on these processes.

Lately, we have discovered that a substantial amount of trinuclei
can exist in the neutrinosphere of the newly born proto--neutron
star. This is a result of the previously neglected interaction
between nuclei in that area. Thus, neutrino interaction with these
nuclei may affect the neutrino spectra.

These facts motivate evaluation of the cross--sections. In this
work, the first {\it ab-initio} calculation of the inelastic
reactions is given, with realistic forces: AV18 NN potential and UIX
3NF. By using the LIT method, all break--up channels are considered.
The vector meson exchange currents are implicitly calculated using
the Siegert theorem. The contribution of the (non--conserved) axial
meson exchange currents to the cross sections has to be calculated
explicitly. I cope with this by using effective field theory of QCD
at low energy, with the nucleons and pions as explicit degrees of
freedom. The axial currents are calibrated to reproduce the triton
half--life, within this nuclear Hamiltonian, thus enabling
parameter--free prediction of the cross--sections. This calculation
procedure, known in the literature as EFT*, has been tested
successfully in the calculation of numerous weak reactions, e.g.
Park {\it et. al} calculated $pp$-fusion and $hep$ process. The
current work is the first application of the method for the
calculation of complete inelastic processes, not only on--threshold.

The contribution of the exchange currents to the $\nu-\alpha$
scattering process is negligible, due to the ``closed shell''
character of this nucleus. However, the effect of meson exchange
currents in the case of neutrino scattering on mass-three nuclei is
about $20\%$ of the cross--section for low neutrino temperature. As
neutrino temperature increases, the effect reduces to less than
$2\%$ for neutrino temperatures above $5\mev$.

The error estimation, for both $\alpha$ and the trinuclei, due to
the effective theory cutoff dependence of the cross--section is less
than a percent, for neutrino temperatures higher than $2\mev$. This
is a strong validation of the calculation.

The neutrino scattering cross--section is found to be rather
sensitive to the properties of the force model. Thus, the predicted
neutrino scattering cross--sections are considered to be of 5\%
accuracy.

The reactions calculated in this thesis make an important step
towards more precise predictions made by nuclear physics. These
predictions have the ability to check the properties of the nuclear
forces on the one hand, and on the other hand to be used as accurate
and reliable microscopic input for simulations of astrophysical
phenomena.

%% file: Introduction.tex
\chapter{Introduction}
\label{Chap:Introduction}

The physics of the atomic nucleus has drawn a large amount of
attention since the dawn of quantum mechanics, fascinating
physicists in its diversity and complexity. Understanding the
nuclear world has led to many practical uses, the examples stretch
from medical application, where radioactive and nuclear magnetic
resonance mappings are already standard procedures; through nuclear
reactors, in which one makes use of the energy released in
stimulated nuclear decays as a power source; and finally to the
destructive power of nuclear weapons.

The nucleus is a complex entity, built of nucleons held together by
the strong force. A nucleon is a common name for both neutrons and
protons, emphasizing the approximated symmetry of their response to
the strong force, and their similar mass. Determining the properties
of the strong force was, and still remains, a scientific challenge.
It is accepted today that the strong force is the low energy
appearance of quantum chromodynamics (QCD). However, deriving the
nuclear force from the fundamental theory is an open question, due
to the non--perturbative character of QCD in low energy.

The modern nuclear physics has two main approaches to meet this
problem, a semi--phenomenological approach and an effective field
theory (EFT) approach. In the semi--phenomenological approach the
potential includes all functional shapes and operators allowed by
Poincar\`{e} and isospin symmetry of QCD. The parameters for this
potential are calibrated by experimental scattering phase shifts. A
clear disadvantage is that the fit of the parameters is not unique
as the theory lacks a clear connection to the underlying theory, a
fact which leads to a questionable predictive measure. However, the
approach has many successes. One of its important consequences is
identifying the need of adding attractive three nucleon forces (3NF)
to fit the spectra of $A \ge 3$ nuclei. The realistic
semi--phenomenological potentials combined with 3NFs successfully
reproduce the low lying spectra of $A<12$
nuclei.\\
The EFT approach uses the chiral symmetry of QCD to describe the
force among nucleons. The approach leads to a consistent
perturbative expansion of the interaction, with numerical
coefficients which contain information about higher energy degrees
of freedom. These coefficients can be in principal calculated from
QCD, but due to the aforementioned difficulties are usually
calibrated using low energy experiments. Among obvious advantages,
in this approach the 3NF are predicted and appear at higher order of
perturbation theory, thus less important - as also found
phenomenologically. However, in order to reach the accuracy achieved
in describing the entirety of problems as the semi--phenomenological
approach one has to expand the potentials up to
next--to--next--to--next--to leading order (N$^3$LO). Such
potentials are still not completely available (see however
\citep{EFT1,EFT2}).

It is of great interest to check the success of both approaches in
predicting dynamical observables of nuclei. This is not a simple
task. The problem arises from the need to solve the
Schr{\"{o}}dinger equation in $3A-3$ dimensions, with substantial
inter-particle correlations. Without any further approximations,
this is an example of the few--body quantum problem, which explores
the response of $A$ bodies to the forces between them.  Even today,
this problem lacks a general mathematical solution, mainly in cases
of scattering to continuum states. For bound state problems, the
last decade has brought immense progress, due to theoretical
advances and increasing available computer power. A variety of
methods solving the nuclear problem microscopically (for bound
states) have evolved, among which are the No Core Shell Model
\citep{NCSM}, Green Function Monte Carlo \citep{GFMC}, Stochastic
Variational Method \citep{SVM} and the Effective Interaction
Hyperspherical Harmonics approaches, which have been used for
calculations of nuclei in the medium range ($A>4$).

Contrary to this, calculations of reactions have not met this
success. Specifically, light nuclei are weakly bound and as such
they have few, if any, bound excited states. Thus, experiments
designed to study nuclear dynamics include a transition between the
nucleus ground state and the continuum. The latter are essentially
fragments interacting with each other in final state interaction
(FSI). It is clear that a complete theoretical evaluation of the
reaction should include an accurate description of both the ground
state and the final state interaction. An exact calculation of a
final state in the continuum for all break--up channels is out of
reach already for $A=4$. In order to avoid a calculation of a
continuum final state one can use an integral transform with a
lorentzian kernel. This novel method, the Lorentz integral transform
(LIT) method \citep{LIT} reduces the full scattering problem to a
Schr{\"{o}}dinger like equation with bound state boundary
conditions. This equation can be solved using the above mentioned
few-body methods, and in this thesis the modern method of effective
interaction hyperspherical harmonics (EIHH) method
\citep{EIHH1,EIHH2,symhh} is used for solving the LIT equations.

The combination of the LIT and EIHH methods allows nuclear theory to
describe reactions of complex nuclei with $A \ge 3$ which were
previously unreachable in {\it ab initio} calculations. The methods
have proven successful for the calculation of the total
photoabsorption cross--section \citep{tri_photo} and inclusive
electron--scattering process of $A=3$ nuclei \citep{3H_dis}, with
realistic nucleon--nucleon potential and 3NF. Calculations of
photoabsorption and electron scattering processes using
semi--realistic forces were done for $A=4-7$ body systems, giving
quite realistic results in comparison with experiments
\citep{bacca1,bacca2,ELO97,ELO97_2}.

The focus of this thesis is the accurate modeling and calculation,
feasible by using the LIT and EIHH methods, of electro--weak
interaction with light nuclei, specifically neutrino scattering and
photoabsorption on nuclei. Electro--weak reactions are of central
importance not only due to their use in experiments which probe
nuclear structure and force properties. Their importance stems also
from the role they play in cosmic and stellar events. This research
field, ``nuclear astrophysics'', evolved in the last century from
the understanding that in many astrophysical phenomena the typical
thermodynamic conditions are so extreme that nuclear reactions can
be induced thermally. By this observation, nuclear physics has
turned to be the microscopic ingredient used to address
astrophysical mysteries. For example, energy released in nuclear
fusion has been discovered to be the fuel of stars, thus nuclear
theory sets the main properties of a star, i.e. its composition,
evolution and way of death. Another key question addressed in
nuclear astrophysics is the origin of the variety of elements in
nature. Big--bang nucleosynthesis, i.e. the synthesis of light
elements in the early stages of the universe, provides a hatch to
physics in the first three minutes of the universe. The rest of the
elements are believed to be produced during the evolution of a star
and mainly in the extreme conditions of its death, in a supernova.
In this work, neutrino reactions with $A=3$ and $^4$He are
calculated. These reactions are of interest in the description of
the exploding death of a massive star, and in the nucleosynthesis
within, as will be explained.

In the next two sections we will elaborate on the specific processes
addressed in this thesis.

\section{Photodisintegration of $^4$He}
Photonuclear reactions are a common experimental method for
investigating the nuclear structure and dynamics. In these
processes, a real photon excites a nucleus. The perturbative
character of the electromagnetic interaction, dictated by the fine
structure constant $\frac{e^2}{\hbar c} \ll 1$, makes it possible to
seperate the scattering process and the nuclear properties affecting
the process. The photon, as a result, is ideal for experimental
study of nuclei.

A fundamental example of an electromagnetic response which drew a
continuous interest in the last three decades, both in theory and in
experiment (see \citep{Lund,Shima} and references therein), is the
$^4$He photodisintegration process. The $\alpha$-particle is drawing
such a great attention because it has some typical features of
heavier systems (e.g. binding energy per nucleon), which make it an
important link between the classical few-body systems, viz deuteron,
triton and $^3$He, and more complex heavy nuclei. For example, in
$^4$He one can study the possible emergence of collective phenomena
typical of complex nuclei. Such a phenomenon is the giant dipole
resonance, an enhanced cross section in the range $10-30\mev$ found
in heavy nuclei, due to an $E^1$ dipole excitation. Furthermore,
$^4$He is the ideal testing ground for microscopic two- and
three-body forces, which are fitted in the two- and three-body
systems. At present the 3NF is not yet well determined, thus it is
essential to search for observables where it plays an important
role. Because of gauge invariance, in electromagnetic processes
nuclear forces manifest themselves also as exchange currents, which
have turned out to be very important in photonuclear reactions and
hence 3NF effects might become significant. For the three-nucleon
systems photonuclear processes have already been studied
\citep{ELOT,Fad_CHH,RepBoKr}, e.g. in \citet{ELOT} it was found that
the 3NF leads to an almost 10\% reduction of the electric dipole
peak and up to 15\% enhancement at higher energy. One expects that
the 3NF is of considerably greater relevance in the four-body
system, since it involves six nucleon pairs and four triplets,
compared to three pairs and just one triplet in the three-nucleon
systems. In addition, the higher density of $^4$He might enhance the
importance of the 3NF.

The $^4$He$(\gamma$) reaction represents a very challenging
theoretical problem due to the fact that the nucleus has no bound
excited states, so the full four-body continuum dynamics and all
possible fragmentation have to be considered. It is of no surprise
that the current theoretical situation of the $^4$He
photodisintegration is not sufficiently settled. Calculations with
realistic nuclear forces have not yet been carried out, and exist
only for semi-realistic nucleon-nucleon (NN) potentials. In Refs.
\citep{ELO97,BELO01,Sofia1} it has been shown that such models lead
to pronounced peak cross sections, in rather good agreement with the
data of \citet{Lund} and much different from what was calculated
earlier \citep{Sandhas}. \\
A similar status holds for the experimental research of the process,
as most of the experimental work has concentrated on the two-body
break-up channels $^4$He$(\gamma,n)^3$He and $^4$He$(\gamma,p)^3$H
in the giant resonance region, but a large disagreement still exists
in the peak. In fact in two very recent $(\gamma,n)$ experiments
\citep{Lund,Shima} one finds differences of a factor of two.

It is evident that the experimental and theoretical situations are
very unsatisfactory. In this thesis I present an important step
forward on the theory side performing a calculation of the total
photoabsorption cross section $\sigma_\gamma$ of $^4$He with a
realistic nuclear force, namely Argonne V18 (AV18) NN potential
\citep{AV18} and the Urbana IX (UIX) 3NF \citep{UIX}.

\section{Neutrino Scattering on Light Nuclei in Supernova}
Core-collapse supernovae (SN) are giant explosions of massive stars,
above $9$ solar masses, that radiate $99\%$ of their energy in
neutrinos. The current theory of core collapse supernova holds some
open questions regarding the two important phenomena related to the
event, i.e. the explosion mechanism and the synthesis of complex
nuclei.

The extreme conditions within the supernova make nuclear reactions
important microscopic phenomena which govern the equation of state.
Therefore, the dynamics and neutrino signals can be sensitive to the
details of neutrino interactions with nucleonic matter. In order to
analyze the mentioned questions, a better understanding of the
involved microscopic processes is needed. In particular, due to the
high abundance of $\alpha$ particles in the supernova environment,
the inelastic neutrino--$^4$He reaction has drawn attention in
recent years. This interest yielded a number of studies trying to
estimate the cross-section and the role of neutrino--$^4$He
reactions in the described phenomena
\citep{HA88a,HA88b,ME95,YO05,YO06,OH06,SU06,WO90}. However to date,
a full {\it ab--initio} calculation that includes a realistic
nuclear Hamiltonian is still missing. Moreover, the contribution of
meson exchange currents (MEC) to this particular scattering process
was never estimated.

Core collapse supernovae are believed to be neutrino driven
explosions of massive stars. As the iron core of the star becomes
gravitationally unstable it collapses until the nuclear forces halt
the collapse and drive an outgoing shock. This shock gradually
stalls due to energy loss through neutrino radiation and
dissociation of the iron nuclei into a mixture of $\alpha$ particles
and free nucleons.

At this stage, the proto-neutron star (PNS) cools mainly by emitting
neutrinos in enormous numbers. These neutrinos are a result of
thermal pair production, and thus are produced in flavor
equilibrium. The characteristic temperatures of the emitted
neutrinos are about $6-10$ MeV for $\nu_{\mu,\tau}$
($\bar{\nu}_{\mu,\tau}$), $5-8$ MeV for $\bar{\nu}_e$, and $3-5$ MeV
for $\nu_e$. The difference in temperature originates from the large
cross-sections for $\nu_e, \bar \nu_e$ electron scattering and
charge current reactions, thus $\nu_{\mu,\tau}$
($\bar{\nu}_{\mu,\tau}$) decouple deeper within the star, where the
temperature is higher. Interactions of neutrino with nuclei in this
location, called neutrinosphere, can change neutrino temperature and
spectra. Clearly, it is essential to estimate the abundances of
nuclei in that region.

Abundances of nuclei are usually predicted using nuclear statistical
equilibrium (NSE) models based on binding energies and the quantum
numbers of nuclei. However, NSE models only treat approximately (or
neglect) strong interactions between nuclei, and consequently break
down near the neutrinosphere. In a recent work \citep{v3H}, we have
shown this results in a substantial amount of $A=3$ nuclei near the
newly born proto-neutron star (PNS), an effect forehand unnoticed.

The hot dilute gas above the PNS and below the accretion shock
contains up to $70\%$ $^4$He nuclei. The high temperature of heavy
flavored neutrinos migrating out of the PNS leads to a considerable
amount of $\mu$ and $\tau$ neutrinos (and anti-neutrinos) which
carry more than $20$ MeV, hence may dissociate the $^4$He nucleus
through inelastic neutral current reactions. If these reactions
deposit enough energy in the matter behind the shock, they can
eventually reverse the flow and revive the shock. This delayed shock
mechanism, originally introduced by \citet{CO66}, has not yet been
proven in full hydro-reactive simulations. \citet{HA88a} has
suggested that inelastic neutral reactions of neutrinos with $^4$He
can lead to an enhanced neutrino energy deposition. This effect is
usually ignored (see however \citep{OH06,WO90}) and was not
considered in full hydrodynamic simulations.

The energy deposition also creates the needed conditions for the
$r$--process, nucleosynthesis by rapid capture of free neutrons,
believed to occur in the material ejected from the PNS. The break-up
of $^4$He by neutrinos is part of the chain of reactions which
determines the amount of free neutrons \citep{FU95} needed for a
successful $r$--process.

A different nucleosynthesis process which is influenced by
$\nu-\alpha$ interaction is $\nu$--nucleosynthesis in the outer
layers of the star. A knock out of a nucleon from a $^4$He nucleus
in the helium rich layer, creates the seed to light element
nucleosynthesis in the supernova environment \citep{WO90}.  Followed
by a fusion of the remaining trinucleus with another $\alpha$
particle, this will result in a $7$--body nucleus. This process is
an important source of $^7$Li, and of $^{11}$B and $^{19}$F through
additional $\alpha$ capture reactions. Due to the high dissociation
energy of the $\alpha$, this mechanism is sensitive to the
high--energy tail of the neutrinos. Thus a correct description of
the process must contain an exact, energy dependent cross-section
for the neutral inelastic $\alpha-\nu$ reaction, which initiates the
process. The relatively low temperature of the $\nu_e$'s and $\bar
\nu_e$'s emitted from the core of the star suppress the probability
for inelastic reactions of these neutrinos with $^4$He in the
supernova scenario. Oscillations of the $\mu$ and $\tau$ (anti)
neutrinos can yield a secondary source of energetic electron
neutrinos. The resulting charge current reactions, which have larger
cross-sections than the neutral cross-sections, would affect the
discussed yields \citep{YO06}.

The work in this thesis predicts, to a percentage level accuracy,
the cross--sections for neutrino interactions with $A=3$ and $^4$He
nuclei. This removes the uncertainty, which exists today, in the
microscopic information needed for a correct evaluation of the
aforementioned processes.

\section{About this thesis}
The work presented in this thesis summarizes the results published
in a series of papers written during my PhD. studies.

Two of these papers investigated the photoabsorption process on
$^4$He:
\begin{description}
  \item[P1] ``Photoabsorption on $^4$He with a Realistic Nuclear Force'',
  D. Gazit, S. Bacca, N. Barnea, W. Leidemann, G. Orlandini, {\it
  Phys. Rev. Lett.} {\bf 96}, 112301 (2006).
  \item[P2] ``Photonuclear sum rules and the tetrahedral configuration of
  $^4$He'',
  D. Gazit, N. Barnea, S. Bacca, W. Leidemann, G. Orlandini, {\it
  Phys. Rev.} {\bf C 74}, 061001(R) (2006).
\end{description}

A substantial mass of the work was done on the subject of
low--energy neutrino scattering on light nuclei, namely triton,
$^3$He and $^4$He, published in:
\begin{description}
  \item[P3] ``Neutrino neutral reaction on $^4$He: Effects of final state interaction and realistic
  $NN$ force'', D. Gazit and N. Barnea, {\it
  Phys. Rev.} {\bf C 70}, 048801 (2004).
  \item[P4] ``Low-Energy Inelastic Neutrino Reactions on $^4$He'', D. Gazit and N. Barnea, {\it
  Phys. Rev. Lett.} {\bf 98}, 192501 (2007).
  \item[P5] ``Neutrino breakup of A = 3 nuclei in supernovae'',
  E. O'Connor, D. Gazit, C. J. Horowitz, A. Schwenk, N. Barnea, {\it
  Phys. Rev.} {\bf C 75}, 055803 (2007).
  \item[P6] ``Few body Calculation of Neutrino Neutral Inelastic scattering on
  $^4$He'', D. Gazit and N. Barnea, {\it Nucl. Phys.} {\bf A 790}, 356 (2007) .
  \item[P7] ``Low energy inelastic neutrino reactions on light nuclei'',
  D. Gazit and N. Barnea, {\it in prep.}
\end{description}

The structure of the thesis is as follows. The first part,
chapters~{2-5}, contains a detailed description of the underlying
physics and the methods of calculations. Chap.~\ref{chap:ew_chapter}
explains the quantum formalism of electro--weak probes scattering on
nuclei. The next chapter uses EFT approach to derive the scattering
operator. The LIT method is discussed in Chap.~\ref{chap:LIT},
followed by a description in Chap.~\ref{chap:EIHH} of the nuclear
wave functions used to solve the LIT equations.

The second part of the thesis presents the calculated
cross--sections and their physical application. In
Chap.~\ref{chap:Photo} the photoabsorption on $^4$He process is
analyzed, based on the work published in {\textbf{P1}} and
{\textbf{P2}}. Chap.~\ref{chap:neu} a complete calculation is given
for neutrino inclusive scattering on $A=3$ and $^4$He nuclei, as was
first presented in {\textbf{P3-P7}}.

The thesis concludes with a summary.

%% file: ew_chapter.tex
\chapter{Scattering of electro-weak probes on Nuclei} \label{chap:ew_chapter}
The unification of the electro-magnetic and weak forces is
considered one of the biggest accomplishments of physics in the
twentieth century. In the forty years since the pioneering work by
Weinberg, Salam and Glashow, the standard model has been a matter of
constant interest and ongoing research.

The application of this theory to the description of photon and
neutrino scattering off matter is a key ingredient in experiments
checking the limits of the standard model or exploring the structure
of baryonic matter. These reactions are also the microscopic engines
of many stellar phenomena, from star evolution to supernova
explosion.

The weakness of the electro-weak interaction, in comparison to the
strong force, makes a perturbative treatment of scattering processes
on nuclei a good approximation. Consequently the interaction between
electro-weak probes and nuclei targets reduces to a current-current
type, as shown in the first section of this chapter.

An important result, presented in this chapter, is that the
observables directly relate to the nuclear matrix element. Not only
does this simplify calculations, but also provides an experimental
separation of the electro-weak effects and the strong force effects.
One cannot overemphasize the importance of the latter fact in the
experimental research of the strong force, as reflected in the
nuclear structure and dynamics. From the point of view of a
theoretician, a construction of the electromagnetic and weak
currents within the nucleus is needed. Chap.~\ref{chap:EFT} is
dedicated to meeting this challenge.

An additional simplification is achieved by using a multipole
decomposition of the nuclear currents. This breaks the current into
a series of operators, each characterized by known parity and
angular momentum $J^\pi$. For low energy probes, whose wavelength
$q^{-1}$ is much larger than the typical size of a nucleus $R$, the
multipole decomposition provides a polynomial dependence in the
small parameter $qR$. In the following discussion, the multipole
decomposition technique is rigourously reviewed for the
electro--weak interaction.

\section{Neutrino Scattering}
Neutrino scattering processes on elementary particles can be easily
calculated using the Feynman diagrams of the standard model. Nuclei,
however, are not elementary particles. On the contrary, they are
built of protons and neutrons, that are built of quarks.

A possible way to cope with this problem, which is common in the
literature, is to treat the nucleons as point fermions, with form
factors indicating their complex quark structure.

\begin{figure*}[t]
\includegraphics[scale=1.,clip=]{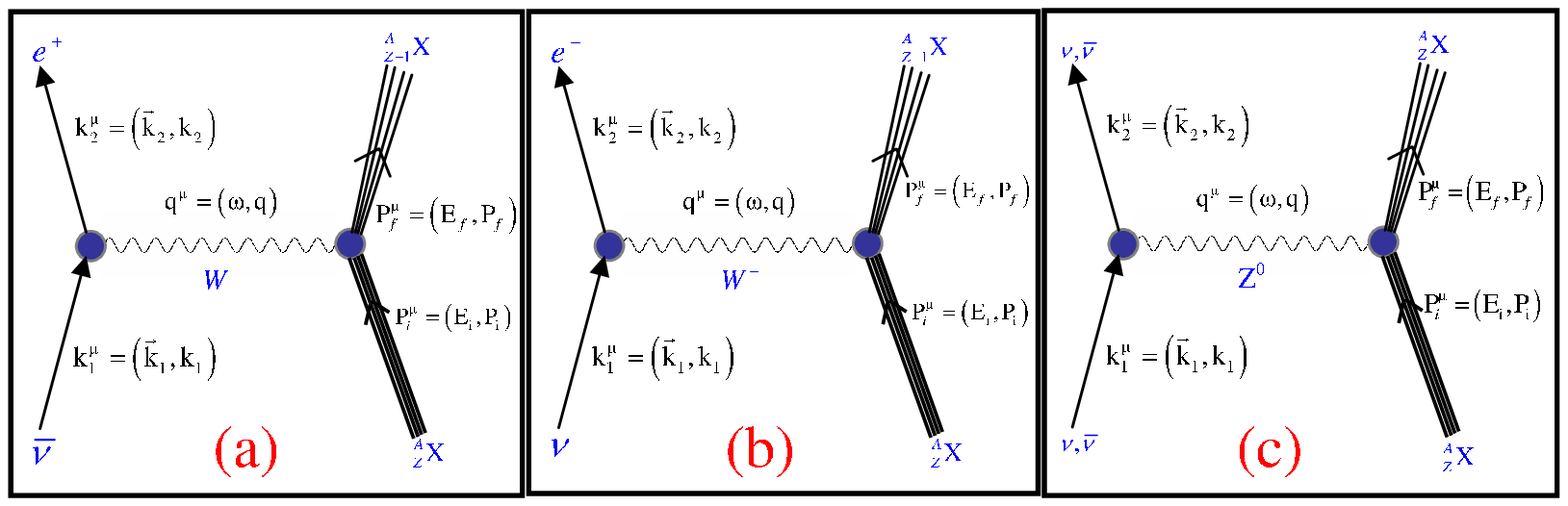}
\caption{Neutrino scattering processes and their kinematics: (a)
charged current $e^+$ production; (b) charged current $e^-$
production; (c) neutral current process} \label{fig:kinematics}
\end{figure*}

The weak coupling constant $G=1.166 \times 10^{-11} \mev^{-2}$
provides a small parameter, so the relevant neutrino scattering
processes include one boson exchange. The first order Feynman
diagrams for neutrino scattering off nucleon appear
in~\citet{WALECKA:BOOK}. The neutrino scattering processes of
interest appear in Fig.~\ref{fig:kinematics}. A neutrino with
momentum $k_i^\mu$ interacts with a nucleus of initial momentum
$P_i^\mu$, through the exchange of heavy boson with momentum
$q^\mu=(\omega,\vec{q})$. The result of the interaction is the
emergence of a lepton with momentum $k_f^\mu$, and nuclear fragments
with momentum $P_f^\mu$. Energy momentum conservation implies:
$q^\mu=k_i^\mu-k_f^\mu=P_f^\mu-P_i^\mu$.

The lepton could be either neutrino of the same flavor in the case
of neutral current reaction, or an electron, muon or tau, for
charged current reactions. In the low energy regime typical to
supernova environment, the possible charged reaction is the creation
of an electron or a positron.

\begin{figure}[h]
\begin{center}
\includegraphics[scale=1.,clip=]{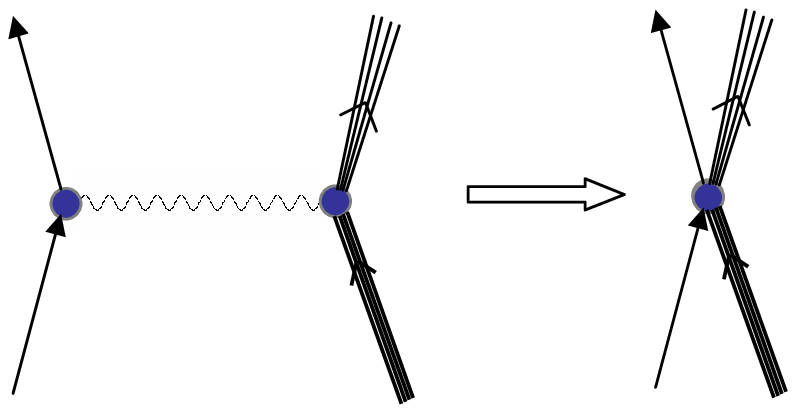}
\caption{Effective vertex for low energy weak interactions.}
\label{fig:weak_eff}
\end{center}
\end{figure}

Thus, for all neutrino scattering processes, one can write the
general form of the transition matrix for both charged and neutral
currents (with one boson exchange):
\begin{equation}
T_{fi}= \left(\text{neutrino current}\right)^\mu \cdot \frac{g_{\mu
\nu} + \frac{q_{\mu}q_{\nu}} {M_B^2}} {q^2+M_B^2} \cdot
\left(\text{nuclear current}\right)^\nu.
\end{equation}
For low energy reactions with respect to the boson masses --
$M_{Z^0} \approx 92 \gev$ and $M_{W^\pm} \approx 80 \gev$ -- the
momentum dependent terms are negligible. The boson propagator is
constant in this limit $\sim g_{\mu\nu}/M_B^2$, and acts as an
effective coupling constant (which explains the weak scale of weak
interaction). This effective vertex, demonstrated in
Fig.~\ref{fig:weak_eff}, leads to the famous current--current
Hamiltonian density: ${\mathcal H}_W \sim j^\mu \mathcal{J}_\mu$
($j^\mu$ is the lepton current, and $\mathcal{J}_\mu$ is the nuclear
current).
The standard model gives additional information regarding the formal
structure of the currents. It is shown in
Appendix~\ref{ap:SM_current}, that the weak neutral current takes
the form,
\begin{align} \label{eq:nu_cu}
\mathcal{{J}}^{(0)}_{\mu} & =
(1-2\cdot\sin^2\theta_W)\frac{\tau_0}{2}
{J}^V_\mu+\frac{\tau_0}{2}{J}^A_\mu-2\cdot\sin^2\theta_W\frac{1}{2}
{J}^V_\mu
\\ \nonumber
 & = \frac{\tau_0}{2} \big({J}^V_\mu+{J}^A_\mu \big)
-2\cdot\sin^2\theta_W {J}^{em}_\mu
\end{align}
Where $J_\mu^V$ and $J_\mu^A$ have vector and axial symmetry,
respectively. The second equality takes advantage of the conserved
vector current hypothesis, by which the weak--vector current is an
isospin rotation of the electro--magnetic current (${J}^{em}_\mu =
\frac{1+\tau_0}{2}{J}^{V}_\mu)$.

A similar calculation for the charged current yields:
\begin{equation} \label{eq:c_cu}
\mathcal{{J}}^{(\pm)}_{\mu}
  = \frac{\tau_{\pm}}{2} \big({J}^V_\mu+{J}^A_\mu \big)
\end{equation}

The vector and axial currents include not only a sum over all
nucleons. It is well known, and will be shown in the next chapter,
that in order to keep consistent with the chiral symmetry one has to
include not only one nucleon currents but also meson exchange
currents, which hold the nucleons bound in the nucleus.
\subsection{Cross section calculation for neutrino scattering
processes} In order to evaluate the probability for a scattering,
the starting point is the Golden Rule (by Fermi). Taking into
account the recoil of the nucleus, the differential cross-section
takes the form,
\begin{eqnarray} \label{eq:Golden_Rule}
{d\sigma = 2\pi\int d\epsilon \delta (\epsilon -\omega +\frac{q^{2}}
{2M_A})  \frac{d^{3}\vec{k}_{f}}{(2\pi )^{3}}} {\sumint_{f}|\langle
f| \hat{\mathcal{H}}_{W}|i\rangle |^{2} \delta (E_{f}-E_{i}-\epsilon
)}
\end{eqnarray}
Here $M_A$ is the initial nucleus mass, $|i\ket$ ($|f\ket$) is the
initial state of the system, and $E_i$ ($E_f$) is the initial
(final) energy of the nuclear system. In the previous chapter it was
shown that, in the limit of low energy-momentum transfer with
respect to the heavy gauge bosons $W^\pm$ and $Z^0$ mass, a proper
approximation for the weak interaction between leptons and baryons
is,
\begin{equation} \label{eq:HW}
\hat{\mathcal{H}}_W=\frac{G}{\sqrt{2}} \int d^3\vec{x}
\hat{j}_\mu(\vec{x}) \hat{\mathcal{J}}^\mu (\vec{x}) .
\end{equation}
The calculation of the matrix element of the Hamiltonian needed in
Eq.~(\ref{eq:Golden_Rule}) is divided accordingly. Keeping the
theory to first order in the weak interaction constant, the lepton
current $\hat{j}_{\mu }(\vec{x})$ is composed only of one lepton,
which is a point Dirac particle. Thus, its matrix element is
$\langle f|j_{\mu }(x)|i\rangle =l_{\mu }e^{-i \vec{q}\cdot
\vec{x}}$, where for neutrino reaction ${l}_{\mu
}=\bar{u}(k_1)\gamma _{\mu }(1-\gamma _{5})u(k_2)$, and for
anti--neutrino reaction ${l}_{\mu }=\bar{v}(-k_2)\gamma _{\mu
}(1-\gamma _{5})v(-k_1)$. Consequently,
\begin{equation} \label{eq:HWme}
\bra f | \hat{\mathcal{H}}_W | i\ket=\frac{G}{\sqrt{2}} \int
d^3\vec{x} e^{-i \vec{q}\cdot \vec{x}} \bigl[
l_0{\mathcal{J}}^0_{fi} (\vec{x}) - \vec{l} \cdot
{\vec{\mathcal{J}}}_{fi} (\vec{x})\bigr] .
\end{equation}
To continue, it is useful to choose the $\hat{z}$ direction parallel
to the momentum transfer $\vec{q}$, so the plane wave can be
expanded as
\begin{equation}
e^{-i\vec{q}\cdot\vec{x}}=\sum_J i^J \sqrt{4\pi (2J+1)} j_J(qx)
Y_{J0}(-\hat x ),
\end{equation}
where $j_J(x)$ is the spherical bessel function of order $J$, and
$Y_{JM} (\hat x)$ are the spherical harmonics.\\
Since the vector $\vec{l}$ is written in the same spherical
orthonormal coordinate system $\hat{e}_{\pm 1},
\hat{e}_0=\frac{\vec{q}}{q}$ it is useful to invert the definition
of the vector spherical harmonics: ${\vec{Y}}_{JLM} \equiv \bigl[Y_L
\otimes \hat{e} \bigr]^{(J)}_M$, to get $Y_{Lm} e_\lambda =
\sum_{JM} \left(
{L}{m}{1}{\lambda}|{J}{M}\right) {\vec{Y}}_{JLM}$.\\
For the nuclear current $\hat{\mathcal{J}}^\mu (\vec{x})$, a
multipole analysis of the current provides natural expansion and
some insight in low energy regime. We define the multipole
operators, viz the Coulomb, electric, magnetic and longitudinal
operators:
\begin{align}
\hat{{C}}_{JM}(q) & =  \int {d\vec{x}
j_{J}(qx)Y_{JM}(\hat{x})\hat{\mathcal{J}}_{0}(\vec{x})}
\\
\hat{{E}}_{JM}(q) & =\frac{1}{q}\int{d\vec{x}\vec {\nabla} \times
[j_J(qx)\vec{Y}_{JJM}(\hat{x})]\cdot
\hat{\vec{\mathcal{J}}}(\vec{x})}
\\
\hat{{M}}_{JM}(q) & =\int{d\vec{x}j_J(qx)\vec{Y}_{JJM}(\hat{x})\cdot
\hat{\vec{{\mathcal{J}}}}(\vec{x})}
\\
\hat{L}_{JM}(q) & =\frac{i}{q}\int{d\vec{x}\vec {\nabla}
[j_J(qx){Y}_{JM}(\hat{x})]\cdot \hat{\vec{\mathcal{J}}} (\vec{x})}
\end{align}
respectively.

 For unoriented and unobserved targets, one has to sum
over all possible final states, and average over all initial states.
For the nuclear sector, this procedure gives the following
expression:
\begin{eqnarray}
\lefteqn {\sum_f \frac{1}{2J_i+1} \sum_{M_i} \sum_{M_f} |\langle f|
\hat{\mathcal{H}}_{W}|i\rangle |^{2} \delta(E_f-E_i-\epsilon) =
\frac{G^2}{2}\frac{4\pi}{2J_i+1} \times }\\  \nonumber & &{ \times
\left\{ \sum_{J \ge 1}\left[ \frac{1}{2}\bigl( \vec{l} \cdot
\vec{l}^{*} - l_3 l^{*}_3\bigr) \bigl(\mathcal{R}_{\hat{E}_J} +
\mathcal{R}_{\hat{M}_J} \bigr) - \frac{i}{2} \bigl(\vec{l} \times
\vec{l}^{*} \bigr)_3 2 Re
\mathcal{R}_{\hat{E}_J \hat{M}_J^{*}} \right] + \right. }\\
\nonumber & &{+ \left. \sum_{J \ge 0}\left[ {l_3 l^{*}_3
\mathcal{R}_{\hat {L}_J}} + l_0 l^{*}_0 \mathcal{R}_{\hat {C}_J} -
2Re \bigl({l}_3 {l}^{*}_0 \mathcal{R}_{\hat{C}_J \hat{L}_J^{*}}
\bigr) \right] \right\} }
\end{eqnarray}
$J_i$ is the total angular momentum of the initial nuclear state
(which is usually the ground state). We use here the nuclear
response functions
\begin{equation}
\mathcal{R}_{\hat{O}_1 \hat{O}_2} \equiv \sumint_f
\delta(E_f-E_i-\epsilon) \bra f | \hat{O}^\dagger_1 | i \ket \bra f
\| \hat{O}_2 \| i \ket,
\end{equation}
which should be calculated using a model for the nucleus. The
$\sumint$ sign in the response function is a sum over discrete bound
states, and an integral over final states which are in the
continuum. For brevity we use the notation
$\mathcal{R}_{\hat{O}} \equiv \mathcal{R}_{\hat{O}\hat{O}}$.\\
For the lepton sector, one sums over all possible helicities of the
ejected lepton. In the case of neutrino scattering, there is no need
in averaging on initial helicities as there is only one (anti)
neutrino helicity in nature that interacts via the electro-weak
force. The lepton summation is simply converted to the calculation
of
traces of Dirac spinors, as explained in Appendix~\ref{app:Lepton_current}. \\
An additional effect, which needs to be taken into account for
charged lepton production, is the deformation of the lepton wave
function due to the long--range electromagnetic interaction with the
nucleus. This is done in an approximated way using the Fermi
function \citep{Fermi34}.

The final result for the differential cross section of a beam of
neutrinos scattering on unpolarized and unobserved targets, into
solid angle $d\Omega$ is thus:
\begin{eqnarray}
\lefteqn {\frac{d\sigma^{(a)}}{d\Omega dk_f} = \frac{2{G^{(a)}}^2}{2J_i+1} F^{(a)}(Z_f,k_f) k_f^2 \times }\\
\nonumber & &{ \times \left\{ \left[ 1-\bigl(\hat{k}_i \cdot \hat{q}
\bigr) \bigl(\vec{\beta}_f \cdot \hat{q} \bigr) \right] \sum_{J \ge
1} \bigl(\mathcal{R}_{\hat{E}_J} + \mathcal{R}_{\hat{M}_J} \bigr)
\mp \hat{q} \cdot \left( \hat{k}_i - \vec{\beta}_f \right) \sum_{J
\ge 1} 2 Re \mathcal{R}_{\hat{E}_J \hat{M}_J^{*}} + \right. }\\
\nonumber & &{+ \left. \sum_{J \ge 0} \left[ \left[ 1-\hat{k}_i
\cdot \vec {\beta_f}+2\bigl(\hat{k}_i \cdot \hat{q} \bigr)
\bigl(\vec{\beta}_f \cdot \hat{q} \bigr) \right] \mathcal{R}_{\hat
{L}_J} + \left( 1+\hat{k}_i \cdot \vec {\beta_f} \right)
\mathcal{R}_{\hat {C}_J} - 2 \hat{q} \cdot \left( \hat{k}_i +
\vec{\beta}_f \right) Re \mathcal{R}_{\hat{C}_J \hat{L}_J^{*}}
\right] \right\} }
\end{eqnarray}
where $Z_f$ is the charge of the final nucleus, and
$\vec{\beta}_f=\frac{\vec{k}_f }{\sqrt{m_f^2+k_f^2}}$ is the
outgoing lepton velocity ($m_f$ is the mass of the lepton). The $-$
sign is for neutrino scattering, and the $+$ sign is for
anti--neutrino. The superscript $(a)=0,\pm 1$ indicates the process:
\begin{description}
  \item[$a=0$] Neutral current, for which the $G^{(0)}=G$ and the
  screening function $F^{(0)}(Z_f,k_f)=1$.
  \item[$a=\pm 1$] Charge changing current. $G^{(\pm)}=G \cos {\theta_C} \approx 0.974
  G$ ($\theta_C$ is the Cabibbo angle, mixing the $u$ and $d$ quarks)
  and the screening function is the Fermi function \citep{Fermi34}:
  \begin{equation}
  F^{(\pm)}(Z_f,k_f)=2(1+\gamma_0) (2 k_f R_f)^{2(\gamma_0-1)}
  \frac{\mid \Gamma(\gamma_0+i\rho) \mid^2}{\mid \Gamma(2\gamma_0+1)
  \mid^2} e^{\pi \rho}
  \end{equation}
  with $\alpha \approx 1/137$ is the fine structure constant, $R_f$ is
  the radius of the final nucleus, $\rho= \mp \alpha Z_f / \beta_f$,
  and $\gamma_0=\sqrt{1-(\alpha Z_f)^2}$ ($\Gamma(x)$ is the Gamma function).
\end{description}

\section{Relations Between Multipoles and the Siegert theorem}
The multipole decomposition is a systematic expansion in the
possible angular momenta exchange in a reaction. This makes it a
powerful and intuitive tool to analyze reactions, in which the
quantum numbers of the final state are given. On a different
perspective, one can view the multipole decomposition essentially as
expansion in powers of the momentum transfer, since $j_J(\rho)
\simeq \frac{\rho^J}{(2J+1)!!}$ for $\rho \ll 1$. Thus, it provides
a tool when investigating low energy reactions with $qR < 1$ ($R$ is
the radius of the nucleus), for which a finite number of multipoles
are sufficient to estimate the cross-section to given accuracy.

The electro--weak reactions discussed in the scope of this work are
investigated up to pion threshold $\sim 140\mev$, thus the multipole
decomposition is applicable and efficient.

In this section we will discuss some important properties of the
multipole operators and relations between multipoles, helpful in
calculations. It is useful to understand that the longitudinal,
Coulomb and electric multipoles are of normal parity, i.e. if one
denotes the parity of the current by $\pi$ then their parity is
$\pi\cdot(-)^{J}$. The magnetic operator parity is
$\pi\cdot(-)^{J+1}$

An important relation exists between the electric and the
longitudinal operators. In order to find it, one uses the following
formulas,
\begin{align} \nonumber
\vec{\nabla}_\rho [j_J(\rho) Y_{JM}(\hat{\rho})] & = & \sqrt
\frac{J+1}{2J+1} j_{J+1}(\rho) \vec{Y}_{J\,J+1\,M}(\hat{\rho}) +
\sqrt \frac{J}{2J+1} j_{J-1}(\rho) \vec{Y}_{J\,J-1\,M}(\hat{\rho})
\\ \nonumber
\vec{\nabla}_\rho \times [j_J(\rho) \vec{Y}_{JJM}(\hat{\rho})] & = &
-i \sqrt \frac{J}{2J+1} j_{J+1}(\rho)
\vec{Y}_{J\,J+1\,M}(\hat{\rho}) + i\sqrt \frac{J+1}{2J+1}
j_{J-1}(\rho) \vec{Y}_{J\,J-1\,M}(\hat{\rho}).
\end{align}
This leads to,
\begin{equation}
\vec{\nabla}_\rho \times [j_J(\rho) \vec{Y}_{JJM}(\hat{\rho})] = i
\sqrt \frac{J+1}{J} \vec{\nabla}_\rho [j_J(\rho) Y_{JM}(\hat{\rho})]
- i\sqrt \frac{2J+1}{J} j_{J+1}(\rho)
\vec{Y}_{J\,J+1\,M}(\hat{\rho}).
\end{equation}
Thus,
\begin{equation} \label{eq:E_with_L_multipoles}
\hat{E}_{JM} = \sqrt \frac{J+1}{J} \hat{L}_{JM} - i \sqrt
\frac{2J+1}{J} \int {d\vec{x} j_{J+1}(qx) \vec{Y}_{J\,J+1\,M}
(\hat{x})\cdot \hat{\vec{{\mathcal{J}}}}(\vec{x})}.
\end{equation}
One notes that since the leading order in $\hat{E}_{JM}$ (and
$\hat{L}_{JM}$) is $O((qR)^{J-1})$, then the second term is smaller
by a factor of $(qR)^2$. Finally, this leads to the following
relation for low momentum transfer:
\begin{equation} \label{eq:E_and_L_multipoles}
\hat{E}_{JM} \approx \sqrt \frac{J+1}{J} \hat{L}_{JM}.
\end{equation}

\subsection{Siegert Theorem} \label{sec:Siegert}
An additional relation between multipole operators exists for
conserved currents, {\it inter alia} the electro--magnetic current
and the vector part of the weak current. For a conserved current
$\vec\nabla \cdot \hat{\vec{J}} =
-i[\mathcal{\hat{H}},\mathcal{\hat{J}}_0]$, and thus its matrix
element (between eigen--states of the Hamiltonian) satisfies:
$\vec{\nabla} \cdot \mathcal{\vec{J}} =-i \omega \mathcal{J}_0$. Let
us first look at the longitudinal operator:
\begin{equation}
\hat{L}_{JM}(q) =\frac{i}{q}\int{d\vec{x}\vec {\nabla}
[j_J(qx){Y}_{JM}(\hat{x})]\cdot \hat{\vec{\mathcal{J}}} (\vec{x})} =
-\frac{i}{q} \int{d\vec{x} [j_J(qx){Y}_{JM}(\hat{x})]\cdot \vec
{\nabla} \cdot \hat{\vec{\mathcal{J}}} (\vec{x})}=-\frac{\omega}{q}
\hat{C}_{JM}(q)
\end{equation}
We use Eq.~(\ref{eq:E_with_L_multipoles}) to connect the electric
multipole to the Coulomb multipole,
\begin{equation} \label{eq:Sie_E_C}
\hat{E}_{JM} = -\frac{\omega}{q} \sqrt \frac{J+1}{J} \hat{C}_{JM} -
i \sqrt \frac{2J+1}{J} \int {d\vec{x} j_{J+1}(qx)
\vec{Y}_{J\,J+1\,M} (\hat{x})\cdot
\hat{\vec{{\mathcal{J}}}}(\vec{x})}.
\end{equation}
These relations are called the Siegert theorem \citep{Siegert}. In
low energy, the second term in Eq.~(\ref{eq:Sie_E_C}) is negligible
with respect to the first one.

The strength of the Siegert theorem is in the connection it makes
between currents and charge. It implies that for conserved currents
in low energy, one can calculate all the corrections due to meson
exchange currents by calculating the single--nucleon Coulomb
operator. This not only simplifies calculation, but also keeps the
operators model--independent.

\section{Photoabsorption on Nuclei Up To Pion Threshold}
\label{sec:photo}

Photoabsorption is the process in which a real photon is absorbed in
a nucleus. The process can result in excitation of the nucleus to a
bound state, but can also result in a break up of the nucleus, a
process called photodisintegration. The latter is the case for all
photoabsorption processes of $^4$He, since it has no bound excited
states.

The interaction Hamiltonian of an electro--magnetic process can be
found by looking at minimal substitution of the momentum in the
kinetic energy operator:
$(\hat{\vec{p}}-e\hat{\vec{A}}(r))^2=\hat{p}^2+e^2\hat{A}^2-
e\{\hat{\vec{p}},\hat{\vec{A}}\}$. The first two terms are the
kinetic energies of the nucleus and the photon field, and the last
is the interaction term. Hence, it is clear that the interaction
Hamiltonian is,
\begin{equation} \label{eq:Hem}
\hat{\mathcal{H}}_{EM}=\frac{4\pi\alpha}{\sqrt{2}} \int d^3\vec{x}
\hat{A}_\mu(\vec{x}) \hat{\mathcal{J}}^\mu (\vec{x}) .
\end{equation}
where $\alpha$ is the fine structure constant, $\hat{A}_\mu$ is the
photon field, and $\hat{\mathcal{J}}^\mu$ is the nuclear
electro--magnetic current. The photon field in the Coulomb gauge
(i.e. $\vec{\nabla} \cdot \hat{A}_\mu (\vec{x})$) is given by,
\begin{equation}
\hat{\vec{A}}=\sum_{\vec{k}} \sum_{\lambda=\pm 1} (2\omega_k)^{-1/2}
\left[ \hat{e}_{\vec{k}\lambda} \hat{a}_{\vec{k}\lambda} e^{i
\vec{k} \cdot \vec{x}} + h.c. \right]
\end{equation}
Due to the gauge only the transverse polarizations, $\lambda=\pm 1$,
participate.

The analogy to the weak Hamiltonian is clear (cf. Eq.~\ref{eq:HW}),
so the calculations are in analogy, but the kinematics is different.

Let us consider a photon of momentum $\vec{q}$ and energy
$\omega=|q|$, as the photon is absorbed. The initial (final) nucleus
state is denoted by $\mid i/f \ket$, with energies $E_{i/f}$
respectively (by definition $\omega=E_i-E_f$). The photon is
absorbed, thus the surviving matrix element is $\bra{\vec{k}\lambda}
\mid \hat{a}^{\dagger}_{\vec{k}^\prime \lambda^\prime} \mid 0 \ket =
\delta_{\vec{k}\vec{k}^\prime} \delta_{{\lambda}{\lambda}^\prime}$.

Due to the possible polarizabilities, only the magnetic and electric
enter in the calculation. By using the golden rule, we get the final
result for the total photoabsorption cross--section:
\begin{equation}
\sigma(\omega)={4\pi^2\alpha} \omega \sum_J
[\mathcal{R}_{\hat{E}_J}(\omega) +\mathcal{R}_{\hat{M}_J}(\omega)].
\end{equation}

%% file: eft_chapter.tex
\chapter{Chiral Effective Field Theory}
\label{chap:EFT} Nuclear theory is believed to be the low--energy
reflection of quantum chromodynamics (QCD). Though QCD properties
are well known, deriving the nuclear forces is not possible
directly, as QCD is non-perturbative in low energy.

Two seminal ideas proposed by \citet{WE79} have pointed the way to a
solution to the problem. The first idea suggests that a Lagrangian
which includes all possible terms consistent with the symmetries of
the underlying theory, would result in the most general S--matrix
consistent with the theory. The second idea introduces a scheme
which organizes the terms to dominant and less dominant
contributions to the S--matrix. This way, the anticipated low energy
expansion for QCD becomes possible. These ideas were used in the
last 15 years to construct a nuclear Lagrangian based on an
effective field theory for low-energy QCD, named chiral perturbation
theory ($\chi$PT) (see for example \citep{NN94,NN94_2,EFT1,EFT2}).

In this chapter, the application of effective field theory to QCD is
presented, and used to get a description of the currents in a
nucleus. In turn, these currents interact with electro-weak probes,
determining their coupling with matter. The effective field theory
is constructed to describe nuclear electro--weak processes of
typical energy of up to $100 \mev$. Thus the relevant degrees of
freedom are pions and nucleons.

\section{QCD symmetries at low energy}
QCD, as a field theory, describes the interaction among quarks and
gluons. Its underlying gauge symmetry is the non--abelian color
$SU(3)$ Lie algebra. The masses of the up and down quarks, which are
the building blocks of nucleons, are small \citep{PDBook}: $m_u=2
\pm 1 \mev$ and $m_d=5 \pm 2 \mev$. Thus, these masses are
negligible with respect to the QCD mass scale ($\sim 1 \gev$). In
the context of nuclear physics, it is natural to treat QCD in the
limit of zero quark masses, with restriction to the up and down
quarks.

Following these restrictions, one writes the quark field composed of
two components: $\quark{u}{d}$. The massless quark field enters in
the QCD Lagrangian as,
\begin{equation}
\mathcal{L}_{QCD}^q=i\bar{q}\gamma_\mu \mathcal{D}^\mu q.
\end{equation}
The quark field has a left and right symmetry. We define right and
left handed components of the quark field:
$q_{R,L}=\frac{1}{2}\big(1 \pm \gamma_5\big)q$, respectively. Using
the properties of the $\gamma$ matrices (see
Apendix~\ref{app:Lepton_current}), the Lagrangian can be separated,
\begin{equation}
\mathcal{L}_{QCD}^q=i\bar{q}_R\gamma_\mu \mathcal{D}^\mu
q_R+i\bar{q}_L\gamma_\mu \mathcal{D}^\mu q_L.
\end{equation}
One may check that this Lagrangian is invariant under the global
transformations:
\begin{equation}
q_R=\quark{u_R}{d_R} \mapsto \exp{\big(-i \vec{\Theta}^R \cdot
\frac{\vec{\tau}}{2}\big)} \quark{u_R}{d_R}
\end{equation}
and,
\begin{equation}
q_L=\quark{u_L}{d_L} \mapsto \exp{\big(-i \vec{\Theta}^L \cdot
\frac{\vec{\tau}}{2}\big)} \quark{u_L}{d_L},
\end{equation}
where $\vec{\tau}$ are pauli matrices generating the flavor symmetry
between up and down. As right and left quarks do not mix, this
implies the {\it{chiral symmetry}} $SU(2)_R \times SU(2)_L$ of QCD.
One can rewrite this to a symmetry on the original quark field:
\begin{equation}
q=\quark{u}{d} \mapsto \exp{\big[-i \frac{1}{2}\big(1 \pm
\gamma_5\big) \vec{\Theta} \cdot \frac{\vec{\tau}}{2}\big]}
\quark{u}{d}.
\end{equation}
The chiral symmetry can be transformed linearly to a vector
symmetry,
\begin{equation}
q=\quark{u}{d} \mapsto \exp{\big[-i  \vec{\Theta}^V \cdot
\frac{\vec{\tau}}{2}\big]} \quark{u}{d},
\end{equation}
and axial symmetry,
\begin{equation}
q=\quark{u}{d} \mapsto \exp{\big[-i \gamma_5 \vec{\Theta}^A \cdot
\frac{\vec{\tau}}{2}\big]} \quark{u}{d}.
\end{equation}
N\"{o}ther's theorem suggests the existence of conserved vector and
axial currents: $\vec{V}^\mu=\bar{q} \gamma^\mu \frac{\vec{\tau}}{2}
q$ and $\vec{A}^\mu=\bar{q} \gamma^\mu \gamma_5 \frac{\vec{\tau}}{2}
q$.

The problem is that this symmetry entails a degeneracy of ground
states. A vector degeneracy is found, the {\textit{isospin
symmetry}}, which couples for example the proton and neutron to a
single entity, the nucleon. This is not the case for the axial
symmetry, as multiplets of opposite parities are not observed in
nature. One concludes that in nature $SU(2)_R \times SU(2)_L$ is
broken to $SU(2)_V$.

This is an example of spontaneous breaking of symmetry, i.e. a
symmetry of the Lagrangian which is not realized in the ground
state. A consequent of thus is the existence of Goldstone--Nambu
bosons, massless particles that posses the quantum numbers of the
broken generators. In the case of the chiral symmetry, the Goldstone
bosons should be pseudoscalar. The pions, pseudoscalar isospin
triplet, are considered to be the Goldstone bosons of the chiral
symmetry. As a result of the non-vanishing up and down small masses,
the pions obtain mass, but they are light
$\left(\frac{m_\pi}{m_N}\right)^2 \approx 0.02$.

The appearance of Goldstone--Nambu bosons is a result of the
Goldstone theorem \citep{Goldstone_Bosons}. This theorem gives an
additional information regarding the structure of the
transformations. The pion field is defined by,
\begin{equation}
\xi=\exp \left(i\frac{\vec{\pi} \cdot \vec{\tau}}{2f_\pi}\right).
\end{equation}
Goldstone theorem leads to the transformation rule for $U \equiv
\xi^2 $
\begin{equation}
U = \exp \left(i\frac{\vec{\pi} \cdot \vec{\tau}}{f_\pi}\right)
\mapsto L U R^\dagger,\, \text{ with } L \in SU(2)_L \text{ and }
R\in SU(2)_R
\end{equation}
Thus, $U$ transforms as the $(\bar{2},{2})$ representation of
$SU(2)_R \times SU(2)_L$. This is a non--linear realization of the
symmetry, as $U^\dagger U=1$ and $\det U =1 $.

This implies the following transformation rules on the nucleon and
pion fields:
\begin{eqnarray} \label{eq:TRANSFORMATION_rules}
& &  N \mapsto h N
\\
& & \xi \mapsto L \xi h^\dagger = h \xi R^\dagger
\end{eqnarray}
with $h \in SU(2)_V$, local matrix.

With these transformation laws, it is possible to build Lorentz
invariant terms to include in the Lagrangian of the effective
theory,
\begin{equation}
\mathcal{L}_{eff}=\sum_A c_A \mathcal{L}_A
\end{equation}
This is an effective theory, thus renormalizability is inserted via
the low energy coefficients. This sum however is infinite. Hence, a
logic in deciding the order of a term in the expansion is needed.

\section{Power Counting of Chiral Perturbation Theory}
EFT of low energy QCD is called Chiral Perturbation Theory
($\chi\text{PT}$). The theory has a natural cutoff scale, the chiral
symmetry breaking scale, which is of the order of the nucleon mass
$\Lambda \sim 1 \gev$. The expansion relies on the small momentum
typical to the needed process $Q$, thus the expansion parameter is
$\big(Q/\Lambda\big)$.

The contribution of an interaction is determined by its power:
$\big(Q/\Lambda\big)^\nu$. Let a scattering process be described by
a diagram. The diagram is built of $C$ separately connected pieces,
$A$ nucleons, $L$ loops. The diagram includes several vertices, each
involves $n_i$ nucleon fields, $d_i$ pion masses or derivatives. The
order $\nu$ is given by \citep{WEINBERG_BOOK_II}:
\begin{equation}
\nu=-2+2A-2C+2L+\sum_i \Delta_i
\end{equation}
with $\Delta_i \equiv d_i + \frac{n_i}{2}-2$.

One has to note that $\nu$ is bounded from below, thus enabling
systematic expansion. This counting holds not only for diagrams, but
also for Lagrangian terms. The counting for a Lagrangian term is
similar to a vertex: $\nu_\mathcal{L}=d+\frac{n}{2}$.

The power counting formalism introduced here is applicable when the
choice of cutoff $\Lambda$ is relevant for the energy-regime of the
process (see discussions by \citet{NvKT,NvKTresp}).
\section{Effective Lagrangian}
As aforementioned, the chiral effective field theory takes into
account pions and nucleons. Not surprisingly, the Lagrangian
includes three parts, a pure pion, a pure nucleon and pion--nucleon
interaction Lagrangian. In this section I will review the different
parts of the Lagrangian, and emphasize those parts needed for
deriving the currents within the nucleus. In general, Lagrangians to
next--to--leading order (NLO) are considered, built using the
formalism introduced in \citet{CWZ,CCWZ} and following the
discussion in \citet{AN02}.

\subsection{Pion Lagrangian}
The Lagrangian has to be a Lorentz invariant scalar term. Thus, it
should include even number of derivatives/pion masses. Since the
expansion used here is up to NLO, the pion Lagrangian term is
expanded to $\nu=2$, in the power counting introduced previously.
Thus,
\begin{equation}
\mathcal{L}_{\pi\pi}^{(2)}=\frac{f_\pi^2}{4} \rm{Tr} \left[ \partial
^\mu U \partial_\mu U^\dagger + m_\pi^2 \left( U + U^\dagger \right)
\right]
\end{equation}
$f_\pi \approx 92.4 \mev$  is the pion decay constant, calibrated
from this process at low pion four momentum. The pion mass is taken
as $m_\pi=139 \mev$.

\subsection{Pion--Nucleon Interaction Lagrangian}
The pion--nucleon interaction, to lowest order, arises from the
nucleon free Dirac Lagrangian. The reason is that the Dirac
Lagrangian $i\bar{N}\big(\gamma_\mu \partial^\mu -M_0 \big)N$ is not
invariant under the transformation rules of
Eq.~(\ref{eq:TRANSFORMATION_rules}). This is not surprising, as the
Yang--Mills theory demands adding gauge term to the derivative. The
additional term has to be of chiral order 1, equivalent to the
derivative. It can be easily checked that this term is:
\begin{equation}
v_\mu \equiv -\frac{i}{2} \left( \xi \partial_\mu \xi^\dagger +
\xi^\dagger \partial_\mu \xi \right).
\end{equation}
The chiral covariant derivative is: $\mathcal{D}_\mu = \partial_\mu
+ iv_\mu$. The notation $v_\mu$ indicates the vector--like parity of
this term. Trivially, a term with axial--like parity of the same
chiral order is,
\begin{equation}
a_\mu \equiv -\frac{i}{2} \left( \xi \partial_\mu \xi^\dagger -
\xi^\dagger \partial_\mu \xi \right).
\end{equation}
This leads to the leading order pion--nucleon Lagrangian,
\begin{equation} \label{eq:L_pi_N_2}
\mathcal{L}_{\pi N}^{(2)}= \bar{N} \left\{i\gamma_\mu
\mathcal{D}^\mu + g_A \gamma^\mu \gamma_5 a_\mu -M_0 \right\} N
\end{equation}
The low energy constant (LEC) introduced here is the axial coupling
constant $g_A = 1.2670 \pm 0.0035$ \citep{PDBook}. $M_0 \approx 890
\mev $ is the nucleon bare mass.

Expanding Eq.~(\ref{eq:L_pi_N_2}) to leading order in the pion
field, one gets the leading interaction Lagrangian:
\begin{equation} \label{eq:L_pi_N_2_perturbative}
\mathcal{L}_{int}^{(2)}=-\frac{1}{4f_\pi^2}\epsilon^{abc}\pi^a\bar{N}(\not
\mspace{-3mu}\partial\pi^b)\tau^cN+\frac{g_A}{2f_\pi}\bar{N}(\not\mspace{-3mu}
\partial\pi^a)\gamma_5
\end{equation}
Using this, one constructs the Feynman diagrams related to this
interaction Lagrangian.

Constructing the next--to--leading order contribution to the
pion--nucleon interaction is easier, as the building blocks are
$a_\mu$ and $v_\mu$, and $U$. It is easily shown that there are two
terms at this order:
\begin{eqnarray}
\label{eq:L1_piN} \delta_1 \mathcal{L}_{\pi N}^{(3)} =
\frac{2\hat{c}_3}{M_N} \bar{N}N \rm{Tr} \left(a_\mu a^\mu \right)
\\
\label{eq:L2_piN} \delta_2 \mathcal{L}_{\pi N}^{(3)} =
i\frac{\hat{c}_4}{M_N} \bar{N} [a_\mu,a_\nu] \sigma^{\mu \nu} N,
\end{eqnarray}
Where $\sigma^{\mu\nu}=\frac{i}{2}[\gamma^\mu,\gamma^\nu]$. The LECs
here are calibrated using $\pi-N$ scattering data $\hat{c}_3=-3.66
\pm 0.08$, and $\hat{c}_4= 2.11 \pm 0.08$ \citep{BE95}.

In view of the need of Feynman diagrams, one gets to this order:
\begin{equation} \label{eq:L_int_3}
\mathcal{L}_{int}^{(3)}=  \frac{i\hat{c}_4}{2f_\pi^2M}\bar{N}
\vec{\tau}\cdot(\partial_\mu\vec\pi\times \partial_\nu\vec\pi)
\sigma^{\mu\nu}N+ \frac{\hat{c}_3}{f_\pi^2 M} \bar{N}N
\partial_\mu\pi^a \partial^\mu\pi^a
\end{equation}

\subsection{Nucleon Contact terms in the Lagrangian}
Nucleon contact terms represent the short--range nuclear force,
usually connected to the exchange of heavy mesons, such as the
$\omega(782)$. This mass is too close to the QCD scale, so it cannot
be described as a multi--pion interaction using $\chi\text{PT}$. The
contact terms are important for the purpose of renormalization and
dimensional regularization of pion--exchange. In case of infinities,
the contact terms act as counter--terms.

The contact terms include two nucleons, without meson fields. Being
the representative of higher energy modes, these terms are not
constrained by chiral symmetry, and required to fulfill parity and
time-reversal symmetries \citep{NN94}. Parity suggests even numbers
of derivatives. The LO has no derivatives, and the NLO consist of
two derivatives/pion masses.

Contact terms appear also when pions, electro-weak leptons, photons
or other probes couple to pairs of nucleons. This gives rise to a
contribution of few nucleon terms to the nuclear force \citep{NN06},
and two nucleon exchange currents. The leading effects are
represented by a term in the chiral Lagrangian \citep{NN94_2},
\begin{equation} \label{eq:contact}
{\mathcal{L}}_4=-2D_1\left(\bar{N}\gamma^\mu\gamma_5 a_\mu
N\right)\left(\bar{N}N\right).
\end{equation}
In general, the LECs in the contact terms should be calibrated by
nuclear matter experiments. In the current case the short--range
LEC is calibrated using the triton half life. This process is
discussed and applied in Sec.~\ref{sec:d_r_cal}.

\section{Currents in the Nucleus}
The axial and vector symmetries of the Lagrangian imply conserved
N\"{o}ther currents. These Currents are,
\begin{equation}
\mathcal{J}^{a\mu} \equiv -\frac{\partial
\mathcal{L}}{\partial(\partial_\mu \epsilon^a(x))}
\end{equation}
where the superscript $a$ indicates isospin components.
$\epsilon^a(x)$ are local, infinitesimal parameters of the symmetry
transformation.

A vector transformation has to be invariant to parity. The
definition of parity, $P:L\rightleftarrows R$, means that for a
vector transformation with infinitesimal group parameters
$\vec{\beta}$,
\begin{equation}
L=R=\exp\left(i\frac{\vec{\beta}\cdot\vec{\tau}}{2}\right).
\end{equation}
The $h(x)$ transformation field can be found using
Eq.~(\ref{eq:TRANSFORMATION_rules}), to get $h(x)=L=R$.

For an axial transformation one gets
\begin{equation}
L=\exp\left(i\frac{\vec{\beta}\cdot\vec{\tau}}{2}\right) ,\,\,
R=L^\dagger=\exp\left(-i\frac{\vec{\beta}\cdot\vec{\tau}}{2}\right).
\end{equation}
In this case $h(x)$ should be solved from its definition in
Eq.~(\ref{eq:TRANSFORMATION_rules}). Putting
$h(x)=\exp\left(i\frac{\vec{\gamma}\cdot\vec{\tau}}{2}\right)$, one
can find,
\begin{equation}
\vec{\gamma}=-2(\hat{\beta}\times\hat{\pi}) \arctan \left[
\frac{\tan \frac{\beta}{2} \tan
\frac{\pi}{2f_\pi}}{1-(\hat{\beta}\cdot\hat{\pi}) {\tan
\frac{\beta}{2} \tan \frac{\pi}{2f_\pi}}}\right] =
-(\vec{\beta}\times\hat{\pi}) \tan \frac{\pi}{2f_\pi} + O(\beta^2)
\end{equation}

Using this, one can derive the currents corresponding to the
Lagrangians of the previous section. The vector currents:
\begin{align} \label{eq:V_pi_pi_2}
{\mathcal{V}}_{\pi \pi}^{(2)a\mu} & =  -i \frac{f_\pi^2}{4}
{\rm{Tr}} \left\{ {\tau}^a \left[U \partial^\mu U^ \dagger + U^
\dagger
\partial^\mu U \right] \right\}=(\vec{\pi}\times\partial^\mu
\vec{\pi})^a \rm{sinc}^2\left(\frac{\pi}{f_\pi}\right)
\\ \nonumber
& = (\vec{\pi}\times\partial^\mu \vec{\pi})^a  + O(\pi^3)
\\ \label{eq:V_pi_N_2}
{\mathcal{V}}_{\pi N}^{(2)a\mu} & =
\frac{1}{4}\bar{N}\gamma^\mu[\xi\tau^a\xi^\dagger+
\xi^\dagger\tau^a\xi]N+\frac{1}{4}g_A\bar{N}\gamma^\mu\gamma_5
[\xi\tau^a\xi^\dagger-\xi^\dagger\tau^a\xi]N =
\\ \nonumber
& = \bar{N}\gamma^\mu\frac{\tau^b}{2}N[\cos(\frac{\pi}{f_\pi})
(\delta^{ab}-\hat{\pi}^a\hat{\pi}^b)+\hat{\pi}^a\hat{\pi}^b]+
g_A\epsilon^{abc}\hat{\pi}^b \sin(\frac{\pi}{f_\pi})
\bar{N}\gamma^\mu\gamma_5\frac{\tau^c}{2}N =
\\ \nonumber
& = \bar{N}\gamma^\mu\frac{\tau^a}{2}N+
\frac{g_A}{f_\pi}\epsilon^{abc}\vec{\pi}^b
\bar{N}\gamma^\mu\gamma_5\frac{\tau^c}{2}N+O(\pi^2)
\\ \label{eq:V_pi_N_3}
{\mathcal{V}}_{\pi N}^{(3)a\mu} & = -\frac{i\hat{c}_4}{2M}
\bar{N}\sigma^{\mu\nu}
[\xi^\dagger\tau^a\xi-\xi\tau^a\xi^\dagger,a_\nu]N-
\frac{2\hat{c}_3}{M}\bar{N}N\mathrm{Tr}
[\xi^\dagger\tau^a\xi-\xi\tau^a\xi^\dagger,a_\mu]
\\ \nonumber
& =\frac{\hat{c}_4}{M f_\pi^2} \bar{N}\sigma^{\mu\nu} \left[
\vec{\pi} \times (\partial_\nu \vec{\pi} \times \vec{\tau})
\right]^a N
\end{align}
where ${\rm{sinc}}(x)=\frac{\sin x}{x}$. The axial currents:
\begin{align} \label{eq:AX_pi_pi_2}
{\mathcal{A}}_{\pi \pi}^{(2)a\mu} & =  -i \frac{f_\pi^2}{4}
{\rm{Tr}} \left\{ {\tau}^a \left[U \partial^\mu U^ \dagger - U^
\dagger
\partial^\mu U \right] \right\}
\\ \nonumber
& = - f_\pi \partial_\mu \pi^b \left[
{\rm{sinc}}\left(\frac{2\pi}{f_\pi}\right) \left(\delta^{ab} -
\hat{\pi}^a\hat{\pi}^b \right)+ \hat{\pi}^a\hat{\pi}^b \right]
\\ \nonumber
& = -f_\pi \partial_\mu \pi^b + O(\pi^2)
\\ \label{eq:AX_pi_N_2}
{\mathcal{A}}_{\pi N}^{(2)a\mu} & =
\frac{1}{4}\bar{N}\gamma^\mu[\xi\tau^a\xi^\dagger+
\xi^\dagger\tau^a\xi]N+\frac{1}{4}g_A\bar{N}\gamma^\mu\gamma_5
[\xi\tau^a\xi^\dagger-\xi^\dagger\tau^a\xi]N =
\\ \nonumber
& = -\epsilon^{abc}\hat{\pi}^b \sin(\frac{\pi}{f_\pi})
\bar{N}\gamma^\mu\frac{\tau^c}{2}N-g_A\bar{N}\gamma^\mu\gamma_5\frac
{\tau^b}{2}N[\cos(\frac{\pi}{f_\pi})(\delta^{ab}-\hat{\pi}^a\hat{\pi}^b)
+\hat{\pi}^a\hat{\pi}^b]
\\ \nonumber
& = -g_A\bar{N}\gamma^\mu\gamma_5\frac{\tau^a}{2}N-
\frac{1}{f_\pi}\epsilon^{abc}\vec{\pi}^b
\bar{N}\gamma^\mu\frac{\tau^c}{2}N+O(\pi^2)
\\ \label{eq:AX_pi_N_3}
{\mathcal{A}}_{\pi N}^{(3)a\mu} & = -\frac{i\hat{c}_4}{2M}
\bar{N}\sigma^{\mu\nu}
[\xi^\dagger\tau^a\xi+\xi\tau^a\xi^\dagger,a_\nu]N-
\frac{2\hat{c}_3}{M}\bar{N}N\mathrm{Tr}
[\xi^\dagger\tau^a\xi+\xi\tau^a\xi^\dagger,a_\mu]
\\ \nonumber
& = \frac{\hat{c}_4}{f_\pi M} \epsilon^{abc}
\partial_\nu \pi^b \bar{N}\tau^c\sigma^{\mu\nu}N-
\frac{2\hat{c}_3}{f_\pi M}\bar{N}N \partial^\mu \pi^a + O(\pi^2)
\\ \label{eq:AX_N_N_3}
{\mathcal{A}}_{N N}^{(3)a\mu} & = -D_1\bar{N}\gamma^\mu \gamma_5
{\tau^a} N \bar{N}N
\end{align}
The notation used in the upper index denotes: $(\nu_\mathcal{L})$ is
the chiral order, $a=0,\pm 1$ is the isospin projection, and
$\mu=0,\,..,\,3$ is the Lorentz vector projection. The expansion is
used to construct the Feynman diagrams for the currents (see
Appendix~\ref{ap:Feynman}).

\section{Nuclear Currents Operators}
Using the equations of the two previous sections, one constructs the
Feynman rules. The rules derived from the Lagrangian are used to
construct nuclear forces form $\chi\rm{PT}$. Let us assume a case,
in which the interaction Lagrangian is
\begin{equation} \label{eq:CC_Lagrangian}
\hat{\mathcal{L}}=-g  S_\mu(\vec{x}) \hat{\mathcal{J}}^\mu
(\vec{x}),
\end{equation}
where $S_\mu$ is an outer source current, and
$\hat{\mathcal{J}}^\mu$ is the nuclear current. This Lagrangian is
relevant for low--energy electro-weak interaction (cf.
Eqs.~(\ref{eq:HW},\ref{eq:Hem})). The Feynman rules derived from the
N\"{o}ther of the Lagrangian symmetry, are used to construct the
nuclear currents.

An extensive description of the Feynman rules and the derivation of
the nuclear currents appear in Appendix~\ref{ap:Feynman}. In view of
the typical momentum of the processes in discussion, one can use a
soft probe approximation and write the currents to first order in
the momentum transfer. It is also clear that any recoil of nucleons
is smaller, and thus the use of heavy baryon expansion is justified,
expanding the nucleon spinors to leading order in $\frac{1}{M_N}$
(see Appendix~\ref{app:spinors})

The leading nuclear current operators are single nucleon operators,
i.e. describing a coupling to only one nucleon within the nucleus,
as if it was a free particle. The next contribution involves two
nucleons, in which case the probe interacts via a contact term of
two nucleons or a nucleon--pion vertex.

The resulting matrix elements are expressed in momentum space, and
as they originate in EFT approach, they are valid up to a cutoff
momentum $\Lambda$. The cutoff momentum is chosen such that $\Lambda
\gg Q$, and lower than QCD scale. A proper choice is thus $\Lambda =
400-800 \mev$, as in \citep{PA03,NN94,NN06}. The renormalization
group theory states that an observable does not depend on the cutoff
momentum. Any cutoff dependence implies excitation of higher energy
degrees of freedom. Thus, the cutoff dependence is a consistency
check for the theory. In the Lagrangians of the previous section,
the only LEC with a connection to higher energy modes is
$\hat{d}_r$, thus one should expect a rather steep cutoff dependence
for this LEC, and its cutoff dependence is chosen in such manner
that the triton half--life reproduces its experimental value.

The nuclear wave functions are written in configuration space, thus
one should use a proper transform to move the operators to this
basis. Fourier transform is of no help as it includes high momentum.
Following \citet{PA03}, let us define the cutoff Fourier transform
of an n-body current as
\begin{eqnarray}
\lefteqn{\hat{\mathcal{J}}^a_{12\ldots
n}(\vec{x}_1,\ldots,\vec{x}_n;\vec{q}) =} \\ \nonumber & & {
=\left[\prod_{i=1}^{n}\int\frac{d^3 k_i}{(2\pi)^3}
e^{i\vec{k}_i\cdot\vec{x}_i}S_\Lambda(\vec{k}_i^2)\right] (2\pi)^3
\delta^{(3)}(\vec{q}+\vec{k}_1+\vec{k}_2+\cdots+\vec{k}_n)
\mathcal{\hat{J}}^a_{12\ldots
n}(\vec{k}_1,\ldots,\vec{k}_n;\vec{q})}.
\end{eqnarray}
The cutoff function, $S_\Lambda(k) = \exp(-\frac{k^2}{2\Lambda^2})$,
does not influence one body currents, due to the momentum conserving
$\delta$ function. For $Q \lesssim \Lambda$ the cutoff function
$S_\Lambda(k)$ is unity, and it approaches zero for $Q \gtrsim
\Lambda$. This choice of cutoff function is convenient both for
numerical and analytical calculations. One should note that this
cutoff function inserts an additional $\frac{k^2}{\Lambda^2}$
dependence to the currents. This is, however, a minor, negligible
effect.

For low energy reactions, where meson exchange currents (MEC) are a
perturbative effect, one can use the long wavelength approximation
$q=0$. Thus, the Fourier transform of a MEC depends only on the
relative coordinate $\vec{x}_{12}=\vec{x}_1-\vec{x}_2$, and
\begin{equation} \label{eq:Fourier}
\mathcal{A}^a_{12}(\vec{x}_{12}) = \int\frac{d^3
k}{(2\pi)^3}e^{-i\vec{k}\cdot\vec{x}_{12}} S^2_\Lambda(\vec{k}^2)
\mathcal{A}^a_{12}(\vec{k}_1=-\vec{k},\vec{k}_2=\vec{k}).
\end{equation}
As the Fourier transform acts on pion propagator, the needed
functions are,
\begin{equation}
\delta_\Lambda^{(3)}(\vec{r}) \equiv \int\frac{d^3
k}{(2\pi)^3}e^{i\vec{k}\cdot\vec{r}} S^2_\Lambda(\vec{k}^2),
\end{equation}
\begin{equation}
y^\pi_{\Lambda 0}(r) \equiv \int\frac{d^3
k}{(2\pi)^3}e^{i\vec{k}\cdot\vec{r}}
S^2_\Lambda(\vec{k}^2)\frac{1}{\vec{k}^2+m_\pi^2},
\end{equation}
\begin{equation}
y^\pi_{\Lambda 1}(r) \equiv -\frac{\partial}{\partial r}
y^\pi_{\Lambda 0}(r),
\end{equation}
\begin{equation}
y^\pi_{\Lambda 2}(r) \equiv
\frac{1}{m^2_\pi}r\frac{\partial}{\partial r}
\frac{1}{r}\frac{\partial}{\partial r} y^\pi_{\Lambda 0}(r).
\end{equation}
One reproduces the usual Yukawa functions in the limit $\Lambda
\rightarrow \infty $.

This process leads to operators acting on one nucleon in the case of
single nucleon operator, or on a pair of nucleons in the case of MEC
operator. To move $A$ nucleons (or $\frac{A(A-1)}{2}$ pairs of
nucleons) to their coordinates in configuration space one uses first
quantization. For the single nucleon operators,
\begin{equation}
\hat{\mathcal{J}}^{(1)\mu}(\vec{x})=\sum_{i=1}^A
{\mathcal{\hat{J}}}^{\mu}_1(\vec{r}_{i})
\delta^{(3)}(\vec{x}-\vec{r}_{i}),
\end{equation}
and for the MEC opeartors,
\begin{equation}
\hat{\mathcal{J}}^{(12)\mu}(\vec{x})=\sum_{i<j}^A
{\mathcal{\hat{J}}}^{\mu}_{12}(\vec{r}_{ij})
\delta^{(3)}(\vec{x}-\vec{r}_{ij}).
\end{equation}

This leads to the final form of the operators. The single nucleon
operators, which are often called the impulse approximation (IA) are
the same as those derived from standard nuclear physics approach:
\begin{align} \label{eq:V0}
\hat{\mathcal{V}}^{(1)0}_i & = \delta^{(3)}(\vec{x}-\vec{r}_i) -
\frac{1}{2}\frac{\vec{\nabla}}{2M_N}\cdot \left\{\vec{\sigma}_i
\times \frac{\vec{p}_i}{2M_N} , \delta^{(3)}(\vec{x}-\vec{r}_i)
\right\}
\\ \label{eq:A0}
\hat{\mathcal{A}}^{(1)0}_i & =-g_A \left(\vec\sigma_i\cdot
\left\{\frac{\vec{p}_i}{2M_N}, \delta^{(3)}(\vec{x}-\vec{r}_i)
\right\} - \frac{1}{2}\left\{\left(\frac{p_i}{2M_N}\right)^2
,\delta^{(3)}(\vec{x}-\vec{r}_i)\right\} \right)
\\ \label{eq:V}
\hat{\vec{\mathcal{V}}}^{(1)}_i & = \left\{\frac{\vec{p}_i}{2M_N}
,\delta^{(3)}(\vec{x}-\vec{r}_i)\right\}
\\ \label{eq:A}
\hat{\vec{\mathcal{A}}}^{(1)}_i &  =- g_A \left(\vec\sigma_i
\delta^{(3)}(\vec{x}-\vec{r}_i) -\frac{\vec{\sigma}}{2} \left\{
\frac{\vec{p}_i}{2M_N}, \left\{\frac{\vec{p}_i}{2M_N},
\delta^{(3)}(\vec{x}-\vec{r}_i) \right\} \right\} \right. \\
\nonumber  & \left. + \left\{
\frac{\vec{p}_i}{2M_N}\cdot\vec{\sigma},
\left\{\frac{\vec{p}_i}{2M_N}, \delta^{(3)}(\vec{x}-\vec{r}_i)
\right\} \right\} + \frac{\vec \nabla \times}{4M_N}
\left\{\frac{\vec{p}_i}{2M_N}, \delta^{(3)}(\vec{x}-\vec{r}_i)
\right\} \right)
\end{align}
Due to the Siegert theorem (Sec.~\ref{sec:Siegert}) for low energy
reactions, the conservation of vector currents implies that neither
the vector current nor the vector MEC should be calculated
explicitly, as they are included implicitly in the calculation of
the single nucleon charge operator.

The axial charge MEC operator is:
\begin{eqnarray}
{ \hat{\mathcal{A}}_{12}^{0,a}(r_{ij})=-\frac{g_A m_\pi}{4f_\pi^2}
(\vec{\tau}^{(1)}\times\vec{\tau}^{(2)})^a
(\vec{\sigma}^{(1)}+\vec{\sigma}^{(2)})\cdot\hat{r}_{ij}
y_{1\Lambda}^\pi(r_{ij})},
\end{eqnarray}
and the axial MEC,
\begin{eqnarray} \label{eq:MEC_operator}
\lefteqn{
\hat{\mathcal{A}}_{12}^{i,a}(r_{ij})=\frac{g_A}{2Mf_\pi^2}\left\{
{\mathcal{O}}_P^{i,a} y_{1\Lambda}^\pi(r_{ij})+
\hat{c}_3({\mathcal{T}}_\oplus^{i,a}-{\mathcal{T}}_\ominus^{i,a})
m^2_\pi y_{2\Lambda}^\pi(r_{ij})-\right.}\nonumber \\ && {\left.
-\hat{c}_3({\mathcal{O}}_\oplus^{i,a}-
{\mathcal{O}}_\ominus^{i,a})(\frac{1}{3}\delta_\Lambda^{(3)}(\vec{r}_{ij})-\frac{1}{3}m_\pi^2
y_{0\Lambda}^\pi)- \right.}\nonumber \\ && {
\left.-(\hat{c}_4+\frac{1}{4}){\mathcal{O}}_\otimes^{i,a} m^2_\pi
y_{2\Lambda}^\pi(r_{ij})- \right.}\nonumber \\ && {
-\left.(\hat{c}_4+\frac{1}{4}) {\mathcal{O}}_\otimes^{i,a}
(\frac{2}{3}\delta_\Lambda^{(3)}(\vec{r}_{ij})- \frac{2}{3}m_\pi^2
y_{0\Lambda}^\pi) \right\}+} \nonumber \\ &&
{+\frac{D_1}{2}({\mathcal{O}}_\oplus^{i,a}+
{\mathcal{O}}_\ominus^{i,a})\delta_\Lambda^{(3)}(\vec{r}_{ij})}
\end{eqnarray}
where
\begin{equation}
{\vec{\mathcal{O}}^{a}_P \equiv -\frac{m_\pi}{4}
(\vec{\tau}^{(1)}\times\vec{\tau}^{(2)})^a
(\vec{P}_1\vec{\sigma}^{(2)}\cdot\hat{r}_{12}+
\vec{P}_2\vec{\sigma}^{(1)}\cdot\hat{r}_{12})}
\end{equation}
\begin{equation}
{{\mathcal{O}}^{i,a}_\odot \equiv
(\vec{\tau}^{(1)}\odot\vec{\tau}^{(2)})^a
(\vec{\sigma}^{(1)}\odot\vec{\sigma}^{(2)})^i},
\end{equation}
\begin{equation}
{{\mathcal{T}}^{i,a}_\odot \equiv
\left(\hat{r}^i_{12}\hat{r}^j_{12}-
\frac{\delta^{ij}}{3}\right){\mathcal{O}}^{i,a}_\odot},
\end{equation}
and $\odot=\times,+,-$.\\
It is important to notice that the contact parts of the current,
i.e. terms proportional to $\delta(r)$, do not vanish only for pairs
in a relative $L=0$ state. This implies $S+T=1$, meaning either
$S=0$ or $T=0$. Consequently, when $\mathcal{O}_\oplus$ acting on
these states, it vanishes.

An additional simplification is achieved by noting that
$\mathcal{O}_\otimes$ and $\mathcal{O}_\ominus$ are connected. The
fermion statistics implies $\Xi^\tau \Xi^\sigma \Xi^r=-1$, where
$\Xi^\tau$ exchanges isospin of the two particles, $\Xi^\sigma$
exchanges spin, and $\Xi^r$ exchanges locations. A trivial identity
for Pauli matrices: $\vec{\sigma}^1 \times \vec{\sigma}^2=
i(\vec{\sigma}^1-\vec{\sigma}^2) \Xi^\sigma $. Thus,
$\mathcal{O}_\otimes=-\mathcal{O}_\ominus \Xi^r$, and as $\Xi^r
\delta(r)= 1$ the connection is clear \citep{PA03}.

Thus, all the coefficients of $\delta^{(3)}(\vec{r})$ are combined
into a single dimensionless LEC\footnote{Different variations of EFT
(e.g. \citep{PA03}) lead to different descriptions of the short
range forces. These, however, renormalize to the same LEC.}
$\hat{d}_r=\frac{2Mf_\pi^2}{g_A}D_1+\frac{\hat{c}_3}{3}+
\frac{2}{3}(\hat{c}_4+\frac{1}{4})$. The $O(\Lambda^{-2})$ appearing
in $\delta_\Lambda^{(3)}(\vec{r})$ has minor effect, which cancels
when renormalizing $\hat{d}_r$. The renormalization
$\hat{d}_r(\Lambda)$ for a range of cutoff momentum $\Lambda=400-800
\mev$, is calibrated to reproduce the triton half--life.

Bearing this in mind, the axial MEC are
\begin{eqnarray}
\lefteqn{\hat{\mathcal{A}}_{12}^{i,a}(\vec{r_{ij}})=
\frac{g_A}{2Mf_\pi^2} \hat{d}_r \mathcal{O}^{i,a}_\ominus
\delta_\Lambda^{(3)}(\vec{r}_{ij}) - \frac{g_A m_\pi^2}{2Mf_\pi^2}
{\mathcal{O}}_P^{i,a} y_{1\Lambda}^\pi(r_{12})- } \\
\nonumber & & {- \frac{g_A m_\pi^2}{2Mf_\pi^2} \left[
\frac{\hat{c}_3}{3} \big({\mathcal{O}}_\oplus^{i,a}+
{\mathcal{O}}_\ominus^{i,a}\big) +\frac{2}{3}
\big(\hat{c}_4+\frac{1}{4}\big)
{\mathcal{O}}_\otimes^{i,a} \right] y_{0\Lambda}^\pi(r_{ij})}- \\
\nonumber & & {- \frac{g_A m_\pi^2}{2Mf_\pi^2} \left[ \hat{c}_3
\big({\mathcal{T}}_\oplus^{i,a}+ {\mathcal{T}}_\ominus^{i,a}\big) -
\big(\hat{c}_4+\frac{1}{4}\big) {\mathcal{T}}_\otimes^{i,a} \right]
y_{2\Lambda}^\pi(r_{ij})}
\end{eqnarray}

\section{Standard Nuclear Physics and EFT}
This chapter has shown the vast advantage in applying EFT approach
for the nuclear problem. The theory provides a perturbative approach
with its symmetries completely under control. This fact gives a
predictive measure to the approach.

These advantages are missing in the standard nuclear physics
approach (SNPA), which is phenomenological. However, SNPA approach
has many successes, using Lagrangians that include Nucleon--Nucleon
potentials and three nucleon forces one can describe the low lying
spectrum of nuclei with $A<12$. SNPA Lagrangians are used to
calculate electro--weak and strong nuclear reactions, to good
accuracy. The Lagrangians are built to reproduce hundreds
nucleon--nucleon scattering processes with energies up to $500
\mev$, with $\chi^2$ per datum close to unity. For example, the
binding energies of light nuclei are reproduced using these forces
to accuracy of tens of keV.

In order to reach this level of accuracy with EFT based Lagrangians,
one has to expand to at least next--to--next--to--next--to--leading
order, which includes more than $20$ parameters. This was done only
lately, by \citet{EFT1,EFT2}, and still has not arrived the level of
$\chi^2$ achieved in SNPA.

Lately, Rho {\it et al.} \citep{rho06,PA03} tried to combine the
advantages of both approaches. The idea relies on the fact that
albeit phenomenological, the SNPA approach holds correct long range
behavior. Thus, one may use the nuclear wave functions predicted by
SNPA, and calculate reactions using the nuclear currents derived in
EFT. The typical accuracy for reactions is in the percentage level,
hence it is sufficient to expand the forces up to next--to--leading
order, with only one unknown LEC needed for renormalization.

This hybrid approach, named EFT*, is of course not consistent, as
the currents and forces are not derived from the same Lagrangian.
Nevertheless, it has been used to calculate reactions of $A=2,3,4$
nuclei \citep{rho06,PA03}. The minor response of the calculated
cross-sections to changes in the cutoff momentum is an evidence for
the power of the method. This method will be used in this work
(Chap.~\ref{chap:neu}) for the calculation of neutrino scattering on
$A=3$ and $^4$He nuclei.

%% file: lit_chapter.tex
\chapter{Lorentz Integral Transform Method}
\label{chap:LIT}

The response of a system to excitations is a key ingredient in
understanding its properties: stability, energy levels, structure,
etc.

For an inclusive reaction with energy transfer $\omega$, described
by an interference between two operators $O_1(\vec{q})$ and
$O_2(\vec{q})$, the response function is defined as:
\begin{equation} \label{eq:response_function}
\mathcal{R}_{\hat{O}_1 \hat{O}_2}(\omega,\vec{q}) \equiv \sumint_f
\delta(E_f-E_i-\omega) \bra i | \hat{O}^\dagger_1(\vec{q}) | f \ket
\bra f | \hat{O}_2(\vec{q}) | i \ket,
\end{equation}
where $|i/f\ket$ and $E_{i/f}$ denote the initial and final wave
functions and energies, respectively. In order to calculate this
observable, one has to consider every possible final state,
including those which are in the continuum. For the light nuclei,
which have few bound excited states, if any, the continuum must be
taken into account if one wants to consider $\omega \ne 0$. This
makes an explicit calculation of the cross--section for such nuclei
a difficult task, not only as one has to keep track of all break--up
channels, but also since a general continuum state wave function is
out of reach already for $A=4$.

However, the response function is an integral quantity, as it
depends only on projections of the initial state. This has been the
motivation for using integral transforms on the response functions,
by which, as shown in the next Section, one can avoid an explicit
calculation of final state.

One of these transforms is Lorentz Integral Transform (LIT) method.
This method has been used in the calculation of various strong and
electro--magnetic nuclear reactions (see for example
\citep{ELO97,ELO97_2,ELO97pol,ELOT,BELO01,bacca2,bacca1}). In this
thesis the LIT method is applied for weak reactions on nuclei
\citep{GA04,GA07a,GA07b,GA07}. Though used here for inclusive
reactions, it has been applied for exclusive reactions as well
\citep{Sofia1}.

\section{Integral Transformations}
Using integral transform transforms the response function to a
continuous function, which depends on a parameter $\sigma$, using an
analytic kernel function $K(\omega,\sigma)$. The integral transform
is a functional of the response function, defined as:
\begin{equation} \label{eq:integral_transform}
\mathcal{I}[\mathcal{R}(\omega,\vec{q})](\sigma,\vec{q}) =
\int_{0^{-}}^{\infty} d\omega \mathcal{R}(\omega,\vec{q})
K(\omega,\sigma)
\end{equation}
The integration starts from $0^-$ to include every possible
scattering, including elastic channels. Finite transform and its
existence depends on the high energy (high $\omega$) behavior of
both the response function and the kernel, and should be checked for
each case.

The advantage in using an integral transform is easily viewed when
inserting the definition of the response function,
Eq.~(\ref{eq:response_function}), in
Eq.~(\ref{eq:integral_transform}). The integration over the energy
conserving $\delta$ function is trivial,
\begin{equation} \nonumber
\mathcal{I}(\sigma,\vec{q}) = \sumint_f \bra i | \hat{O}^\dagger_1
(\vec{q}) | f \ket \bra f | \hat{O}_2(\vec{q}) | i \ket
K(E_f-E_i,\sigma).
\end{equation}
As $K(E_f-E_i,\sigma)$ is a number, one can insert it to the matrix
element. Using then the fact that the states $|f\ket$ are
eigenstates of the Hamiltonian, $\hat{\mathcal{H}}$, namely
$\mathcal{\hat{H}} | f \ket = E_f | f \ket$, yields the expression
\begin{equation} \nonumber
\mathcal{I}(\sigma,\vec{q}) = \sumint_f \bra i | \hat{O}^\dagger_1
(\vec{q}) K(\mathcal{\hat{H}}-E_i,\sigma) | f \ket \bra f |
\hat{O}_2(\vec{q}) | i \ket .
\end{equation}

One notes that in this expression the final state appears only
through the projection operator $| f \ket \bra f |$. The hermiticity
of the Hamiltonian implies closure: $\sumint_f | f \ket \bra f |=1
$. Thus,
\begin{equation} \label{eq:integral_transform_ME}
\mathcal{I}(\sigma,\vec{q}) = \bra i | \hat{O}^\dagger_1(\vec{q})
K(\mathcal{\hat{H}}-E_i,\sigma) \hat{O}_2(\vec{q}) | i \ket.
\end{equation}

There is no doubt that the use of integral transforms simplifies the
calculation, as it depends only on the ground state properties.
However, this is merely a disguise as the kernel operator depends
explicitly on the Hamiltonian $\mathcal{\hat{H}}$. The latter acts
on $\hat{O}_2(\vec{q}) | i \ket$, which is not an eigenstate, hence
depends on all basis states. An additional unresolved question is
the inversion of the transform.

Obviously, an intelligent choice of the kernel can solve both these
problems. Nevertheless, the solution can never be perfect. A choice
of a broad kernel smears out narrower structures, which when
inverting the transform results in numerical instability. Choosing
narrow kernel leads back to the solution of the continuum state
problem.

In the literature, different kernels have been proposed. Among which
is the Laplace kernel $K(\omega,\sigma)=e^{-\sigma\omega}$. The
calculation of the transform with this kernel is relatively easy, as
it can be presented as propagation in imaginary time. This fact has
made the Laplace kernel appropriate for Quantum Monte Carlo
calculation (e.g. \citep{CAR94}). The disadvantage is the fact that
an inversion of Laplace transformation demands an analytic
continuation of the integral transform, which is usually
unavailable.
%The common method
%\citep{CAR94} is to use a Laplace transformation on the experimental
%results, to facilitate a comparison. This is possible only when
%measurements exist up to very high energy, to exhaust the
%integration to infinity, or when theoretical extrapolation is
%possible. Of course when predictiveness is needed, the inversion
%problem holds.

In view of the needed narrowness of an optimal kernel, it is
valuable to use a kernel which depends on an additional parameter
$\sigma_W$, with the following properties:
\begin{itemize} \label{kernel_properties}
  \item $K(\omega,\sigma;\sigma_W)=K(\omega-\sigma;\sigma_W)$.
  \item $\lim_{\sigma_W \rightarrow 0} K(\omega-\sigma;\sigma_W) =
  \delta(\omega-\sigma)$.
\end{itemize}
The advantages are clear. $\sigma_W$ is chosen in view of the
anticipated width of the reaction cross-section. Photodissociation
and neutrino scattering reactions are dominated by giant resonance
excitations, with a typical width of $10-30 \mev$. Thus a reasonable
choice is $\sigma_W \approx 10 \mev$.

Possible examples for kernels which satisfy the aforementioned
properties are the lorentzian and the gaussian. The Lorentzian turns
to be simple to calculate, hence attractive.

\section{Integral Transform with a Lorentzian Kernel}
A Lorentzian kernel is defined as $K(\omega,\sigma)=
\frac{1}{(\omega-\sigma_R)+\sigma_I^2}$, where
$\sigma=-\sigma_R+i\sigma_I$ is a complex continuous parameter, with
a real part indicating the transform variable and its imaginary part
indicates the Lorentzian width. The Lorentz Integral Transform (LIT)
is defined as,
\begin{equation} \label{eq:LIT_definition}
\mathcal{L}(\sigma,\vec{q}) = \int_{0^{-}}^{\infty} d\omega
\frac{\mathcal{R}(\omega,\vec{q})}{(\omega-\sigma_R)+\sigma_I^2}.
\end{equation}

Putting the LIT kernel definition in
Eq.~(\ref{eq:integral_transform_ME}) yields:
\begin{equation} \label{eq:LIT_ME_base}
\mathcal{L}(\sigma,\vec{q}) = \bra i \big |
\hat{O}^\dagger_1(\vec{q})
\frac{1}{\mathcal{\hat{H}}-E_0+\sigma^{*}}
\frac{1}{\mathcal{\hat{H}}-E_0+\sigma} \hat{O}_2(\vec{q}) \big| i
\ket,
\end{equation}
which has the form of an internal product:
\begin{equation} \label{eq:LIT_ME}
      \mathcal{L}(\sigma,\vec{q})= \bra \tilde{\Psi_1} \big|
      \tilde{\Psi_2} \ket
\end{equation}
with,
\begin{equation} \label{eq:Psi_i}
\big|\tilde{\Psi_i} \ket
\equiv\frac{1}{\mathcal{\hat{H}}-E_0+\sigma} \hat{O}_i(\vec{q})
\big| i \ket , \text{    for }i=1,2.
\end{equation}

This important result implies that the evaluation of the LIT
requires the solution of the following equation:
\begin{equation} \label{eq:Psi_i_eq}
(\mathcal{\hat{H}}-E_0+\sigma)\big | \tilde{\Psi}_i(\sigma)\rangle =
 \hat{O}_i\big | i \rangle .
\end{equation}
These equations have several appealing properties:
\begin{description}
  \item[Uniqueness and existence of the solution] The hermiticity of the
  Hamiltonian implies real energies, hence $E_f-E_0-\sigma \ne
  0$ for $\sigma_I \ne 0$. Thus the homogenous equation has only
  the trivial solution. As a result, there could only be a unique
  solution dictated by the source term in the right--hand--side of the
  equation, and the boundary conditions are dictated by it.
  The source term has a finite
  norm, i.e. the boundary conditions of a bound state problem. This
  means that the inner product of Eq.~(\ref{eq:LIT_ME}) is finite
  and thus the transform exists, and unique.
  \item[LIT is Schr\"{o}dinger--like problem] In view of the bound state
  boundary conditions, Eq.~(\ref{eq:Psi_i_eq}) is a
  Schr\"{o}dinger equation with a source term, thus can be solved
  using the variety of methods available for such problems. One of
  these methods is the Effective Interaction in the Hyperspherical
  Harmonics (EIHH) discussed in the following chapter.
  \item[LIT is a Green--function--like matrix element] It is easy to rewrite
  Eq.~(\ref{eq:LIT_ME_base}) as,
  \begin{equation} \label{eq:LIT_LANCZOS}
   \mathcal{L}(\sigma,\vec{q}) = \frac{1}{\sigma_I} {\textrm{Im}} \left\{ \bra i \big |
\hat{O}^\dagger_1(\vec{q}) \frac{1}{\mathcal{\hat{H}}-E_0+\sigma}
\hat{O}_2(\vec{q}) \big| i \ket \right\}.
  \end{equation}
  Thus, a matrix element of a green--function. Several methods exist to calculate such a
  matrix--element, one numerically effective method is the Lanczos
  algorithm, described in the following subsection.
\end{description}

The Lorentz integral transform clearly simplifies calculation of
response functions. The method avoids the difficulty of calculating
continuum wave functions. Instead a bound--state like equation has
to be solved, enabling the theoretical evaluation of previously
unreachable reactions.

\subsection{Calculation of LIT using the Lanczos Algorithm}
\label{sub:LIT_Lanczos} In this subsection an efficient algorithm
for the calculation of the LIT presentation of
Eq.~(\ref{eq:LIT_LANCZOS}) is reviewed, based on the Lanczos
Algorithem \citep{Lanczos}. First, let us focus on the
$\hat{O}_1(\vec{q}) =\hat{O}_2(\vec{q})=\hat{O}$ case.

The Lanczos basis $\{\big | \varphi_{n} \ket ; n\ge 0 \}$ is an
orthonormal set of states starting with a known state $\big |
\varphi_{0} \ket$. The rest of the basis can be defined recursively,
\begin{equation} \label{eq:Lanczos_basis_def}
b_{n+1} \big | \varphi_{n+1} \ket = \mathcal{\hat{H}} \big |
\varphi_{n} \ket - a_n \big | \varphi_{n} \ket - b_n \big |
\varphi_{n-1} \ket, \textrm{   for } n \ge 1,
\end{equation}
With the condition $b_0=0$. From Eq.~(\ref{eq:Lanczos_basis_def}) it
is trivial that the Lanczos coefficients $a_n$ and $b_n$ are given
by:
\begin{equation} \label{eq:Lanczos_coef_def}
a_n=\bra  \varphi_{n} \big| \mathcal{\hat{H}} \big| \varphi_{n} \ket
\text{ and } b_n=\bra  \varphi_{n} \big| \mathcal{\hat{H}} \big|
\varphi_{n-1} \ket.
\end{equation}
Choosing the Lanczos pivot as
\begin{equation}
 \big | \varphi_{0} \ket = \frac{\hat{O} \big | i \ket}{\sqrt{\bra i \big |
 \hat{O}^\dagger\hat{O} \big | i \ket}},
\end{equation}
allows to write the LIT as
\begin{equation}
\mathcal{L}(\sigma)= \frac{1}{\sigma_I}{\bra i \big |
 \hat{O}^\dagger\hat{O} \big | i \ket} \textrm{Im} \{x_{00}(E_0 -
 \sigma)\},
\end{equation}
Where
\begin{equation}
x_{n0}(z) \equiv \bra \varphi_{n} \big|
\frac{1}{\mathcal{\hat{H}}-z} \big| \varphi_{0} \ket.
\end{equation}
In order to calculate this , one uses the identity
$\big({\mathcal{\hat{H}}-z} \big)
\big({\mathcal{\hat{H}}-z}\big)^{-1}=I$, which leads to linear set
of equations,
\begin{equation}
\sum_{n} \big({\mathcal{\hat{H}}-z}\big)_{mn} x_{n0}(z) =
\delta_{m0}.
\end{equation}
Applying Kramers rule, one can express $x_{n0}(z)$ as continuous
fractions, in particular
\begin{equation} \label{eq:x00}
x_{00}(z)=\frac{1}{(z-a_0)-\frac{b_1^2}{(z-a_1)-\frac{b_2^2}{(z-a_2)-\frac{b_3^2}{(z-a_3)-\ldots}}}}.
\end{equation}
In typical examples a few hundreds of Lanczos coefficients are
sufficient to reach convergence in the LIT.

After defining $x_{n0}(z)$, a formulation of a LIT calculation with
$\hat{O}_1(\vec{q}) \ne \hat{O}_2(\vec{q})$ is available. In this
formulation, the Lanczos pivot defined as $\big | \varphi_{0} \ket =
\frac{\hat{O}_2 \big | i \ket}{\sqrt{\bra i \big |
\hat{O}_2^\dagger\hat{O}_2 \big | i \ket}}$. One inserts an identity
operator $\sum_i \big | \varphi_i \ket \bra \varphi_i \big | $ to
get:
  \begin{align} \label{eq:LIT_LANCZOS_al}
   \mathcal{L}(\sigma,\vec{q}) = \frac{{\sqrt{\bra i \big
| \hat{O}_2^\dagger\hat{O}_2 \big | i \ket}}}{\sigma_I}
{\textrm{Im}}  \sum_n  \bra i \big | \hat{O}^\dagger_1(\vec{q})
\big | \varphi_n \ket x_{n0}(z).
  \end{align}

\section{Inversion of the LIT}
The success of an integral transform highly relies on the existence
of a stable inversion. An analytic inversion for the Lorentz
transform does not exist.

One can observe the problem in inverting the LIT in the following
reason. Let us perform a Fourier transform on
Eq.~(\ref{eq:LIT_definition}). The r.h.s of that equation is a
convolution of lorentzian and the response function. Bearing in mind
that the fourier transform of a lorentzian is exponential decay, one
gets
\begin{equation}
\mathcal{F}[\mathcal{L}(\sigma,\vec{q})](k)=\frac{\pi}{\sigma_I}
e^{-\sigma_I |k|} \mathcal{F}[\mathcal{R}(\omega,\vec{q})](k).
\end{equation}
As a result, high energy (short wavelengths) are suppressed when
performing the LIT. Clearly, this fact creates instability in the
inversion, which enhances short wavelength oscillations.

However, a hope lies in the fact that mathematically the inversion
is unique. Moreover, there are several numerical approaches for the
inversion in the literature, which have provided stable and
successful results. The method used in this scope is expanding the
wanted inversion (i.e. the nuclear response function) in a set of
functions $f_n(\omega)$,
\begin{equation} \label{eq:inv_expansion}
\mathcal{R}(\omega,\vec{q})= \sum_{n=1}^{N} c_n(\vec{q})f_n(\omega).
\end{equation}
Since the transform is linear, it is obvious that the LIT of
Eq.~(\ref{eq:inv_expansion}) $\mathcal{L}(\sigma,\vec{q})=
\sum_{n=1}^{N} c_n(\vec{q})\mathcal{L}_n(\sigma)$, where
$\mathcal{L}_n(\sigma)$ is the known LIT of $\{f_n(\omega)\}$. One
obtains the coefficients $c_n$ by solving the minimal squares
problem, on a grid $\{\sigma_R^i,i=1,\ldots,K\}$:
\begin{equation}
\sum_{k=1}^{K} \left| \mathcal{L}(\sigma^k,\vec{q}) - \sum_{n=1}^{N}
c_n(\vec{q})\mathcal{L}_n(\sigma^k) \right|^2 = min
\end{equation}
The problem of the inversion is hence reduced to the solution of a
set of linear equations. It has been shown \citep{EF99a,EF99b} that
fitting on a grid such that $K \gg N$ regularizes the inversion,
thus solves the stability problem. \\
In order to increase both accuracy and control over the inversion,
one can take some important measures:
\begin{description}
  \item[Physical considerations] Often there are physical
  considerations on the asymptotic and threshold of the response
  function. The choice of a physical shape for $f_n$ can diminishes
  the number of needed expansion functions, while
  increasing the stability and accuracy of the inversion. Such a set of
  functions is:
  \begin{equation}
  f_n(\omega)= \omega^{n_0}e^{-\frac{\omega}{n \Delta E}}.
  \end{equation}
  This choice uses the fact that the threshold behavior is a power law, with
  $n_0 \thicksim 1.5$, and that the high energy tail decays exponentially
  with characteristic width $\Delta E$.
  \item[Sum Rules] Integration over the entirety of the response
  with a specific weight function depends on ground state
  properties. Thus, the inversion can be checked for consistency.
  In Chap.~\ref{chap:Photo} an extensive discussion on the subject of sum
  rules in photodisociation processes is given.
  \item[Stability with respect to $\rm{\sigma_I}$] The LIT can be
  calculated for various widths $\sigma_I$. A stability of the
  inversion with respect to $\sigma_I$ shows that there are no
  hidden structures of width less than $\sigma_I$.
\end{description}

%% file: HH_chapter.tex
\chapter{Nuclear Wave functions}
\label{chap:EIHH}

The nuclear problem regards the properties of an $A$--body quantum
mechanical non--relativistic system subjected to strong
correlations. As a result, the calculation of exact solutions to the
problem is a difficult task, demanding elaborate theoretical and
numerical methods. In order to solve the LIT equations, one needs a
good knowledge of the ground state wave function, and the ability to
calculate matrix elements.

In the non--relativistic approximation the nuclear Hamiltonian,
\begin{equation} \label{eq:HAMILTONIAN}
\mathcal{H}=\sum_{i=1}^{A} \frac{p_i^2}{2M_N} + \sum_{i<j=1}^{A}
V^{(2)}(\vec{r}_{ij})  + \sum_{i<j<k=1}^{A}
V^{(3)}(\vec{r}_i,\vec{r}_j,\vec{r}_k),
\end{equation}
includes not only a nucleon--nucleon (NN) interaction
$V^{(2)}(\vec{r}_{ij})$, which depends on the distance between the
nucleons $\vec{r}_{ij} \equiv \vec{r}_{i}-\vec{r}_{j}$, but also
three nucleon force (3NF) $V^{(3)}(\vec{r}_i,\vec{r}_j, \vec{r}_k)$.
The need in 3NF is observed already in the trinuclei binding
energies, where state of the art NN potentials lead to underbindings
of about $10\%$.

In this chapter I will present the use of hyperspherical coordinates
and functions for the evaluation of nuclear wave functions and
matrix elements. The effective interaction in hyperspherical
harmonics (EIHH) approach is introduced at the end of the chapter.

\section{Few--Body problems in terms of the Hyperspherical
Harmonics}

The most simple $A$--body problem is the $2$--body problem. One of
the conclusions from the solution of the $2$--body problem is that a
natural selection for representing the problem is to separate the
center of mass movement and the relative motion among the bodies. In
the case of an isolated system the center of mass movement is
trivial and does not affect the system. The center of mass
coordinate, for a system of $A$ equal masses is $\vec{R}_{cm}=
\frac{1}{A} \sum_{i=1}^{A} \vec{r}_i$. The internal coordinates are
defined as $\vec{r}^{\hspace*{0.3mm}{\prime}}_i=\vec{r}_i -
\vec{R}_{cm} $. This way, when working in the center of mass frame,
$\sum_i \vec{r}_i^{\hspace*{0.3mm}{\prime}}=0$.

It is clear that a good choice of internal coordinates can help
representing the nuclear problem. The Jacobi coordinates provide
such a choice, as the logic of their construction is the separation
of a subset of particles from the center of mass of the other
particles.

For a one body operator, one chooses
\begin{equation} \label{eq:JAC_1b}
\vec{\eta}_{k-1}=\sqrt{\frac{k-1}{k}} \left( \vec{r}_k-
\frac{1}{k-1} \sum_{i=1}^{k-1} \vec{r}_i \right) \text{ ; }
k=2,\ldots,A.
\end{equation}
In this set, the last Jacobi coordinate is proportional to the
coordinate of the last particle,
$\vec{\eta}_{A-1}=\sqrt{\frac{A}{A-1}}
\vec{r}^{\hspace*{0.3mm}{\prime}}_{A}$.

For a two--body operator, one chooses
\begin{equation} \label{eq:JAC_2b}
\vec{\eta}_{k}=\sqrt{\frac{A-k}{A+1-k}} \left( \vec{r}_k-
\frac{1}{A-k} \sum_{i=k+1}^{A} \vec{r}_i \right) \text{ ; }
k=1,\ldots,A-1.
\end{equation}
In this set, the last Jacobi coordinate is proportional to the
relative distance between the two last coordinates,
$\vec{\eta}_{A-1}=\sqrt{\frac{1}{2}} \big (\vec{r}_{A-1}
-\vec{r}_{A} \big)$.

As a result, the matrix element depends only on the last Jacobi
coordinate. One can use the invariance of the matrix element to the
other Jacobi coordinates by defining a generalization of the
spherical coordinates, {\it{the hyperspherical coordinates}}.

\subsection{Hyperspherical Coordinates}
The hyper--radius of a $k$--body internal system is defined as
\citep{EFROS72},
\begin{equation}
\rho_k^2=\sum_{i=1}^k \eta_i^2=\frac{1}{k}\sum_{i<j}^{k+1}
(\vec{r}_i-\vec{r}_j)^2.
\end{equation}
From this definition it is clear that $\rho_k^2=\rho_{k-1}^2 +
\eta_k^2$, so that $(\rho_{k-1},\eta_k,\rho_k)$ is a pythagorean
triplet, with a hyper--angle $\varphi_k \in
\left[0,\frac{\pi}{2}\right]$:
\begin{align} \label{eq:phi_k}
\rho_{k-1} & = \rho_k \cos \varphi_k,
\\ \nonumber
\eta_{k} & = \rho_k \sin \varphi_k.
\end{align}
With this, the $A-1$ internal degrees of freedom are described by
one hyper--radius $\rho_{A-1}$, $A-2$ hyper--angles $\varphi_{(A-2)}
\equiv \{ \varphi_2,\ldots,\varphi_{A-1} \}$, and $2(A-1)$ degrees
of freedom related to the solid angles of the Jacobi coordinates
$\Omega_{(A-1)} \equiv \{ \hat{\eta}_1,\ldots,\hat{\eta}_{A-1} \}$.

The volume element related to the internal $A$--body system is
\citep{BARNEA_thesis}
\begin{align}
dV_{3(A-1)} & = \rho_{A-1}^{3A-4} \rho_{A-1} dS_{3A-4}
\\ \nonumber
& = \rho_{A-1}^{3A-4} \rho_{A-1} \sin^2 (\varphi_{A-1}) \cos
^{3A-7}(\varphi_{A-1}) d\varphi_{A-1} d\hat{\eta}_{A-1} dS_{3A-7},
\end{align}
where $dS_N$ is a hyper--sphere surface element in $N$ dimensions.

One still has to write the kinetic energy operator using
hyperspherical coordinates. It is again useful to start with an
analogy to the $2$--body problem, where the Laplace operator for one
relative coordinate is
\begin{equation} \label{eq:LAPLACIAN_1}
\triangle_{\vec{\eta}_1}=\triangle_{{\eta}_1}-\frac{\hat{l}_1^2}{{\eta}_1^2}
\end{equation}
here the radial part is
$\triangle_{{\eta}_1}=\frac{1}{\eta_1^2}\frac{\partial}{\partial
\eta_1} \left({\eta_1^2}\frac{\partial}{\partial \eta_1}\right)$.
For a set of $A-1$ Jacobi coordinates, the Laplace operator is a sum
$\triangle_{(A-1)}= \sum_{i=1}^{A-1}\triangle_{\vec{\eta}_i}$. After
a transformation to $\rho_{A-1},\varphi_{A}$, one gets
\citep{EFROS72}:
\begin{equation} \label{eq:LAPLACIAN_N}
\triangle_{(A-1)}=\triangle_{\rho_{A-1}}-\frac{\hat{K}_{A-1}^2}{\rho_{A-1}^2}
\end{equation}
with the hyper--radial part
\begin{equation}
\triangle_{\rho_{A-1}}= \frac{\partial^2}{\partial \rho_{A-1} ^2} +
\frac{3(A-1)-1}{\rho_{A-1}} \frac{\partial}{\partial \rho_{A-1}}.
\end{equation}
The grand--angular--momentum operator $\hat{K}_{A-1}$ is built
recursively, starting with $\hat{K}_1 \equiv \hat{l}_1$. The
construction is done by choosing two subsets of Jacobi coordinates,
and using Eq.~(\ref{eq:phi_k}). If one chooses $A-2$ coordinates as
a subset, the recurrence relation is
$\triangle_{(A-1)}=\triangle_{(A-2)}+\triangle_{\vec{\eta}_{A-1}}$.
Putting Eqs.~(\ref{eq:LAPLACIAN_1},\ref{eq:LAPLACIAN_N}) in this
relation, one gets:
\begin{align}
\hat{K}_{A-1} & =- \frac{\partial^2}{\partial \varphi_{A-1}^2} +
\frac{3[(A-1)-2]- [3(A-1)-2] \cos 2\varphi_{A-1}}{\sin
2\varphi_{A-1}} \frac{\partial}{\partial \varphi_{A-1}} + \\
\nonumber
 & + \frac{\hat{K}_{A-2}^2}{\cos^2 \varphi_{A-1}} +
\frac{\hat{l}_{A-1}^2}{\sin^2 \varphi_{A-1}}
\end{align}
If we define the internal angular momentum operator
$\hat{\vec{L}}_{A-1}=\sum_{i=1}^{A-1} \hat{\vec{l}}_i$, it is clear
that the operators $\hat{K}_{A-1}$, $\hat{l}_{A-1}$,
$\hat{K}_{A-2}$, $\hat{{L}}^2_{A-1}$ and $\hat{\vec{L}}^Z_{{A-1}}$
are commutative. Thus, an eigenstate of the Laplacian can be labeled
by the quantum numbers $K_{A-1},\,\ldots,\, K_1$, $l_1,\,\ldots,\,
l_{A-1}$, $L_{A-1},\,\ldots,\, L_1$ and $M_{A-1}$.

\subsection{The Hyperspherical Harmonics Functions}
The hyperspherical harmonics functions are the eigen--functions of
the Laplace operator. As explained in the previous section the
operator is built recursively, a fact which reflects upon the
construction of its eigen--functions.

In general, when treating the nuclear problem as $A$ identical
particles (i.e. when identifying protons and neutrons in the isospin
symmetry) the internal wave functions belong to an irreducible
representation ({\it irrep}) of the symmetric group $\mathcal{S}_A$.
This simplifies the calculation of matrix elements, as for a generic
$N$--body operator $\hat{O}_{(N)}=\sum_{i_1<i_2<\ldots<i_N=1}^{A}
\hat{O}_{i_1,i_2,\ldots,i_N}$ the problem reduces to a matrix
element of an operator acting on one $N$--body subsystem $\bra \Psi
\left| \hat{O}_{(N)} \right| \Psi^\prime \ket = {{A}\choose{N}} \bra
\Psi \left| \hat{O}_{A-N+1,\ldots,A} \right| \Psi^\prime \ket $. The
relevant cases of one--, two--, and three--body operators are
written explicitly as,
\begin{align} \nonumber
\bra \Psi \left| \hat{O}_{(1)} \right| \Psi^\prime \ket & = A \bra
\Psi \left| \hat{O}_A \right| \Psi^\prime \ket
\\ \nonumber
\bra \Psi \left| O_{(2)} \right| \Psi^\prime \ket & = \frac{A
(A-1)}{2} \bra \Psi \left| O_{A,A-1} \right| \Psi^\prime \ket
\\ \nonumber
\bra \Psi \left| \hat{O}_{(3)} \right| \Psi^\prime \ket & = \frac{ A
(A-1) (A-2)}{6} \bra \Psi \left| \hat{O}_{A,A-1,A-2} \right|
\Psi^\prime \ket
\end{align}

In order to construct a hyperspherical function with such a
property, one starts in the $2$--body problem. In this case there
exists one Jacobi coordinate, and the hyper--spherical formalism
coincides with the usual spherical formalism, thus the
eigen--functions are the spherical harmonics, $Y_{l_1
m_1}(\hat{\eta}_1)$. The latter also belongs to a well defined {\it
irrep} of the symmetry group $\mathcal{S}_2$. Adding an additional
particle is done by the usual coupling of angular momenta,\\
$\Phi_{L_2,M_2;l_1,l_2} (\Omega_{(2)})= \sum_{m_1,m_2} (l_1 m_1 l_2
m_2|L_2,M_2) Y_{l_1 m_1}(\hat{\eta}_1) Y_{l_2 m_2}(\hat{\eta}_2)$.
This does not exhaust the eigen--function of the $3$--body
grand--angular momentum $\hat{K}_2$, as there is an additional
dependence on the hyper--angle coordinate $\varphi_2$. The resulting
differential equation is separable, bringing to the hyperspherical
function for three particles,
\begin{equation}
\mathcal{Y}_{K_2,L_2,M_2,l_1,l_2}(\Omega_{(2)},\varphi_2) = \Psi
_{K_2;l_1 l_2} (\varphi_2) \Phi_{L_2,M_2;l_1,l_2}.
\end{equation}
By a linear combination of these functions with common $K_2,L_2,M_2$
one can construct basis functions that belong to well defined {\it
irrep}s of the orthogonal group $\mathcal{O}(2)$. Consequently, the
hyperspherical function depends on the Gel'fand--Zetlin (GZ) symbol
${\bm{\Lambda}}_2$, and a multiplicity label $\alpha_2$. So the
hyperspherical function is labeled by five good quantum numbers,
$\mathcal{Y}_{[K_2]}(\Omega_{(2)},\varphi_2)$, with $[K_2] \equiv
K_2,L_2,M_2,{\bm{\Lambda}}_2,\alpha_2$.

As mentioned, the construction of the hyper--spherical function for
the $A$--body, $A-1$ Jacobi coordinates, problem is done
recursively. Let us assume the construction led to
$\mathcal{Y}_{[K_{A-2}]}(\Omega_{(A-2)},\varphi_{(A-2)})$ with
$[K_{A-2}] \equiv K_{A-2},L_{A-2},M_{A-2}, K_{A-3},L_{A-3}, l_{A-2}
,{\bm{\Lambda}}_{A-3},\alpha_{A-3}$. By coupling this to $Y_{l_{A-1}
m_{A-1}}(\hat{\eta}_{A-1})$ one gets
\begin{eqnarray}
\lefteqn{ \Phi_{L_2,M_2;l_1,l_2} (\Omega_{(2)})= } \\
 & & \nonumber {= \sum_{M_{A-2},m_{A-1}} (L_{A-2} M_{A-2} l_{A-1}
m_{A-1} |L_{A-1},M_{A-1})
\mathcal{Y}_{[K_{A-2}]}(\Omega_{(A-2)},\varphi_{(A-2)}) Y_{l_{A-1}
m_{A-1}}(\hat{\eta}_{A-1})}.
\end{eqnarray}
Next, one constructs the orthonormalized eigen--functions of the
grand angular momentum operator $\hat{K}_{A-1}^2$ \citep{EFROS72}:
\begin{eqnarray} \label{eq:HF}
\lefteqn{\Psi _{K_{A-1};l_{A-1} K_{A-2}} (\varphi_{A-1}) =} \\
\nonumber & &{\mathcal{N}_{A-1} (K_{A-1};l_{A-1} K_{A-2}) (\sin
\varphi_{A-1})^{l_{A-1}} (\cos \varphi_{A-1})^{K_{A-2}}
P_{n_{A-1}}^{(l_{A-1}+\frac{1}{2}, K_{A-2}+\frac{3A-8}{2})} (\cos
2\varphi_{A-1})},
\end{eqnarray}
where $P_n^{(\alpha,\beta)}$ are the Jacobi polynomials,
$n_{A-1}=\frac{K_{A-1}-K_{A-2}-l_{A-1}}{2}$, and the normalization
factor is \citep{EFROS72}:
\begin{equation}
\small {\mathcal{N}_{A-1} (K_{A-1};l_{A-1} K_{A-2})=\sqrt
{\frac{(2K_{A-1} +3A -5 ) n_{A-1}! \Gamma \big( n_{A-1} + K_{A-2} +
l_{A-1} + \frac{3A-5}{2}\big)}{\Gamma \big( n_{A-1} + l_{A-1} +
\frac{3}{2}\big) \Gamma \big( n_{A-1} + K_{A-2} +
\frac{3A-6}{2}\big)}}}.
\end{equation}
The function $\Psi$ is an eigen--function of $\hat{K}_{A-1}^2$ with
an eigen--value $K_{A-1}(K_{A-1} +3A -5)$. By construction, $K_{A-1}
- (K_{A-2}+l_{A-1}) \ge 0$, and even. As a result $n_{A-1}$ is a
non--negative integer.

The result of the complete separation between hyper--angular
coordinate and the Jacobi coordinates, means that one can construct
the hyper--spherical function
\begin{equation}
\mathcal{Y}_{[K_{A-1}]}(\Omega_{(A-1)},\varphi_{(A-1)})= \Psi
_{K_{A-1};l_{A-1} K_{A-2}} (\varphi_{A-1})
\Phi_{L_{A-1},M_{A-1};[K_{A-2}] l_{A-1}} (\Omega_{(A-2)}
\varphi_{(A-2)})
\end{equation}
where $[K_{A-1}]$ stands for
$K_{A-1},L_{A-1},M_{A-1},K_{A-2},L_{A-2},l_{A-1},{\bm{\Lambda}}_{A-2},\alpha_{A-2}$.
These functions are a complete set, orthonormal in each of the
quantum numbers.

However, using this construction, the hyperspherical harmonics
functions still do not possess the symmetry property for quantum
symmetry, i.e. they do not belong to an {\it irrep} of the symmetry
group $\mathcal{S}_A$. For this purpose we use an algorithm
developed by \citet{symhh}, which uses an extension of the
Novoselsky, Katriel, Gilmore (NKG) algorithm \citep{NOV87}, and
yields an $A$ particle state belonging to well defined {\it irrep}s
of both the orthogonal and the symmetry groups.

The process starts with a part overlooked in the previous
explanation, the recursive construction of the hyperspherical
functions as an {\it irrep} of $\mathcal{O}(A-1)$. This step is done
by understanding that the GZ states can be expressed recursively
${\bm{\Lambda}}_N=[{\bm{\lambda}}_N {\bm{\Lambda}}_{N-1}]$, where
${\bm{\lambda}}_N$ is an {\it irrep} of $\mathcal{O}(N)$ containing
the {\it irrep}s represented by the GZ state ${\bm{\Lambda}}_{N-1}$.
Using this relation, one gets
\begin{eqnarray}
\lefteqn{\big| K_N L_N M_N {\bm{\Lambda}}_N \alpha_N \ket = } \\
\nonumber & &{ = \sum_{K_{N-1} L_{N-1} \alpha_{N-1} l_N } [(K_{N-1}
L_{N-1} {\bm{\lambda}}_{N-1} \alpha_{N-1}; l_N) K_N L_N | \} K_N L_N
{\bm{\lambda}}_{N-1} \alpha_{N-1}] \times } \\
\nonumber & &{ \times \big | (K_{N-1} L_{N-1} {\bm{\Lambda}}_{N-1}
\alpha_{N-1}; l_N) K_N L_N) K_N L_N M_N \ket }.
\end{eqnarray}
The transformation coefficients connect $N-1$ coordinates with the
$N^{th}$ coordinate, thus called the hyperspherical orthogonal
coefficients of fractional parentage (hsocfps). in \citet{symhh} it
is shown that the hsocfps depend only on the $\mathcal{O}(N)$ and
$\mathcal{O}(N-1)$ {\it irrep}s, and are independent of $M_N$.

The next step is the reduction of the kinematical symmetry to the
permutation symmetry, i.e. $\mathcal{O}(A-1) \rightarrow
\mathcal{S}_A$ in the case of $A$ identical particles. The NKG
algorithm is based on the recursive construction of a state of an
{\it irrep} of $\mathcal{S}_N$ from a state of an {\it irrep} of
$\mathcal{S}_{N-1}$. The algorithm uses the transposition
representation of a permutation, by which any contained permutation
is represented as a subset of the transpositions. By construction, a
state with $N$ particles is characterized by the symmetric group
{\it irrep} that correspond to the group--subgroup chain
$\mathcal{S}_1 \subset \mathcal{S}_2 \subset \ldots \subset
\mathcal{S}_N$, represented by Young diagrams
$\Gamma_1,\,\ldots,\,\Gamma_N$. For short, one uses the Yamanouchi
symbol $Y_{N}=[\Gamma_N \Gamma_{N-1} \ldots \Gamma_1]=[\Gamma_N
Y_{N-1}]$. The symmetry adaptation of the $\mathcal{O}_{N-1}$
symmetry to include the permutation symmetry has its price, letting
go of the kinematical symmetry chain of group--subgroup
${\bm{\Lambda}}_{N-1}$ and holding only the $\mathcal{O}_{N-1}$
representation invariant, i.e. ${\bm{\lambda}}_{N-1}$. Also, a
possibility of degeneracy of the permutation states in the
kinematical group, suggests an additional label $\beta_N$. This is a
kind of parentage transformation, thus its coefficients are called
orthogonal coefficients of fractional parentage (ocfps).

Summing this up, we have the recursive relation:
\begin{eqnarray} \label{eq:HH_state}
\lefteqn{\big| K_{A-1} L_{A-1} M_{A-1} {\bm{\lambda}}_{A-1}
\alpha_{A-1} \Gamma_{A} Y_{A-1} \beta_A \ket = } \\
\nonumber & &{ = \sum_{{\bm{\lambda}}_{A-2} \beta_{A-1}}
[({\bm{\lambda}}_{A-2} \Gamma_{A-1} \beta_{A-1} )
{\bm{\lambda}}_{A-1} | \}
{\bm{\lambda}}_{A-1} \Gamma_{A} \beta_{A} ] \times } \\
\nonumber & &{  \times \sum_{K_{A-2} L_{A-2} \alpha_{A-2} l_{A-1} }
[(K_{A-2} L_{A-2} {\bm{\lambda}}_{A-2} \alpha_{A-2}; l_{A-1})
K_{A-1} L_{A-1} | \} K_{A-1} L_{A-1}
{\bm{\lambda}}_{A-1} \alpha_{A-1}] \times } \\
\nonumber & &{ \times \big | (K_{A-2} L_{A-2} {\bm{\lambda}}_{A-2}
\alpha_{A-2} \Gamma_{A-1} Y_{A-2} \beta_{A-1} ; l_{A-1}) K_{A-1}
L_{A-1} M_{A-1} {\bm{\lambda}}_{A-1} \ket }.
\end{eqnarray}
Where of course,
\begin{equation} \label{eq:HH_function}
{\bra \Omega \big| K_{A-1} L_{A-1} M_{A-1} {\bm{\lambda}}_{A-1}
\alpha_{A-1} \Gamma_{A} Y_{A-1} \beta_A \ket = \mathcal{Y}_{ K_{A-1}
L_{A-1} M_{A-1} {\bm{\lambda}}_{A-1} \Gamma_{A} Y_{A-1} \alpha_A^K}
(\Omega) },
\end{equation}
with $\Omega=(\Omega_{(A-1)},\varphi_{(A-1)})$ for the angular
coordinates, and $\alpha_A^K=(\alpha_{A-1},\beta_A)$ is the
multiplicity label.

\subsection{The Hyperspherical Formalism with Internal Degrees of
Freedom}

Pauli principle suggests that a set of identical particles would
have good symmetry properties with respect not only to the
configurational exchange of two particles, but to all the degrees of
freedom of the particle. In the case of nucleons, this means that
the wave function of the $A$--body nucleus can be written as,
\begin{eqnarray} \label{eq:HHTS}
{\Phi(\vec{\eta}_1 \ldots \vec{\eta}_N, s_1\ldots s_A,t_1\ldots
t_A)=\sum_{[K]\nu} R_{[K];\nu}(\rho) H_{[K]}(\Omega,s_1\ldots
s_A,t_1\ldots t_A)}.
\end{eqnarray}
$R_{[K];\nu}(\rho)$ are solutions to the hyper--radial equation,
with quantum number $\nu$, and $[K] \equiv J_A,J^z_A,[K_A], S_A,
T_A, T_A^z, \alpha_A^{ST}$. The function $H_{[K]}(\Omega;s_1\ldots
s_A;t_1\ldots t_A)$ possesses the correct symmetry property for a
set of fermions, i.e. antisymmetric in any exchange of particles.
The nuclear states are described by total isospin, $T$, and total
angular momentum $J$. The latter is the sum of the total spin $S$
and orbital angular momentum $L$.

It is clear that in order to couple the hyperspherical harmonics to
a spin--isospin state, one needs to create the latter as a
representation of the permutation group.

This is done, once again, by means of the NKG algorithm. One uses
spin--isospin coefficients of fractional parentage (stcfps), to
build the state
\begin{eqnarray}
\lefteqn {\chi_{S_A S_A^z T_A T_A^z {\Gamma}_A {Y}_{A-1}
\alpha_A^{ST}}(s_1,\ldots,s_A;t_1,\ldots,t_A)= \sum_{S_{A-1} T_{A-1}
\alpha_{A-1}^{TS}}} \\ \nonumber & & {
[(S_{A-1};s_A)S_A(T_{A-1};t_A)T_A \Gamma_{A-1} \alpha_{A-1}^{TS}|\}
S_AT_A\Gamma_A \alpha_A^{ST}] \times}
\\ \nonumber & & {[\chi_{S_{A-1} T_{A-1} {\Gamma}_{A-1} {Y}_{A-2}
\alpha_{A-1}^{ST}} \otimes s_A t_A]^{S_A S_A^z T_A T_A^z}},
\end{eqnarray}
where $S_A$ ($T_A$) is the total spin (isospin), with projection
$S_A^z$ ($T_A^z$), and the label $\alpha_A^{ST}$ is needed to remove
the degeneracy.

Consequently, the function
$H_{[K]}(\Omega;s_1,\ldots,s_A;t_1,\ldots, t_A)=\bra \Omega
,s_1\ldots s_A,t_1\ldots t_A \mid [K] \ket$ is just,
\begin{eqnarray}
\lefteqn {H_{[K]}(\Omega;s_1,\ldots,s_A;t_1,\ldots,t_A)=
\sum_{Y_{A-1}}\frac{\Lambda_{\Gamma_A,Y_{A-1}}}{\sqrt{\mid \Gamma_A
\mid}} \times } \\ & & \nonumber {\times \left[
{\mathcal{Y}}_{K_{A-1} L_{A-1} {\bm{\lambda}}_{A-1} \Gamma_A Y_{A-1}
\alpha_N^K}(\Omega) \otimes \chi_{S_A T_A T_A^z \tilde{\Gamma}_A
\tilde{Y}_{A-1} \alpha_A^{ST}}(s_1,\ldots,s_A;t_1,\ldots,t_A)
\right]^{J_A}_{J_A^z}},
\end{eqnarray}
where $\Lambda_{\Gamma_A,Y_{A-1}}$ are phase factors, which in the
case of fermionic system are positive (negative) when the number of
boxes in the diagram $\Gamma_A$ below the row of the $A^{th}$
particle is odd (even). $\mid \Gamma_A \mid$ is the dimension of the
young diagram. This function couples the total spin and orbital
angular momentum to angular momentum $J_A$, with projection $J_A^z$,
i.e. an $LS$ scheme. However, in microscopic problems, where the
forces are not necessarily central, a $JJ$ coupling scheme is
usually preferred.

In the $JJ$ coupling scheme, one first couples the spin and angular
momentum of the last particle $s_A=\frac{1}{2}$ and $l_A$ to $j_A$,
which is then coupled to the total angular momentum $J_{A-1}$ of the
residual $A-1$ particles, given by the coupling of $S_{A-1}$ and
$L_{A-2}$. Finally, the hyperspherical spin--isospin basis state is
written as
\begin{eqnarray}  \label{eq:JJ_1b}
\lefteqn \nonumber & & {|[K]\rangle_{(1)} = \sum_{Y_{A-1}} \frac
{{{\Lambda}}_{\Gamma_A,Y_{A-1}}} {\sqrt{\mid \Gamma_A \mid}}
\sum_{{\bm{\lambda}}_{A-2} \beta_{A-1}^{\bm{\lambda}}}
[({\bm{\lambda}}_{A-2} \Gamma_{A-1} \beta_{A-1}^{\bm{\lambda}} )
{\bm{\lambda}}_{A-1} |\} {\bm{\lambda}}_{A-1} \Gamma_A
\beta_{A}^{\bm{\lambda}}] \times }\\ \nonumber & &
{\sum_{K_{A-2}L_{A-2} \beta_{N-1}^K l_{A-1}} [(K_{A-2} L_{A-2}
{\bm{\lambda}}_{A-2} \beta_{A-2}^K ; l_{A-1}) K_{A-1}
L_{A-1} |\} K_{A-1} L_{A-1} {\bm{\lambda}}_{A-1} \beta_{A-1}^K] \times}\\
& & {\sum_{S_{A-1} T_{A-1} \alpha_{A-1}^{ST}}
[(S_{A-1};s_A)S_A(T_{A-1};t_A)T_A \Gamma_{A-1}
\alpha_{A-1}^{ST}|\} S_AT_A\Gamma_A \alpha_A^{ST}] \times}\\
\nonumber & & {\sum_{j_A J_{A-1}}
\sqrt{(2S_A+1)(2L_{A-1}+1)(2j_A+1)(2J_{A-1}+1)} \ninej
{S_{A-1}}{s_A}{S_A}{L_{A-2}}{l_{A-1}}{L_{A-1}}{J_{A-1}}{j_A}{J_A} \times}  \\
\nonumber & & {|(T_{A-1},t_A)T_A T_A^z \rangle \times}  \\
\nonumber & & {|\left((S_{A-1} ; K_{A-2}L_{A-2} {\bm{\lambda}}_{A-2}
\beta_{A-2}^K \Gamma_{A-1} Y_{A-2} \beta_{A-1}^{\bm{\lambda}})
J_{A-1} ; (s_A ; l_{A-1}) j_A \right) J_A J_A^z \rangle}
\end{eqnarray}
This is relevant mainly when the choice of Jacobi coordinates is
according to Eq.~(\ref{eq:JAC_1b}), in which the last coordinate is
proportional to the last particle coordinate.

A different $JJ$ coupling scheme is used when the choice of Jacobi
coordinates is according to Eq.~(\ref{eq:JAC_2b}), i.e. the last
coordinate is proportional to the relative coordinate of the last
two particles. In this scheme, one couples the last two particles
total spin $\hat{\vec{s}}=\hat{\vec{s}}_A+\hat{\vec{s}}_{A-1}$ and
orbital angular momentum $\hat{\vec{l}}=\hat{\vec{l}}_A
+\hat{\vec{l}}_{A-1}$ to a total angular momentum $\hat{j}$. The
latter is then coupled to the total angular momentum of the residual
$A-2$ particle system, which is the coupling of its spin $S_{A-2}$
and orbital angular momentum $L_{A-3}$. This leads to the following
representation of the basis state,
\begin{eqnarray}  \label{eq:JJ_2b}
\lefteqn \nonumber & & {|[K]\rangle_{(2)} = \sum_{Y_{A-1} Y_{A-2}}
\frac {{{\Lambda}}_{\Gamma_A,Y_{A-1}}} {\sqrt{\mid \Gamma_A \mid}}
\frac {{{\Lambda}}_{\Gamma_{A-1},Y_{A-2}}} {\sqrt{\mid \Gamma_{A-1}
\mid}} \sum_{{\bm{\lambda}}_{A-2} \beta_{A-1}^{\bm{\lambda}}}
[({\bm{\lambda}}_{A-2} \Gamma_{A-1} \beta_{A-1}^{\bm{\lambda}} )
{\bm{\lambda}}_{A-1} |\} {\bm{\lambda}}_{A-1} \Gamma_A
\beta_{A}^{\bm{\lambda}}] \times }\\ \nonumber & &
{\sum_{K_{A-2}L_{A-2} \beta_{N-1}^K l} [(K_{A-2} L_{A-2}
{\bm{\lambda}}_{A-2} \beta_{A-2}^K ; l) K_{A-1} L_{A-1} |\} K_{A-1}
L_{A-1} {\bm{\lambda}}_{A-1} \beta_{A-1}^K] \times}\\ \nonumber & &
{\sum_{S_{A-1} T_{A-1} \alpha_{A-1}^{ST}}
[(S_{A-1};s_A)S_A(T_{A-1};t_A)T_A \Gamma_{A-1}
\alpha_{A-1}^{ST}|\} S_AT_A\Gamma_A \alpha_A^{ST}] \times}\\
& & {\sum_{S_{A-2} T_{A-2} \alpha_{A-2}^{ST} s t} [(S_{A-2};s
)S_{A-1}(T_{A-2}; t ) T_{A-1} \Gamma_{A-2} \alpha_{A-2}^{ST}|\}
S_{A-2} T_{A-2} \Gamma_{A-2} \alpha_{A-2}^{ST}] \times}\\\nonumber &
& {\sum_{j s t J_{A-2}} (-)^{S_{A-2}+S_A} \sqrt{(2S_{A-1}+1)(2s+1)}
\sixj{S_{A-2}}{\frac{1}{2}}{\frac{1}{2}}{S_A}{S_{A-1}}{s}} \times
\\\nonumber & & { (-)^{T_{A-2}+T_A}
\sqrt{(2T_{A-1}+1)(2t+1)}
\sixj{T_{A-2}}{\frac{1}{2}}{\frac{1}{2}}{T_A}{T_{A-1}}{t}} \times \\
\nonumber & & {\sqrt{(2S_A+1)(2L_{A-1}+1)(2j_A+1)(2J_{A-1}+1)}
\ninej
{S_{A-2}}{s}{S_A}{L_{A-2}}{l}{L_{A-1}}{J_{A-2}}{j}{J_A} \times}  \\
\nonumber & & {|(T_{A-2},t)T_A T_A^z \rangle \times}  \\
\nonumber & & {|\left((S_{A-2} ; K_{A-2} L_{A-2}
{\bm{\lambda}}_{A-2} \beta_{A-2}^K \Gamma_{A-1} Y_{A-2}
\beta_{A-1}^{\bm{\lambda}}) J_{A-2} ; (s ; l) j \right) J_A J_A^z
\rangle }
\end{eqnarray}

With these results, one can now proceed to the calculation of matrix
elements.

\subsection{Matrix Elements within the Hyperspherical Formalism}
In nuclear problems one has to consider matrix elements of operators
of pure isospin $\mathcal{T}$, and its projection $\mathcal{T}_z$,
and spin--orbital coupling to a total angular momentum $\mathcal{J}$
with projection $\mathcal{M}$. This operator can act either on a
single nucleon, or more. In the general case of an operator acting
on $N$ nucleons, one considers the following operator:
\begin{equation}
\hat{\mathcal{O}}_{(N)}^{\mathcal{JM;TT}_z} = \sum_{i_1<i_2< \ldots
<i_N=1}^A \hat{\mathcal{O}}^{\mathcal{JM}} (\vec{r}_{i_1},\ldots,
\vec{r}_{i_N}; s_{i_1},\ldots,s_{i_N})
\hat{\mathcal{O}}^{\mathcal{TT}_z} (t_{i_1},\ldots,t_{i_N})
\end{equation}
In the case of $N=1$ or $2$, i.e. one and two body matrix elements,
one can use hyper--spherical wave functions of Eqs.~(\ref{eq:JJ_1b},
 \ref{eq:JJ_2b}), which lead to explicit dependence on one Jacobi
coordinate. The advantage is that usually one can separate the
hyper--radial part from the hyper angular part, i.e.
\begin{equation}
\hat{\mathcal{O}}^{\mathcal{JM}}=\mathcal{R}^\mathcal{J} (L
\eta_{A-1}) \hat{\tilde{\mathcal{O}}}^{\mathcal{JM}}
(\hat{\eta_{A-1}},spins)
\end{equation}
where $L=\sqrt{2}$ for two--body operator, and $L=\sqrt
{\frac{A}{A-1}}$ for one--body operator. The matrix element of the
scalar operator $\mathcal{R}^\mathcal{J} (L\eta_{A-1})$ includes two
integrations. The first integration is over the hyper--angle,
\begin{eqnarray}
\lefteqn{\mathcal{M}^\mathcal{J}(\rho)=\bra K_{A-2} l;K_{A-1} \mid
\mathcal{R}^\mathcal{J} (L\rho\sin \varphi_A) \mid K_{A-2}
l^{\prime};K^{\prime}_{A-1} \ket =} \\ \nonumber & &
{\mathcal{N}(K_{A-1};l K_{A-2}) \mathcal{N}(K^\prime_{A-1};l^\prime
K_{A-2}) \times} \\ \nonumber & & { \times \int_{0}^{\frac{\pi}{2}}
d\varphi (\sin \varphi)^{\alpha+\alpha^\prime} (\cos
\varphi)^{\beta-1} P_n^{(\alpha,\beta)}(\cos2\varphi)
\mathcal{R}^\mathcal{J} (L\rho\sin \varphi_A)
P_n^{(\alpha^\prime,\beta)}(\cos2\varphi)}
\end{eqnarray}
where $\alpha=l+\frac{1}{2}$, $\alpha^\prime=l^\prime+\frac{1}{2}$,
and $\beta=K_{A-2}+\frac{3A-8}{2}$ (see Eq.~(\ref{eq:HF}) for the
other notations). This integration can be performed analytically by
expanding the interaction to power series \citep{BARNEA_thesis}. Due
to the orthonormality of the Jacobi polynomials, this integration
includes a finite number of elements.

One now integrates over the hyperradius, by expanding the
hyperradial part of the wave function (cf. Eq.~(\ref{eq:HHTS})) as,
\begin{equation}
R_{[K];\nu}(\rho)=\sum C_{n_\rho}^{[K];\nu} \phi^{(a,b)}_{n_\rho}
\left(\frac{\rho}{b} \right).
\end{equation}
Here, $\phi^{(a,b)}_{n_\rho}(x)$ are the ortho--normalized
generalized Laguerre polynomials of order ${n_\rho}$ and parameters
$a,\,b$. The parameter $b$ is a length scale for the integration.
Explicitly, we take
\begin{equation}
\phi^{(a,b)}_{n}(x)=\sqrt{\frac{n!}{(n+a)!}} b^{-\frac{3(A-1)}{2}}
x^{\frac{a-(3A-4)}{2}} L_n^a(x) e^{-\frac{x}{2}}
\end{equation}
As a result, the hyperradial matrix element is written as
\begin{equation}
\bra R_{[K];\nu}(\rho) \mid \mathcal{M}^\mathcal{J}(\rho) \mid
R_{[K]^\prime;\nu^\prime}(\rho) \ket = \sum
C_{n^\prime_\rho}^{[K]^\prime;\nu^\prime} C_{n_\rho}^{[K];\nu}
\int_0^\infty d\rho \rho^{3A-4} \phi^{(a,b)}_{n}
\left(\frac{\rho}{b} \right)  \mathcal{M}^\mathcal{J}(\rho)
\phi^{(a^\prime,b)}_{n^\prime}\left(\frac{\rho}{b} \right)
\end{equation}
This integration is done numerically, using Gauss--Laguerre
quadrature.

In the current work, we confront also three--body matrix elements,
which include the coupling of two Jacobi coordinates
\citep{EIHH_3NF}.

%\subsubsection{Matrix Elements of One Body Operators}
%A one--body operator is written as,
%\subsubsection{Matrix Elements of Two Body Operators}
%\subsubsection{Matrix Elements of Three Body Operators}

\section{Effective Interaction with the Hyperspherical Harmonics}
The Hyperspherical Harmonics expansion has the correct asymptotic
behavior, thus provides a promising channel to the solution of the
nuclear problem defined by the Hamiltonian in
Eq.~(\ref{eq:HAMILTONIAN}). Two of the advantages in using this
expansion are the removal of the center of mass coordinate; and the
localization of the quantum description, provided by the explicit
dependence in the hyperradius, which holds within information
regarding the entire nucleus.

Alas, in order to reach convergence in the expansion of the nuclear
wave function one has to expand up to very high grand angular
momentum $K$. In order to cope with this disadvantage it was
proposed to replace the bare potential with an effective
interaction, which truncates the model space. The concept of
effective interaction has been used traditionally in the framework
of harmonic oscillator basis. However, lately
\citep{EIHH1,EIHH2,EIHH_3NF} it was extended to the hyperspherical
harmonics basis. This method, the Effective Interaction with the
Hyperspherical Harmonics (EIHH), has been shown to give excellent
results both in comparison to other few--body methods, and in
confronting previously unreachable nuclear reactions. In this
section I will present a brief overview of the method, as used for
the calculations shown in the following chapters.

\subsection{Lee--Suzuki Approach for Effective Interactions}

The effective interactions are projections of the bare interactions
to a subspace of the entire model space. Defining $P$ as the
projection operator onto the model space, then $Q=1-P$ is the
projection operator onto the complementary space. By definition, the
effective interaction leaves unchanged the low lying eigen-energies.

The first step in the construction of the effective interaction is
choosing the model space. It is helpful to note that the
hyper--radius, by definition, limits the inter--nucleon distance.
This fact makes it possible to choose the model space as the product
of the entire hyper--radial space and the set of HH basis functions
with $K\le K_{max}$. It is clear that $V_{eff} \rightarrow V$ as
$K_{max} \rightarrow \infty$, and the same follows for the
eigen--energies and states of the Hamiltonian.

In order to continue, one method is to use the Lee-Suzuki procedure
\citep{LS80,LS82,LS83}, in which one uses a similarity
transformation $X$. The effective Hamiltonian is given by
\begin{equation}
\mathcal{H}_{eff}=PX^{-1}\mathcal{H}XP.
\end{equation}
The effective interaction acts only in the $P$--space. The operator
$X=e^\omega$ re-projects the information contained in the $Q$--space
back to the model space, i.e. $\omega=Q \omega P$ (which also leads
to $\omega^2=\omega^3=\ldots=0$, and thus $X=1+\omega$, and
$X^{-1}=1-\omega$). One can summarize this method in the following
way (we use the notation $ O_{AB} \equiv AOB$) \citep{SONIA_THESIS}:
\begin{equation}
\mathcal{H} = \left( \begin{array}{c|c}
                 \mathcal{H}_{PP} & \mathcal{H}_{QP} \\ \hline
                 \mathcal{H}_{PQ} & \mathcal{H}_{QQ}
              \end{array}  \right) \longrightarrow
\mathcal{\tilde{H}} = X^{-1}\mathcal{H}X =
                 \left( \begin{array}{c|c}
                 \mathcal{\tilde{H}}_{PP} & 0 \\ \hline
                 \mathcal{\tilde{H}}_{PQ} & \mathcal{\tilde{H}}_{QQ}
              \end{array}  \right) \longrightarrow
\mathcal{H}_{eff}=\mathcal{\tilde{H}}_{PP}.
\end{equation}
The implicit condition that $\omega$ has to fulfill
$\mathcal{\tilde{H}}_{QP}=0$ can be rearranged to the equation:
\begin{equation} \label{eq:Bloch_Horowitz}
Q(\mathcal{H}+[\mathcal{H},\omega]-\omega \mathcal{H}\omega)P=0.
\end{equation}

The strength of the effective interaction approach is that the low
energy solutions to the original problem can be exhausted from the
eigen--vectors of the effective Hamiltonian, $\mathcal{H}_{eff}$,
which by construction are in the $P$ space. The theorem states: let
$\tilde{\epsilon}_P$ be the eigen--vectors of this subspace, then
${\epsilon}_P \equiv X\tilde{\epsilon}_P$ are eigen--vectors of the
true Hamiltonian with the same eigen--value.

In order to prove this theorem, let $\mid \tilde{\Psi}_\mu \ket \in
\tilde{\epsilon}_P$, i.e, $\mathcal{H}_{eff} \mid \tilde{\Psi}_\mu
\ket = \epsilon_\mu \mid \tilde{\Psi}_\mu \ket$ and $P\mid
\tilde{\Psi}_\mu \ket =\mid \tilde{\Psi}_\mu \ket$. Then
$\mathcal{H}\mid {\Psi}_\mu \ket = \mathcal{H}X\mid \tilde{\Psi}_\mu
\ket = XX^{-1} \mathcal{H}X\mid \tilde{\Psi}_\mu \ket =
X\mathcal{H}_{eff} \mid \tilde{\Psi}_\mu  \ket = \epsilon_\mu X \mid
\tilde{\Psi}_\mu \ket = \epsilon_\mu \mid {\Psi}_\mu \ket$, which
concludes the proof.

The subspace $\epsilon_P$ contains information regarding both the
$P$ and the $Q$ spaces. Furthermore, the similarity transformation
depends only on this space. Let $\mid \alpha \ket$ ($\mid \beta
\ket$) be a basis for the $P$ ($Q$) space, and $(A)_{\alpha
\mu}=\bra \alpha \mid {\Psi}_\mu \ket$, $(B)_{\beta \mu}=\bra \beta
\mid {\Psi}_\mu \ket$. Then, it is easily proven that a solution to
Eq.~(\ref{eq:Bloch_Horowitz}) is
\begin{equation} \label{eq:sim_tra}
\omega=A \cdot (B)^{-1}.
\end{equation}

This similarity transformation is not hermitian, however an
equivalent effective interaction can be written as,
\begin{equation}
\mathcal{H}_{eff}=\frac{P+\omega}{\sqrt{P(1+\omega^{\dagger}\omega)P}}
\mathcal{H} \frac{P+\omega}{\sqrt{P(1+\omega^{\dagger}\omega)P}}.
\end{equation}

It is clear that the effective interaction stores information
regarding correlations in the $Q$--space, while solving only in the
$P$ space. Alas, this procedure produces, in general, $A$--body
interactions, which solution is as difficult as finding the full
space solutions. To cope with this problem, one approximates the
effective interaction, while keeping the property of converging to
the bare interaction as $P \rightarrow 1$.

A good approximation is used in the ``No Core Shell Mode'' (NCSM)
approach \citep{NCSM,NCSM2,NCSM3}. The approximation is to include
in the effective potential only two--body effective interactions:
\begin{equation}
V_{eff}=PX^{-1} \left[ \sum_{i<j=1}^A V_{ij} \right] XP
\longrightarrow V_{eff}^{app} = \sum_{i<j=1}^A V_{eff,ij}^{(2)}.
\end{equation}
The two--body effective potential is the non--trivial part of the
two--body effective Hamiltonian, i.e. one decomposes
$\mathcal{H}^{(2)}={\rm{H}}_0+\sum_{i<j=1}^A V_{ij}^{(2)}$ with
$[{\rm{H}}_0,P]=[{\rm{H}}_0,Q]=0$ and $Q{\rm{H}}_0 P=P{\rm{H}}_0
Q=0$. Then the two--body effective Hamiltonian is built using a
Lee--Suzuki transformation:
\begin{equation} \label{eq:LS2}
\mathcal{H}^{(2)}_{eff}=\frac{P_2+\omega_2}
{\sqrt{P_2(1+\omega_2^{\dagger}\omega_2)P_2}} \mathcal{H}^{(2)}
\frac{P_2+\omega_2}{\sqrt{P_2(1+\omega_2^{\dagger}\omega_2)P_2}},
\end{equation}
thus,
\begin{equation}
\sum_{i<j=1}^A
V_{eff,ij}^{(2)}=\mathcal{H}^{(2)}_{eff}-P{\rm{H}}_0P.
\end{equation}

\subsection{EIHH method}
In the EIHH method, one solves the hyper--radial part of the
Hamiltonian. Then, for each hyper--radius $\rho$, the two--body
hamiltonian is clearly
\begin{equation} \label{eq:H2}
\mathcal{H}^{(2)}=\frac{1}{2M_N} \frac{\hat{K}^2}{\rho^2}+V_{A,A-1}.
\end{equation}
Let us note that:
\begin{itemize}
  \item This Hamiltonian includes the total
  hyper--angular kinetic energy.
  \item The only interacting particles are the ``last'' two.
  \item The residual $A-2$ system acts as a mean--field force, entering only
  through the collective hyper--radius. The hyper--radius enters not
  only in the kinetic energy, but also in the potential matrix
  element, as the relative distance between the last two particles
  is  $\vec{r}_{A,A-1}=\sqrt{2}\vec{\eta}_{A-1}=\left( \sqrt{2} \rho \sin
  \varphi_A, \hat{\eta}_{A-1} \right)$.
  \item The calculation includes only one--dimensional integration.
\end{itemize}
Choosing the model space within the HH formalism is quite natural,
as the two--body Hamiltonian of Eq.~(\ref{eq:H2}) is diagonal in the
quantum numbers of the $A-2$ residual system. Thus, the $P_2$--space
is chosen as $P_2=\left\{ [K] : K \le K_{max}, [K_{A-2}] \,\, fixed
\right\}$. It is clear that $Q_2=\left\{ [K] : K > K_{max},
[K_{A-2}] \,\, fixed \right\}$. A numerical calculation of the $Q_2$
space includes a numerical cutoff $K_{MAX}$ which assures
convergence. This depends on the orbitals of the potential, and was
checked to be $K_{MAX}\sim 60$ for $S$--wave interaction and
$K_{MAX} \sim 180$ for $P$--wave interaction. The construction of
$\epsilon_P$ space is now trivial, using Eq.~(\ref{eq:sim_tra}).

By construction, the effective interaction is ``state dependent''.
This means that for every fixed value of $K_{A-2}$ one has to
calculate a similarity transformation. In \citet{MIN04} it was
demonstrated that the ``no--core'' approach is equivalent to the
 bare Hamiltonian (through a unitary transformation), in the limit
$P\rightarrow 1$. It was also shown that any effective two body
operators constructed through the Lee--Suzuki similarity
transformation of Eq.~(\ref{eq:LS2}) can be regarded accurate to
second order in the limit $P\rightarrow 1$.

The EIHH method accelerates the convergence of the HH expansion, and
has been shown to do so in the calculation of nuclear bound states
and reactions of nuclei in the mass range $3 \le A \le 7$ (for
example \citep{EIHH1,EIHH2,BELO01,bacca1,bacca2,GA04,GA06,GA07}).

%% file: Photo_chapter.tex
\chapter{Photoabsorption on $^4$He with a realistic nuclear force}
\label{chap:Photo}

In previous chapters I outlined a detailed description of the
theoretical background for {\it{ab-initio}} calculation of few--body
nuclear structure and reactions. In the first section of this
chapter, the application of these methods to the $^4$He total
photoabsorption cross section is presented with the realistic
nucleon-nucleon (NN) potential Argonne $v_{18}$ and the
three-nucleon force (3NF) Urbana IX, as published in \citet{GA06}.
In the second part of the chapter, based on \citep{GA06a},
sum--rules of this cross--section are investigated, emphasizing
relations to ground state properties.
\section{Photoabsorption Calculation} \label{sec:Photo_crs}
\subsection{The Unretarded Dipole Approximation}

For low energy reactions, below the pion production threshold, a
good approximation for the photoabsorption cross--section,
$\sigma_\gamma$, is using only the leading electric dipole ($E1$) in
its non--relativistic long wavelength approximation. In addition,
since the Siegert theorem (Sec.~\ref{sec:photo}) applies for the
conserved electro--magnetic current, the dipole operator is
proportional to the Coulomb multipole. Thus, the effects of MEC are
implicitly included.

As a result, the total photoabsorption cross section is given by
\begin{equation}
\sigma_\gamma(\omega)=\frac{4\pi^{2}\alpha}{2J_0+1}\omega
R(\omega)\,,
\end{equation}
\noindent where $\alpha$ is the fine structure constant, $J_0$
denoting the nucleus total angular momentum, and
\begin{equation} \label{1}
   R(\omega )=\sumint \left| \left\langle \Psi _{f}\right| \hat{D}_z \left|
\Psi _{0}\right\rangle \right| ^{2}\delta (E_{f}-E_{0}-\omega)
\end{equation}
is the response function in the unretarded dipole approximation with
$\hat{D}_{z}=\sum_{i=1}^{A}\frac{\tau^{3}_{i}z'_{i}}{2}$.

The wave functions of the ground and final states are denoted by
$\left| \Psi_{0/f} \right>$ and the energies by $E_{0/f}$,
respectively. The operators $\tau^{3}_{i}$  and $z'_{i}$ are the
third components of the $i$-th nucleon isospin and center of mass
frame position.

We refer back to Sec.~\ref{sec:photo} and recall the Siegert
theorem, which ensures that the dominant part of the exchange
current contribution is included. In the classical few-body nuclei
this has proven to be an excellent approximation
\citep{Fad_CHH,SaA},
particularly for photon energies below 50 MeV. \\
For triton, $\sigma_\gamma(^3$H) \citep{Fad_CHH} the contributions
of retardation and other multipoles lead to an enhancement of
$\sigma_\gamma$ by less than 1\% for $\omega \le 40 \mev$. The
contribution grows with energy transfer. This result is rather
insensitive to the nuclear force model. When calculating with AV18
NN potential only, the contribution is 5\% for $\omega=60\mev$, 16\%
for $\omega=100\mev$, and 26\% at pion threshold ($\omega=140
\mev$). When including UIX 3NF the numbers change to 5\%, 18\%, and
33\% at $\omega= 60$, 100, and 140 MeV, respectively. If one bares
in mind that the cross--section is almost exhausted in the giant
dipole resonance (GDR), below $40 \mev$, the approximation is
appropriate for the photoabsorption process below pion--threshold.

\subsection{Calculation of The Response Function}

In order to find the Response function, the LIT method is used,
calculated with the Lanczos algorithm (see Chap.~\ref{chap:LIT}).
\begin{table}
\caption{ \label{tb:4he_gs} Convergence of HH expansion for the
$^4$He binding energy $E_b$ [MeV] and root mean square matter radius
$\langle r^2\rangle^{\frac{1}{2}}$ [fm] with the AV18 and AV18+UIX
potentials. Also presented are results of other methods (see text).}
\begin{center}
\begin{tabular}
{| c | cc | cc|} \hline
                & \,\,\,\,\,\,\,\,\,\,\,\,\,\,\,\,\,\,\, AV18 &
                & \,\,\,\,\,\,\,\,\,\,\,\,\,\,\,\, AV18+UIX & \\
     $K^0_{m}$ & $E_b$ & $ \langle r^2\rangle^{\frac{1}{2}} $
%              & $\sigma_D$
               & $E_b$ & $ \langle r^2\rangle^{\frac{1}{2}} $ \\
%              & $\sigma_D$
\hline \hline
   6 &  25.312  &  1.506  & 26.23   &  1.456  \\
   8 &  25.000  &  1.509  & 27.63   &  1.428  \\
  10 &  24.443  &  1.520  & 27.861  &  1.428  \\
  12 &  24.492  &  1.518  & 28.261  &  1.427  \\
  14 &  24.350  &  1.518  & 28.324  &  1.428  \\
  16 &  24.315  &  1.518  & 28.397  &  1.430  \\
  18 &  24.273  &  1.518  & 28.396  &  1.431  \\
  20 &  24.268  &  1.518  & 28.418  &  1.432  \\
\hline \hline
FY1 & 24.25 &   -   & 28.50   &   -     \\
FY2 & 24.22 & 1.516 &    -    &   -     \\
HH      & 24.21 & 1.512 & 28.46   & 1.428      \\
GFMC    &   -   &   -  & 28.34   & 1.44        \\
%\citep{NO02}  & 24.25 &       & 28.50   &         \\
%\citep{LaC}   & 24.22 & 1.516 &         &         \\
%\citep{VI05}  & 24.21 & 1.512 & 28.46   & 1.428      \\
%\citep{GFMC}  &       &       & 28.34   & 1.44        \\
\hline
\end{tabular}
\end{center}
\end{table}

The EIHH expansions of the ground state $\big| \Psi_{0} \ket$ and
the LIT vector $\big |\widetilde {\Psi}\ket$ are performed with the
full HH set up to maximal values of the HH grand-angular momentum
quantum number $K$ ($K\leq K^0_{m}$ for $\Psi_0$, $K\leq K_{m}$ for
$\widetilde\Psi$).

The ground state properties of $^4$He calculated using the EIHH
approach are shown in Tab.~\ref{tb:4he_gs}. The convergence of the
binding energy, $E_b$, and matter radius are presented as a function
of $K^0_{m}$. Since the EIHH method is not variational the
asymptotic $E_b$ value can be reached from below or above, in fact
both cases are realized in Tab.~\ref{tb:4he_gs}. Both convergence
patterns are not sufficiently regular to allow safe extrapolations
to asymptotic values. The calculated ground state properties of
$^4$He agree quite well with those of other methods, presented in
Tab.~\ref{tb:4he_gs}. The other calculations are done with the FY
equations (in \citet{NO02,LaC}), the HH expansion (in \citet{VI05})
and the GFMC method \citep{GFMC}, and appear in this order in
Tab.~\ref{tb:4he_gs}. One should compare the binding energies to the
experimental value of $28.296 \mev$.

\begin{figure} [t]
\begin{center}
\rotatebox{270}{ \resizebox{9cm}{!}
{\includegraphics{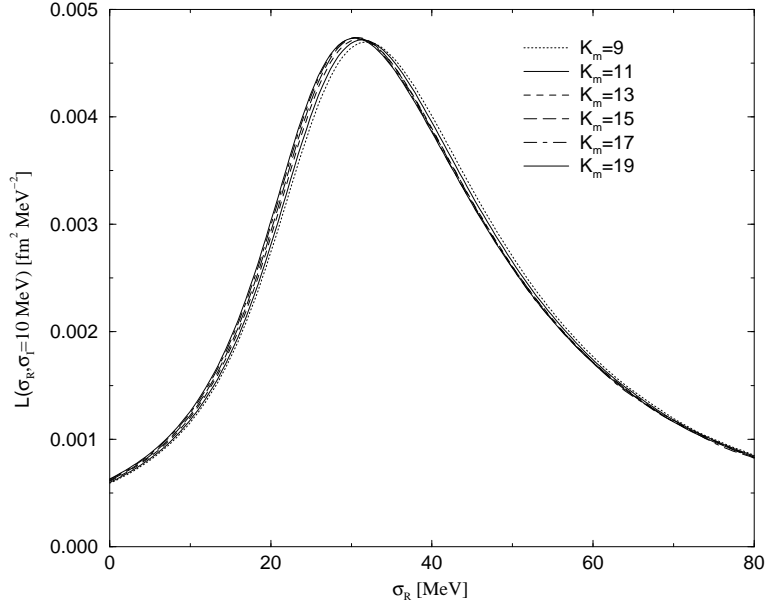}}} \caption{\label{Fig1}
Convergence of $\mathcal{L}_{K_{m}}$ with $\sigma_I=10$ MeV
(AV18+UIX). }
\end{center}
\end{figure}

\begin{figure} [t]
\begin{center}
\rotatebox{0}{ \resizebox{9cm}{!}{
\includegraphics{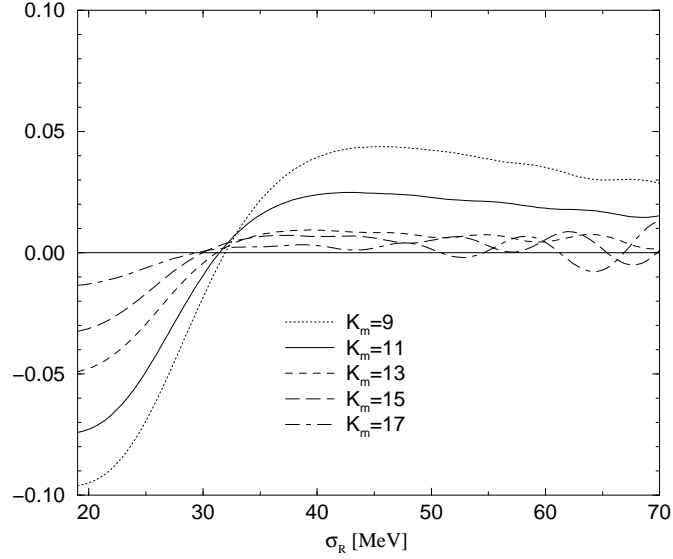}
} } \caption{\label{Fig2} Convergence of
$\Delta_{K_{m}}=(\mathcal{L}_{K_m}-\mathcal{L}_{19})/\mathcal{L}_{19}$
(AV18+UIX).}
\end{center}
\end{figure}

\begin{figure} [t]
\begin{center}
\rotatebox{270}{ \resizebox{9cm}{!}{
\includegraphics{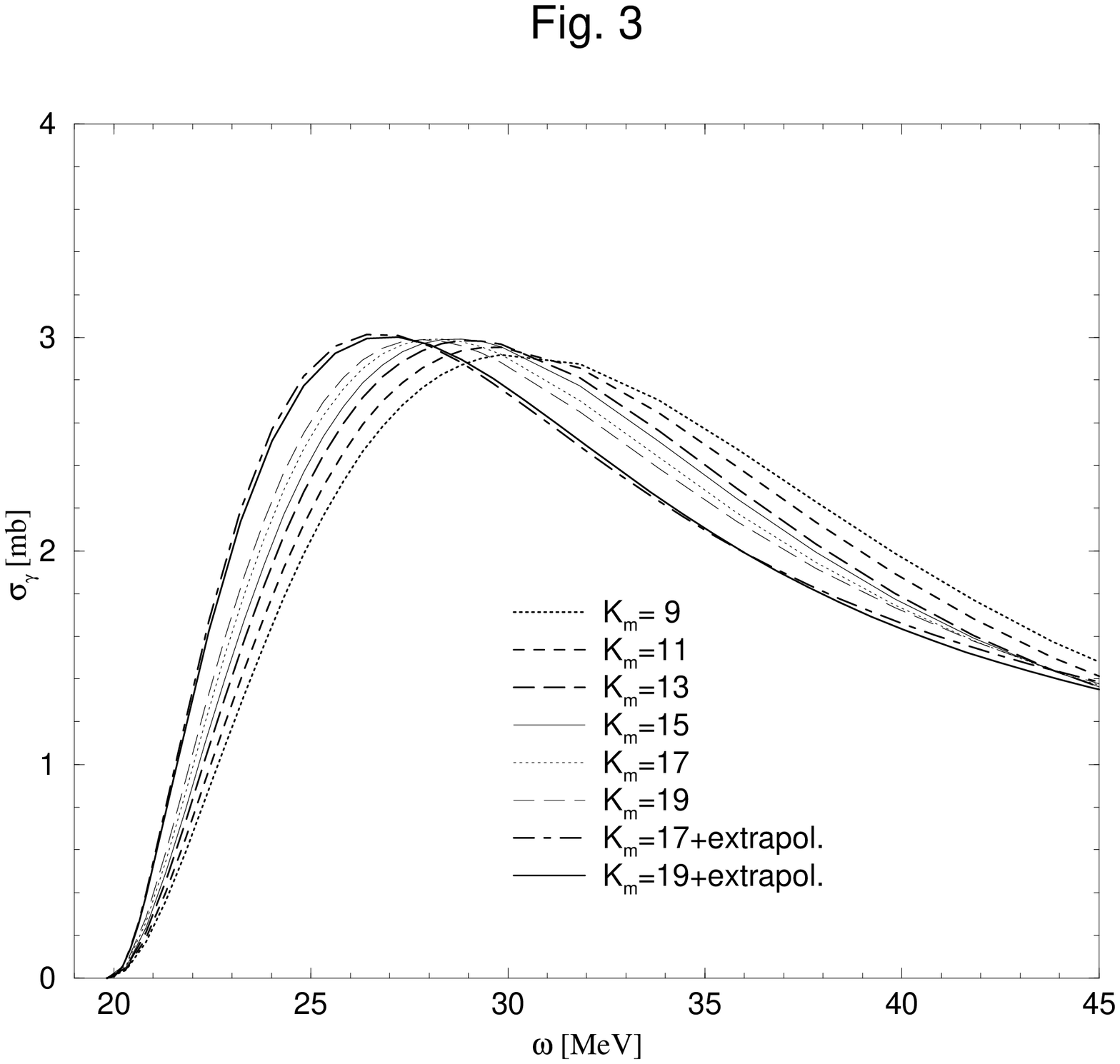}
} } \caption{\label{Fig3} Convergence of $\sigma_{\gamma,K_m}$ (AV18+UIX), also
shown  $\sigma^\infty_{\gamma,17}$ and $\sigma^\infty_{\gamma,19}$.}
\end{center}
\end{figure}

The EIHH convergence of the transform $\mathcal{L}(\sigma)$ is
excellent for the AV18 potential. Here we discuss in detail only the
case AV18+UIX, where the convergence is quite good,  but not at such
an excellent level. The reason is that in our present EIHH
calculation an effective interaction is constructed only for the NN
potential, while the 3NF is taken into account as bare interaction.

\begin{figure}
\begin{center}
\rotatebox{0}{ \resizebox{10cm}{!}{
\includegraphics{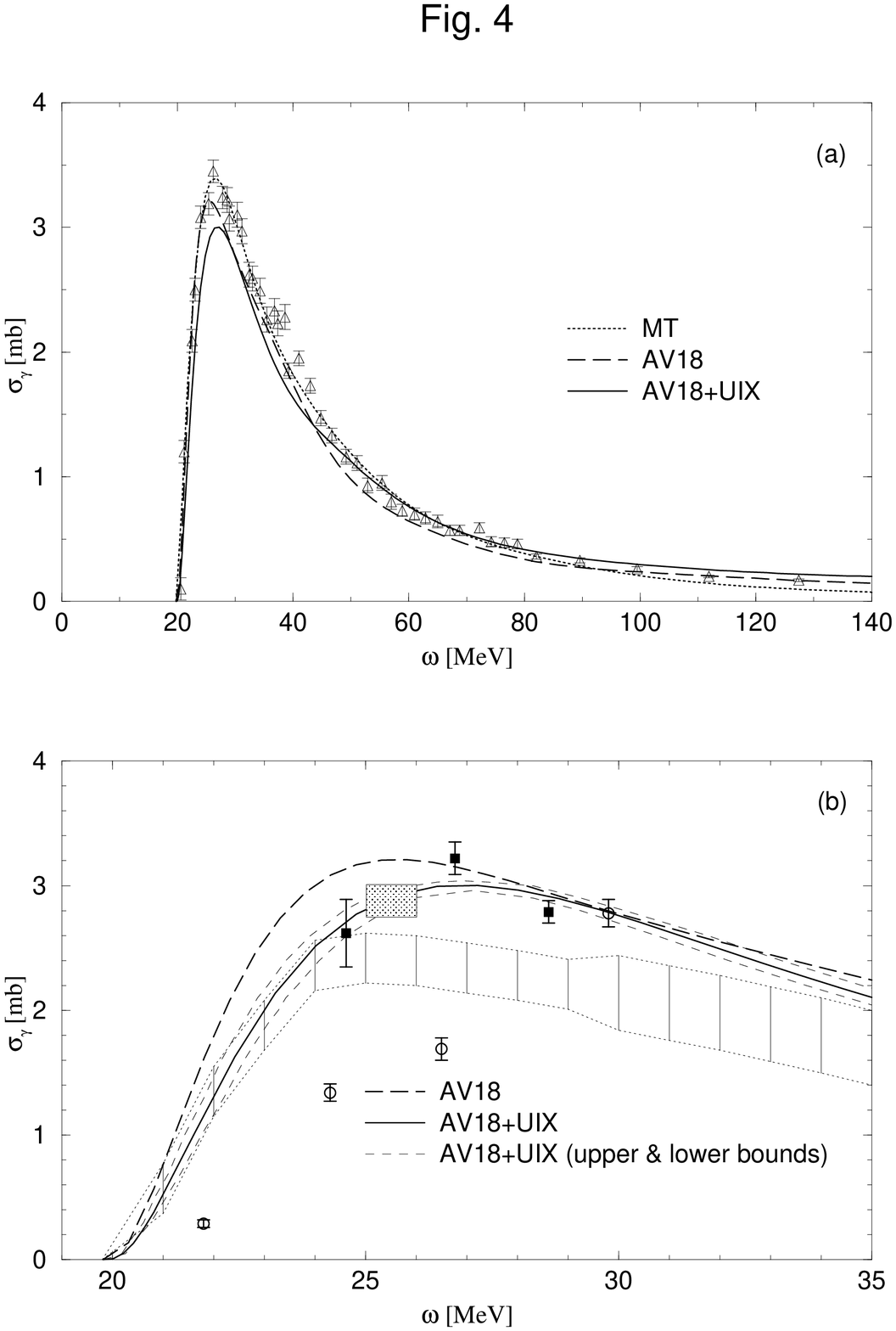}
} }
\end{center}
\caption{\label{Fig4} Total $^4$He photoabsorption cross section:
(a) $\sigma_\gamma$ (MT,AV18) and $\sigma^\infty_{\gamma,19}$
(AV18+UIX); experimental data from \citep{Ark}. (b) as (a) but also
included upper/lower bounds and various experimental data (see
text), area between dotted lines \citep{Berman,Feldman}, dotted box
\citep{Wells}, squares \citep{Lund}, and circles \citep{Shima}.}
\end{figure}

In Fig.~\ref{Fig1} we show results for the transform $\mathcal{L}$
obtained with various $K_m$ and $K^0_m$ values. Since we take
$K^0_{m}=K_{m}-1$, the corresponding transform can be denoted by
$\mathcal{L}_{K_{m}}$. One sees that there is a very good
convergence beyond the peak, that the peak height is very well
established, but that the peak position is not yet fully converged
in view of the fact that we aim for a percentage level accuracy. In
fact with increasing $K_{m}$ the peak is slightly shifted towards
lower $\sigma_R$.

By observing the relative error of the transform,
$\Delta_{K_{m}}=\mathcal{L}_{K_{m},19}/\mathcal{L}_{19}$ with
$\mathcal{L}_{\alpha,\beta}=\mathcal{L}_\alpha-\mathcal{L}_\beta$,
this point is illustrated better, see Fig.~\ref{Fig2} (the chosen
$\sigma_R$-range starts at the $^4$He$(\gamma)$ break-up threshold).
One again notes the very good convergence for $\sigma_R
> 30$ MeV with almost identical results from $\mathcal{L}_{13}$ to $\mathcal{L}_{19}$.
Altogether we consider our result for $\sigma_R > 30$ MeV as
completely sufficient. However, it is obvious that convergence is
not entirely reached for lower $\sigma_R$.

A trivial solution for this is an additional increase in $K_{m}$.
However, $\mathcal{L}_{19}$ includes already 364000 states in the HH
expansion. Thus, a further increase is beyond our present
computational capabilities. Fortunately a closer inspection of
Fig.~\ref{Fig2} shows that the convergence proceeds with a rather
regular pattern: (i) $\mathcal{L}_{13,11} \simeq \mathcal{L}_{11,9}$
and $\mathcal{L}_{17,15} \simeq \mathcal{L}_{15,13}$ and (ii)
$\mathcal{L}_{19,17} \simeq \mathcal{L}_{17,15}/1.5 \simeq
\mathcal{L}_{13,11}/(1.5)^2$. Therefore it is possible to obtain an
extrapolated asymptotic result.

The extrapolation used is a Pad\'e approximation
\citep{FaA76,Besd72}
\begin{equation}
\label{Pade} \mathcal{L}^\infty_{K_{m}} = \mathcal{L}_{K_{m-8}}+
\mathcal{L}_{K_{m}-4,K_{m}-8}/
                         (1- \frac{ \mathcal{L}_{K_{m},K_{m}-4} }
                              { \mathcal{L}_{K_{m}-4,K_{m}-8} } ) \,.
\end{equation}

\subsection{Results and Discussion} In Fig.~\ref{Fig3} the results
for $\sigma_{\gamma,K_m}$ obtained from the inversions of the
transforms $\mathcal{L}_{K_{m}}$ are presented. Due to the Lorentz
kernel the $\sigma_{\gamma}$ presents the same features as
$\mathcal{L}$ itself: stable peak height with a value very close to
3 mb, and not yet completely converged peak position.

In Fig.~\ref{Fig3} the asymptotic Pad\'e approximations are given,
where $\sigma^\infty_{\gamma,17}$ and $\sigma^\infty_{\gamma,19}$
result from the inversions of $\mathcal{L}^\infty_{17}$ and
$\mathcal{L}^\infty_{19}$. Unquestionably, the extrapolated
$\mathcal{L}^\infty_{K_{m}}$ have a lower numerical quality than the
calculated $\mathcal{L}_{K_m}$ and consequently the stability for
the corresponding inversions worsens. Therefore, an additional
constraint in the inversion is imposed by fixing the peak cross
section to the already converged value of $3$ mb. In Fig.~\ref{Fig3}
it is evident that $\sigma^\infty_{\gamma,17}$ and
$\sigma^\infty_{\gamma,19}$ are very similar, hence establishing a
very good approximation for the asymptotic $\sigma_\gamma$. One also
notices that compared to $\sigma_{\gamma,19}$ they show a shift of
the peak position by about 1 MeV towards lower energy.

The final results for the cross--section are presented in
Fig.~\ref{Fig4}a. Due to the 3NF one observes a reduction of the
peak height by about 6\% and a shift of the peak position by about 1
MeV towards higher energy. Very large effects of the 3NF are found
above 50 MeV with an enhancement of $\sigma_\gamma$ by e.g. 18, 25,
and 35\% at $\omega=60$, 100, and 140 MeV, respectively. The
discussion in the previous section indicates that the 3NF effect
could change somewhat if all multipole contributions are considered.

It is very interesting to compare the 3NF effects on
$\sigma_\gamma(^4$He) to those found for $\sigma_\gamma(^3$H/$^3$He)
\citep{ELOT,Fad_CHH}. Surprisingly, the reduction of the peak height
is smaller for $^4$He. For $^3$H/$^3$He the size of the reduction is
similar to the increase of $E_b$ (10\%), whereas for $^4$He the 3NF
increases $E_b$ by 17\%, but reduces the peak by only 6\% and thus
cannot be interpreted as a simple binding effect.

One can find additional differences at higher energy--transfer. The
enhancement of $\sigma_\gamma(^4$He) due to the 3NF is significantly
stronger, namely about two times larger than for the three-body
case. Interestingly this reflects the above mentioned different
ratios between triplets and pairs in three- and four-body systems.
An integration of $\sigma_\gamma$ up to pion threshold yields 93.0
(AV18) and 97.4 MeVmb (AV18+UIX) thus showing a rather small
difference. If one considers also contributions beyond pion
threshold the difference becomes more pronounced, in fact
integrating up to 300 MeV one finds 103 (AV18) and 114 MeVmb
(AV18+UIX). The latter values correspond to enhancements of 72 \%
(AV18) and 91 \% of the classical Thomas-Reiche-Kuhn sum rule.
Different sum--rules will be discussed in the next section.

In Fig.~\ref{Fig4}a we compare the cross--section with the one
achieved using the semi-realistic Malfliet-Tjon (MT) NN potential
\citep{ELO97,BELO01}. Similar to $\sigma_\gamma(^3$H/$^3$He)
\citep{ELOT} one finds a rather realistic result in the giant
resonance region (overestimation of the peak by about 10-15\%) and
quite a correct result for the the peak position; however, at higher
energy $\sigma_\gamma$ is strongly underestimated for $^4$He, by a
factor of three at pion threshold. Also shown are data from
\citet{Ark}, the only measurement of $\sigma_\gamma(^4$He) in the
whole energy range up to pion threshold. In the peak region the data
agree best with the MT potential, while for the high-energy tail one
finds the best agreement with the AV18 potential.

A comparison of the low-energy results to further data is presented
in Fig.~\ref{Fig4}b. For the AV18+UIX case upper/lower bounds are
included to account for possible errors in the extrapolation,
Eq.~(\ref{Pade}). As bounds we take $\pm (\sigma^\infty_{\gamma,19}
- \sigma_{\gamma,19}$)/2; we believe that this is a rather safe
error estimate.

For a better understanding of the data in Fig.~\ref{Fig4}b a few
comments are needed:
\begin{enumerate}
  \item  In \citet{Wells} the peak cross section is determined from
Compton scattering via dispersion relations.
  \item  The dashed curve corresponds to the sum of cross sections
for $(\gamma,n)$ from \citet{Berman} and $(\gamma,p)^3$H from
\citet{Feldman} as already shown in \citet{ELO97}.
  \item The data from the above mentioned recent
$(\gamma,n)$ experiment \citep{Lund} are included only up to about
the three-body break-up threshold, where one can rather safely
assume that $\sigma_\gamma \simeq 2\sigma(\gamma,n)$ (see also
\citet{Sofia1}).
  \item In \citet{Shima} all open channels are
considered. One sees that the various experimental $\sigma_\gamma$
are quite different, exhibiting maximal deviations of about a factor
of two.
\end{enumerate}
The theoretical $\sigma_\gamma$ agrees quite well with the
low-energy data of \citep{Berman,Feldman}. In the peak region,
however, the situation is very unclear. There is a rather good
agreement between the theoretical $\sigma_\gamma$ and the combined
data of \citep{Lund} and \citep{Wells}, while those of
\citep{Berman,Feldman} are quite a bit lower. Very large
discrepancies  are found in comparison to the recent data of
\citet{Shima}. It is evident that the experimental situation is
rather unsatisfactory and further improvement is urgently needed.

\section{Photonuclear Sum Rules and the Tetrahedral Configuration of $^4$He}
Sum rules (SR) are related to moments of different order of the
photonuclear cross section and reflect important electromagnetic
properties of nuclei. In fact they can often be expressed in terms
of simple ground state properties in a model independent or quasi
model independent way. Well known examples are the
Thomas-Reiche-Kuhn (TRK) sum rule~\citep{TRK1,TRK2,TRK3}, which
gives information about the importance of exchange effects in
nuclear dynamics via the so-called TRK enhancement factor
$\kappa^{\rm{TRK}}$, the bremsstrahlung sum rule (BSR)
~\citep{LevingerBethe,Brink,Foldy,DellafioreBrink}, which is
connected to the nuclear charge radius and to the mean distance
between protons~\citep{DFlippa}, and the polarizability sum rule
(PSR)~\citep{Friar},  related to the electric nuclear
polarizability, which is proportional to the shift in energy levels
due to external electric field.

These SR assume that the dominant contribution to the cross section
comes from unretarded electric dipole (E1UR) transitions. Two- and
three-body studies~\citep{SaA,Fad_CHH} indeed confirm that other
contributions are much smaller. Much discussed is also the
Gerasimov-Drell-Hearn (GDH) sum rule~\citep{GDH1,GDH2}, which is
related to the nuclear anomalous magnetic moment.

In this chapter, the  TRK, BSR and PSR of $^4$He are considered
within a realistic nuclear potential model consisting of two- and
three-body forces (AV18 and UIX~\citep{AV18,UIX}). The GDH sum rule
is trivial for $^4$He: it vanishes, since the $^4$He total angular
momentum is equal to zero. Also investigated are the related moments
by integrating explicitly the properly weighted total
photoabsorption cross section calculated in the previous section.

The aim of this study is to show that in some cases sum rules can
allow an experimental access to two-body properties of the nuclear
ground state, like the proton-proton, neutron-neutron and
proton-neutron distances. In the case of $^4$He this allows to test
the validity of the configuration tetrahedral symmetry of this
nucleus and at the same time to ``measure'' the amount of symmetry
breaking.

This section aims also at providing a guideline for experiments, for
which a direct test of SR is difficult, as only lower bounds for the
SR can be determined, as well as at giving an idea of the
reliability of the SR approach to heavier systems, where the direct
theoretical determination of the cross section, and therefore its
integration, is presently out of reach. The advantage to perform
this kind of study in $^4$He, compared to analogous ones in the
two-~\citep{2bodySR} and three-body systems~\citep{ELO97pol}, is
that $^4$He is a rather dense and compact nucleus, resembling
heavier systems more closely. Only now that realistic theoretical
results for the photonuclear cross section are available such a
study is possible and one  can put the extrapolation of the results
to heavier systems on safer grounds.

\subsection{SR and the Nucleus Ground State}

We start by recalling the formalism of the photonuclear SR. The
various moments of the photonuclear cross section are defined as
\begin{equation}
m_n(\bar\omega) \,\,\, \equiv
\int_{\omega_{th}}^{\bar\omega}\,d\omega \,\omega^n\,
\sigma_\gamma^{\rm E1UR}(\omega)\,,\label{moments}
\end{equation}
where $\omega$ is the photon energy and $\omega_{th}$ and
$\bar\omega$ indicate threshold energy and upper integration limit,
respectively. Where $\sigma_\gamma^{\rm E1UR}(\omega)$ is the
unretarded dipole cross section, which can be written as
\begin{equation}
\sigma_\gamma^{\rm E1UR}(\omega)={\cal G}\, \omega\,
R(\omega)\,,\label{sigma}
\end{equation}
where ${\cal G}= 4\pi^2 \alpha/(3(2 J_0+1))$. $R(\omega)$ is the
nuclear response function to excitations of the operator ${\vec D}$
is the unretarded dipole operator:
\begin{equation}
{\vec D}=\sum_{i=1}^A {\vec r}_i \tau_i^{3}/2,
\end{equation}
where $A$ is the number of nucleons and $\tau^3_i$ and ${\vec r}_i$
are the third component of the isospin operator and the coordinate
of the $i$th particle in the center of mass frame, respectively. For
$n=0$ the SR is the TRK SR, $n=-1$ is the BSR, and $n=-2$ the PSR.

Assuming that $\sigma_\gamma^{\rm E1UR}(\omega)$ converges to zero
faster than $\omega^{-n-1}$ and applying the closure property of the
eigenstates of the hamiltonian $H$, it is straight forward to get
the following identities for SR with $n=0,-1,-2$:
\begin{eqnarray}
\Sigma^{\rm TRK}&\equiv&m_0(\infty)\,
 = \,\frac{{\cal G}}{2}\,
\langle 0| \left[{\vec D}  ,\left[ H, {\vec D}
\right]\right]|0\rangle
\label{mTRK}\\
\Sigma^{\rm BSR}&\equiv& m_{-1}(\infty)\,
 = \,{\cal G}\,\langle 0|  {\vec D} \cdot  {\vec D}|0
\rangle \,
\label{mBSR} \\
\Sigma^{\rm PSR}&\equiv& m_{-2}(\infty)\,
 = \,{\cal G}\,\sum_n (E_n-E_0)^{-1}|\langle n|{\vec D}|0\rangle|^2\,,\label{mPSR}
\end{eqnarray}

Working out the expressions in Eqs.~(\ref{mTRK}~-~\ref{mPSR}), one
finds that those moments are related to interesting properties of
the system under consideration. In fact the TRK sum rule is also
given by the well known relation~\citep{TRK1,TRK2,TRK3}
\begin{equation}
\Sigma^{\rm TRK}\,=\,{\cal G} \frac{3NZ}{2mA}
\left(1+\kappa^{TRK}\right),\label{TRK}
\end{equation}
where $N$ and $Z$ are the neutron and proton numbers, respectively,
$m$ is the nucleon mass and $\kappa^{TRK}$ is the so-called TRK
enhancement factor defined as
\begin{equation}
\kappa^{TRK} \equiv \frac{mA}{3NZ} \langle 0|[ {\vec D},[ V, {\vec
D}]]|0\rangle\,. \label{kappa}
\end{equation}
From this expression it is evident that $\kappa^{TRK}$  embodies the
exchange effects of the nuclear potential $V$, since the double
commutator in (\ref{kappa}) vanishes for systems like atoms, where
no exchange effects are present.

In the literature one finds a few interesting equivalences for the
bremsstrahlung sum rule. Rewriting the dipole operator as ${\vec D}
= (NZ/A){\vec R}_{PN}$, where ${\vec R}_{PN}$ denotes the distance
between the proton and neutron centers of mass, one
has~\citep{Brink}
\begin{equation}
\Sigma^{\rm BSR}\,={\cal G}\,\left(\frac{N Z}{A}\right)^2\langle
0|R_{PN}^2|0\rangle\,. \label{BSRBrink}
\end{equation}
\citet{Foldy} demonstrated that
\begin{equation}
\Sigma^{\rm BSR}\,={\cal G}\,\frac{N  Z }{A-1}\langle r_p^2
\rangle\,, \label{BSRFoldy}
\end{equation}
where $ \langle r_p^2 \rangle$ is the mean square (m.s.) point
proton radius
\begin{equation}
\langle r_p^2 \rangle\equiv \frac{1}{Z}\langle 0|\sum_{i=1}^Z r_i^2
|0\rangle\,. \label{rpsquare}
\end{equation}
However, this relation is valid only under the assumption that the
ground state wave function is symmetric in the space coordinates of
the nucleons.

\citet{DellafioreBrink} have found that, in the framework of the
oscillator shell model, one has
\begin{equation}
\Sigma^{\rm BSR}\,={\cal G}\,\left(Z^2\langle r_p^2 \rangle -
Z\langle r_p^{'2}\rangle\right)\,, \label{BSRDellafioreBrink}
\end{equation}
where $\langle r_p^{'2}\rangle$ is the m.s. distance of protons with
respect to the proton center of mass ${\vec R}_P$
\begin{equation}
\langle r_p^{'2} \rangle\equiv \frac{1}{Z}\langle 0|\sum_{i=1}^Z
({\vec r}_i-{\vec R}_P)^2 |0\rangle\,. \label{rpprimesquare}
\end{equation}
Later, in \citet{DFlippa}, it was shown that the validity of
Eq.~(\ref{BSRDellafioreBrink}) is not limited to the oscillator
shell model, but it is a model independent relation, which can also
be written as
\begin{equation}
\Sigma^{\rm BSR}\,={\cal G}\,\left(Z^2\langle r_p^2 \rangle -
\frac{Z(Z-1)}{2}\langle r_{pp}^2 \rangle\right)\,,
\label{BSRDFLippa}
\end{equation}
where $\langle r_{pp}^2\rangle$ is the m.s. proton-proton distance
\begin{equation}
\langle r_{pp}^2 \rangle \equiv \frac{1}{Z(Z-1)}\langle
0|\sum_{i,j=1}^{Z} ({\vec r}_i- {\vec
r}_j)^2|0\rangle\,.\label{r_pp}
\end{equation}

For the BSR two additional relations exist, which are easy to prove,
but, to our knowledge, have not been considered in the literature,
viz
\begin{equation}
\Sigma^{\rm BSR}\,  =  \,{\cal G}\,\left(N^2\langle r_n^2 \rangle -
\frac{N(N-1)}{2}\langle r_{nn}^2 \rangle\right)\, \label{BSR3}
\end{equation}
and
\begin{equation}
\Sigma^{\rm BSR}\,  =  {\cal G}\, \frac{NZ}{2} \left( \langle
r_{pn}^2 \rangle\,-\, \langle r_p^2 \rangle \,-\, \langle
r_{n}^2\rangle \right) \,,\label{BSR4}
\end{equation}
where $\langle r_n^2\rangle$ is the m.s. point neutron radius and
$\langle r_{\alpha\beta}^2 \rangle$ are the m.s. nucleon-nucleon
(NN) distances, i.e.
\begin{eqnarray}
\langle r_{nn}^2 \rangle &\equiv& \frac{1}{N(N-1)}\langle
0|\sum_{i,j=1}^{N}
({\vec r}_i- {\vec r}_j)^2|0\rangle\,,\label{r_nn}\\
\langle r_{pn}^2 \rangle &\equiv& \frac{1}{NZ}\langle 0|\sum_{i=1}^Z
\sum_{j=1}^N ({\vec r}_i- {\vec r}_j)^2|0\rangle\,.\label{r_pn}
\end{eqnarray}
It is interesting to note that Eqs.~(\ref{BSRDFLippa}),~(\ref{BSR3})
and~(\ref{BSR4}) express $\Sigma^{\rm BSR}$ via a one-body ($\langle
r_\alpha^2 \rangle$) as well as  a two-body quantity ($\langle
r_{\alpha\beta}^2\rangle$).

Finally, regarding the polarizability sum rule, one has
\begin{equation}
\Sigma^{PSR}= 2 \pi^2 \alpha_D\,,
\end{equation}
where $\alpha_D$ denotes the nuclear polarizability in the E1UR
approximation.

\subsection{Calculation of SR using the LIT method}
There are various approaches for the calculation of moments and sum
rules. The obvious one is to obtain the moments by integrating the
$^4$He total photoabsorption cross section of
Sec.~\ref{sec:Photo_crs}. However, a more direct approach to obtain
the SR exist, as explained in the following.

One way to evaluate the LIT is by using the Lanczos technique
described in Subsec.~\ref{sub:LIT_Lanczos}. We recall that the LIT
can be re-expressed as
\begin{equation} \label{loreins}
{\cal  L}(\sigma_R, \sigma_I)=\frac{1}{\sigma_I}\,
            \langle 0 | {\vec D}\cdot {\vec D}|0\rangle\,
            \mbox{Im} \, x_{00}(z) \mbox{}
\end{equation}
with $z=E_0+\sigma_R+ i\sigma_I$ and
\begin{equation} \label{start}
  |\phi_0\rangle =\frac{{\vec D}|0\rangle }{\sqrt{\langle
  0|{\vec D}\cdot {\vec D}|0 \rangle }} \,.
\end{equation}

\noindent $x_{00}(z)$ can be expressed (cf. Eq.~(\ref{eq:x00}) as a
continued fraction containing the Lanczos coefficients $a_i$ and
$b_i$, defined in Eq.~(\ref{eq:Lanczos_basis_def}).

Therefore, the implementation of the Lanczos algorithm leads to
${\cal L}(\sigma_R,\sigma_I)$. While the inversion of the LIT
\citep{EF99a} gives access to $R(\omega)$, and thus to the moments
of Eq.~(\ref{moments}), the normalization of the Lanczos "pivot"
$|LP\rangle= {\vec D}|0 \rangle$ and the Lanczos coefficients allow
to obtain the SR of Eqs.~(\ref{mTRK}-\ref{mPSR}). In fact one has:
\begin{eqnarray}
\Sigma^{\rm PSR} & = & {\cal G}\,x_{00} (E_0)\,,\label{LPSR}\\
\Sigma^{\rm BSR} & = & {\cal G}\,\langle LP |LP\rangle\,,\label{LBSR}\\
\Sigma^{\rm TRK} & = & (a_0- E_0)\,\Sigma^{\rm BSR} \,.\label{LTRK}
\end{eqnarray}

\subsection{Results and Discussion}
We use a HH basis, therefore the ground state $|0\rangle$, the
Lanczos "pivot" $|LP\rangle$ and the Lanczos coefficients $a_n$ are
given in terms of HH expansions.  While for the ground state the
expansion is characterized by an even hyperspherical grand-angular
quantum number $K$ and total isospin T=0, $|LP\rangle$ has to be
expanded on $\{K'=K+1, T=1\}$  states (we neglect the AV18 isospin
mixing, which is very small, as shown for the $^4$He ground state
in~\citet{NO02,VI05}). The rate of convergence of the various SR
results from Eqs.(\ref{LPSR}-\ref{LTRK}) is given in
Tab.~\ref{Table1} as a function of the hyperspherical grand-angular
quantum number $K$. One observes sufficiently good convergence
patterns for $\Sigma^{\rm TRK}$ and $\Sigma^{\rm BSR}$. For the
latter an additional test of the convergence is performed,
calculating $\Sigma^{\rm BSR}$ directly as mean value of the
operator ${\vec D} \cdot {\vec D}$ on the ground state
(Eq.~(\ref{mBSR})). In this way the expansion of $|LP\rangle$ on
$\{K'=K+1, T=1\}$ states is avoided. We obtain practically identical
results.

From Tab.~\ref{Table1} one sees that the convergence of $\Sigma^{\rm
PSR} $ is slower when the three-nucleon force (3NF) is included.
This is not a surprise, in view of the problem found for the cross
section itself in the previous section. In fact it has been shown
that the peak of the giant dipole resonance is slightly shifting
towards lower energies with increasing $K$. The sum rule
$\Sigma^{\rm PSR}$, which has the strongest inverse energy
weighting, is more sensitive to this shift than the other two SR.
\begin{table}
\begin{center}
\begin{tabular}
{|c | ccc|} \hline\hline &&
                \,\, AV18+UIX\,\, & \\
\hline
&$\,\Sigma^{\rm PSR}\,$ &$\,\Sigma^{\rm BSR}\,$ &$\,\Sigma^{\rm TRK}\,$\\
K &$\,10^{-2}$[mb MeV$^{-1}$]\, & \,[mb]\, & \,$10^{2}$[mb MeV]\,\\
\hline\hline
  8 &  6.230 & 2.398 & 1.430 \\
 10 &  6.277 & 2.396 & 1.448 \\
 12 &  6.331 & 2.394 & 1.451 \\
 14 &  6.382 & 2.401 & 1.458 \\
 16 &  6.434 & 2.406 & 1.460 \\
 18 &  6.473 & 2.410 & 1.462 \\
\hline\hline
&&AV18 &\\
\hline &7.681 & 2.696 & 1.383 \\
\hline\hline
&& \,\, AV18+UIX\,\, &\\
\hline
 $\bar\omega$ [MeV]&$\,m_{-2}\,(\bar\omega) \,$ & $\,m_{-1}\,(\bar\omega) \,$ & $\,m_{0}\,(\bar\omega) \,$\\
\hline
 $135$ &  6.55 & 2.27 & .944 \\
 $300$  &  6.55 & 2.37 & 1.14 \\
\hline\hline
\end{tabular}
\caption{ \label{Table1} Convergence in K of the SR for AV18+UIX
potentials. The converged AV18 results are also shown. The last two
lines of the table show the convergence in $\bar\omega$ of the
various moments. }
\end{center}
\end{table}

We have used $\sigma_\gamma^{\infty}$ extrapolated in the previous
section, to determine from Eq.~(\ref{moments}) the various moments
for $\bar\omega=$ 300 and 135 MeV. These results are also listed in
Table~\ref{Table1}.

For $\Sigma^{\rm TRK}$ one sees that the SR is not yet exhausted at
300 MeV. In fact more than 20\% of the strength is still missing. At
pion threshold, one has only about 2/3 of $\Sigma^{\rm TRK}$. As was
already discussed in~\citep{ELOT} for the triton case, the rather
strong contribution from higher energies seems to be connected to
the strong short range repulsion of the AV18 potential. As to the
TRK enhancement factor the present calculation gives $\kappa^{\rm
TRK}=1.31$ for AV18 and 1.44 for AV18+UIX. These numbers are
somewhat larger than older results obtained either with a
variational wave function and AV14+UVII potential ($\kappa^{\rm
TRK}=1.29$ ~\citep{Schiavilla}) or with more approximated wave
functions and various soft and hard core NN potentials ($\kappa^{\rm
TRK}=0.9 - 1.30$ ~\citep{Weng,Horlacher,Gari}).

For $\Sigma^{\rm BSR}$, as expected, the contribution at high energy
is much smaller than for $\Sigma^{TRK}$, in fact at $\bar\omega=
300$ MeV the missing sum rule strength is less than 2\%. For
$\Sigma^{\rm PSR}$ the strength beyond 300 MeV is even more
negligible. Actually in this case the explicit integration leads to
an even higher result than the sum rule evaluation of
Eq.~(\ref{LBSR}). The seeming contradiction is explained by the
already discussed fact that $\Sigma^{\rm PSR} $ is not yet
convergent, while for the explicit integration an  extrapolated
cross section is used. Indeed, integrating the K=18 cross section we
obtain a value of 6.46 mb, which is consistent with the
corresponding sum rule result of 6.47 mb, within the numerical error
of the calculation. For the AV18+UIX force the value of the
polarizability $\alpha_D$ that we deduce from the extrapolated
$\Sigma^{\rm PSR}$ is $0.0655$ fm${^3}$. The AV18 result, which
already shows a good convergence for $K=16$, is  0.0768 fm${^3}$.
This means that the 3NF reduces the polarizability by 15\%. It would
be very interesting to measure this nuclear polarizability by
Compton scattering, as a test of the importance of the three-body
force on such a classical low-energy observable. We find  that
$\Sigma^{\rm PSR}$ is the SR that is affected most by the 3NF. In
fact $\Sigma^{\rm BSR}$ is reduced by only 10\% and one has an
opposite effect on $\Sigma^{\rm TRK }$ with a 5\% increase. The
quenching or enhancement of SR due to the 3NF is the reflection of
its effects on the cross section i.e a decrease of the peak and an
increase of the tail.

This work allows a few conclusions regarding a very old question,
already discussed in~\citet{LevingerBethe}, of the ``existence'' of
the SR (finiteness of $m_n(\bar\omega)$ for
$\bar\omega\rightarrow\infty)$, which is connected to the
high-energy fall-off of the E1UR cross section. Since there is a
rather good consistency  between the SR and the moment values, it
can be stated with a rather high degree of confidence that
$\Sigma^{\rm TRK} $ (and consequently the other two SR) ``exist''.
Therefore one can try to extract some information about the high
energy behavior of the cross section and hence about the
``existence'' of SR with higher $n$. With an  $\omega^{-p}$ ansatz
for the fall-off of the cross section above pion threshold, and
requiring that $\Sigma^{\rm TRK} $ is 146 mb MeV (see
Table~\ref{Table1}), one gets a rather weak energy fall-off, i.e. $
p\simeq 1.5$. This value is also consistent with $\Sigma^{\rm BSR}$.
In fact, adding such a tail contribution to $m_{-1}(135)$ one gets
$\Sigma^{\rm BSR}=2.39 $ mb, to be compared with 2.41 mb in
Table~\ref{Table1}. The value of $p$ might be somewhat different for
other potentials, but probably it will not change much. An
additional, rather safe, conclusion is that higher order SR do not
exist for realistic nuclear potential models.

A very interesting aspect can be viewed by observing the BSR. As
already mentioned $\Sigma^{\rm BSR}$ contains information about one-
and two-body densities via $\langle r_\alpha^2 \rangle$ and $\langle
r_{\alpha\beta}^2\rangle$, respectively. This means that a
measurement of $\Sigma^{\rm BSR}$ and the knowledge of the
experimental m.s. radius allow to determine $\langle
r_{\alpha\beta}^2\rangle$ via Eqs.~(\ref{BSRDFLippa}),(\ref{BSR3})
and~(\ref{BSR4}). In this way one gets information about the
internal configuration of $^4$He as it is explained in the
following.

In his derivation of Eq.~(\ref{BSRFoldy}), Foldy assumed a totally
symmetric $^4$He spatial wave function, which corresponds to a
configuration where the four nucleons are located at the four
vertexes of a tetrahedron. For such a configuration one has $\langle
r_p^2\rangle=\langle r_n^2\rangle= \langle r^2\rangle$ and $\langle
r_{pp}^2\rangle=\langle r_{nn}^2\rangle= \langle r_{np}^2\rangle$
with $Q_T\equiv\langle r_{\alpha\beta}^2\rangle/\langle
r^2\rangle=8/3$. Foldy's assumption is a very good approximation for
$^4$He, but other spatial symmetries (mixed symmetry, antisymmetric)
are also possible. \\
{\it{What can be learned from $\Sigma^{\rm BSR}$ with respect to
this question?}}\\
For $^4$He, which is a T=0 system one can safely assume that
$\langle r^2_p\rangle = \langle r^2\rangle= \langle r_n^2\rangle$
(isospin mixing is tiny~\citep{VI05,NO02}). Using in
Eqs.~(\ref{BSRDFLippa}),~(\ref{BSR3}) and~(\ref{BSR4}) the AV18+UIX
value of $\langle r^2\rangle = 2.04$ fm$^2$ from
Sec.~\ref{sec:Photo_crs} (which coincides with the experimental
one~\citep{radiusexp}, corrected for the proton charge radius) and
$\Sigma^{\rm BSR}$ from Table~\ref{Table1}, one obtains $\langle
r^2_{pp}\rangle=\langle r^2_{nn}\rangle=5.67$ fm$^2$ and $\langle
r^2_{pn}\rangle=5.34$
fm$^2$, i.e. two values which differ by about 6\%. \\
The ratios $Q_{pp(nn)}\equiv\langle r^2_{pp(nn)}\rangle/\langle
r^2\rangle$ and $Q_{np}\equiv\langle r^2_{np}\rangle/\langle
r^2\rangle$ are not much different from the correspondent value
$Q_T$ of a classical tetrahedral configuration. We obtain
$Q_{pp}=2.78$ and $Q_{np}=2.62$ instead of  $Q_T=2.67$. One notices
that $Q_{pp(nn)}-Q_T \simeq 2 (Q_{np}-Q_T)$. This reflects the
different numbers of proton-proton and neutron-neutron pairs (2)
with respect to  proton-neutron pairs (4). Using
Eq.~(\ref{BSRBrink}) one can also derive the distance between the
proton and neutron centers of mass. One has $R_{pn}=$ 1.58 fm
instead of 1.65 fm for the tetrahedral configuration.

Notice that with $\langle r^2\rangle = 2.04$ fm$^2$ the
"tetrahedral" BSR would be 2.62 mb, the same value that one obtains
using Eq.~(\ref{BSRFoldy}). This value is  9\% larger than our
result. The distortions that I find from the 9\% smaller BSR are the
consequence of the different effects of the potential on isospin
triplet and isospin singlet pairs.

A rather intriguing conclusion arises: when considered in its body
frame $^4$He should look like a slightly {\it deformed} tetrahedron.
Of course this statement has to be interpreted in a quantum
mechanical sense, regarding the mean square values of the
nucleon-nucleon distances on the two-body density. It is clear that
one cannot measure this {\it deformation} "directly", since it is
not a deformation of the one-body charge density (the $^4$He charge
density has only a monopole). On the other hand such a {\it
deformation} is accessible experimentally in an indirect way via the
measurements of the charge radius and of $\Sigma^{\rm BSR}$.

This leads to the question how exactly $\Sigma^{\rm BSR}$ can be
measured in a photonuclear experiment. Two points have to be
addressed: i) the contributions of E1 retardation and higher
multipoles, which are not contained in $\Sigma^{\rm BSR}$, but which
will contribute to the experimental cross section and ii) the
contribution of the high energy tail. \\
Regarding point i), additional effects of the E1 retardation and
higher multipoles have been calculated in Ref.~\citep{Fad_CHH} for
the AV18+UIX potentials in a Faddeev calculation. Using those
results one finds that E1 retardation and higher multipoles increase
$m_{-1}(135)$ by about 1\% only. In fact according to
Gerasimov\citep{Gerasimov} there is a large cancelation of E1
retardation and other multipoles. There is no reason for a larger
effect in $^4$He. On the contrary, since the leading isovector
magnetic dipole (M1) transition is suppressed in this nucleus (it is
zero for an S-wave) one can expect an even smaller contribution. As
to point ii),  considering the fall-off of the triton E1UR cross
section around pion threshold from Ref.~\citep{Fad_CHH} one obtains
a result very similar to our $^4$He case, namely $p \simeq 1.5$.
However, including the other multipole contributions one gets a
considerably smaller value, namely $p \simeq 1.1$. This is no
contradiction with the small increase of 1\% for $\Sigma^{\rm BSR}$,
since the inverse energy weighted cross section integrated from 100
to 135 MeV gives only a rather small contribution to the sum rule.
On the other hand one would overestimate $\Sigma^{\rm BSR}$ taking
the full cross section. Thus it is suggested that the tail
contribution to an experimental BSR should be estimated using the
theoretically established fall-off $\omega^{-p}$ with $p \simeq
1.5$.

%% file: nu_scattering.tex
\chapter{Neutrino Scattering on Light Nuclei in Supernovae}
\label{chap:neu}

The inelastic scattering of neutrino off $A=3$ nuclei and $^{4}$He
is calculated microscopically at energies typical for core collapse
supernova environment. The calculation is carried out with the
Argonne $v_{18}$ nucleon--nucleon potential and the Urbana IX three
nucleon force. Full final state interaction is included via the
Lorentz integral transform (LIT) method. The contribution of axial
meson exchange currents to the cross sections is taken into account
from effective field theory of nucleons and pions to order
${\cal{O}}(Q^3)$. The main results of the chapter are based on a
series of papers published in \citet{GA04,GA07a,GA07b,GA07,v3H}.

The chapter starts with an overview on core collapse supernovae,
emphasizing the abundances and role of $^4$He and A=3 nuclei in the
processes taking place in the SN environment. The cross-sections are
then presented and discussed. Finally, neutrino mean-free-paths in
the neutrinosphere region are evaluated.

\section{Core-collapse Supernovae}
A massive star, above about $10$ solar masses, evolves over millions
of years, to create an onion--like structure of layers, with lighter
elements in the atmosphere and heavier in the center, where the
gravitational pressure is highest. The cycle of fusion stops when
the core burns Si nuclei to iron. Further burning can not be a
source of energy, as the binding energy per nucleon is peaked in
iron. Consequently, the evolution of the star comes to an end.
Observations have shown that this death is not peaceful, it is
followed by an enormous explosion, shining in a magnitude of a
galaxy, believed to occur in the following chain of events.

When the core reaches the Chandrasekhar mass it cannot support its
own gravity and collapses. The collapse halts only when the core
reaches nuclear density, in which short--range nuclear forces
dominate the dynamics. This initiates an outgoing shock wave. As the
shock travels through the outer layers of the core it losses energy
due to thermal neutrino radiation, and dissociation of iron nuclei
(bound by about 8 MeV/nucleon) to $\alpha$ particles and free
nucleons.

The fate of this shock is a matter of debate. In the ``prompt
shock'' mechanism, the shock has enough initial energy to burst
through the core and cause an explosion. However, in modern
calculations, about $100$ miliseconds after bounce the shock stalls
at about $200$ km from the center and becomes an accretion shock,
with matter from the outer layers of the star falling through it.
During this period the newly born proto-neutron star in the center
of the star cools by neutrino emission. This huge burst of
neutrinos, in total energy of almost $10^{53}$ ergs, is believed to
deposit enough energy in the matter below the shock to revive the
shock. This so called ``delayed shock'' mechanism has not been
quantitatively proved in numerical calculations, but received a
qualitative proof in the observation of SN 1987A. Comparing the
energy of the neutrino--burst to the gravitational well of the star,
it is easy to realize that one needs only 1\% of neutrino energy to
be transferred to matter. This turns to be a highly delicate
problem, in which the accuracy of the microscopic input is a crucial
factor. The heating is achieved through elastic scattering on
nucleons and electrons. \citet{HA88a} has suggested that inelastic
scattering on nuclei can increase neutrino--matter coupling. The
higher energy of the heavy flavored neutrinos\footnote{The
characteristic temperatures of the emitted neutrinos are about
$6-10$ MeV for $\nu_{\mu,\tau}$ ($\bar{\nu}_{\mu,\tau}$), $5-8$ MeV
for $\bar{\nu}_e$, and $3-5$ MeV for $\nu_e$} favors neutral
scattering on nuclei as an energy transfer method. Thus, neutrino
interaction with nuclei, which exist below the shock, is believed to
have an important part in the explosion process. This will be
discussed thoroughly in Subsec.~\ref{sub:shock}.

Another outcome of the huge neutrino flux is changing the chemical
evolution of the star, mainly by break--up of nuclei into fragments
of nuclei which act as seeds for different nucleosynthesis
processes. $^4$He break--up by neutrinos has a role in this process,
which will be reviewed in Subsec.~\ref{sub:nucleo}.

\subsection{Shock revival and Light Nuclei} \label{sub:shock}
\begin{figure*}
\begin{center}
\includegraphics[scale=0.4,clip=0.1]{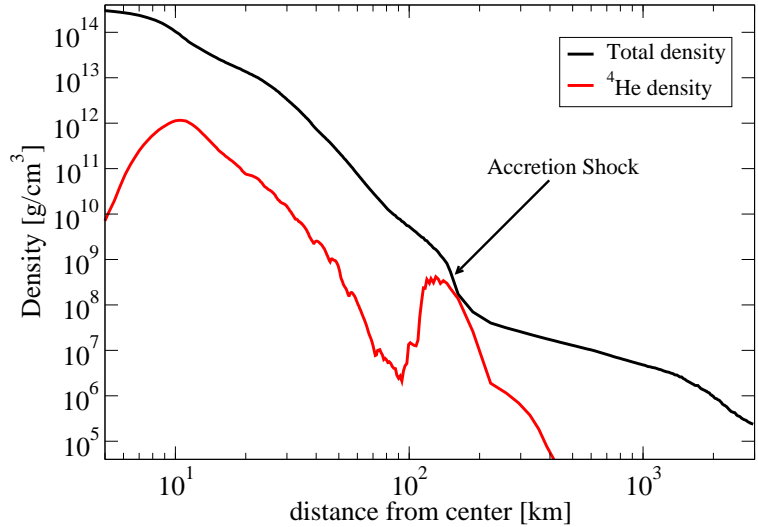}
\caption{Density profile of $11 \text{M}_{\odot}$ star, 120
miliseconds after bounce. The black line is the total density, and
the red line is the $\alpha$ mass fraction. One can view the
accretion shock at about 200 km off center.}
\label{fig:density_profiles}
\end{center}
\end{figure*}

In order to estimate how significant is neutrino interaction with
light nuclei in the shocked region, one needs to know not only the
neutrino luminosity but also the mean--free--paths of neutrinos in
the region. Thus one needs the composition of the shocked regions
and the cross--sections for neutrino scattering on light nuclei. In
Fig.~\ref{fig:density_profiles} one can view the density profile of
an $11 \text{M}_{\odot}$ progenitor, 120 miliseconds after bounce
\citep{eli_private_communication}. The calculation is 1D simulation,
with full neutrino--transport, and using the equation of state
developed by \citet{Shen}. The red line in the Figure is the
$\alpha$ mass--fraction, which is the most abundant nuclei in the
shocked region by this calculation. The rest of the composition are
nuclei and free electrons. One can see that in this simulation the
amount of $^4$He is about 1\% up to 100 km off center and rises to
around $10-20\%$ just below the shock. Thus neutrino scattering off
$^4$He could play a role in the revival of the shock, as first
indicated by \citet{HA88a}. Moreover, the energy transfer due to
elastic scattering is low, $\omega \sim T^2/m$, therefore inelastic
scattering could be as important as elastic scattering on nucleons.
One has to keep in mind that with every inelastic scattering there
is a substantial release of energy -- larger than the break--up
energy of $20 \mev$. A complete study on the role of $^4$He
excitation on the shock revival was not done, and is called for.
However, a first study was carried out by ~\citep{OH06}, and showed
a small effect on accretion shock instabilities.

Nuclear statistical equilibrium (NSE) models, like those used by
\citet{Shen}, predict abundances based on binding energies and the
quantum numbers of nuclei. However, NSE models only treat
approximately (or neglect) strong interactions between nuclei, and
consequently break down as the density increases. Recently a
description of low-density nuclear matter (composed of neutrons,
protons and alpha particles) in thermal equilibrium based on the
virial expansion was developed\citep{vEOSnuc,vEOSneut}. The virial
equation of state (EoS) systematically takes into account
contributions from bound nuclei and the scattering continuum, and
thus provides a framework to include strong-interaction corrections
to NSE models. This formalism makes model-independent predictions
for the conditions near the neutrinosphere, i.e. for densities of
$\rho \sim 10^{11-12} \gcq$ and high temperatures $T \sim 4
\mev$~\citep{sn1987a1,sn1987a2}. In particular, the resulting alpha
particle concentration differs from all equations of state currently
used in SN simulations, and the existence of trinuclei in this area
is expected. For example, in Fig.~\ref{fig:density_profiles} the
region of about $10-30$ km off center is characterized by this
density range. Thus, the calculation substantially underestimates
the abundances of $A=3,\,4$ nuclei in this region.

To determine the abundance of $A=3$ nuclei near the neutrinosphere
in supernovae, we presented in \citet{v3H} an extension of the
virial approach to the EoS to explicitly include neutrons, protons,
$\alpha$ particles, $^3$H and $^3$He nuclei; deuterons are included
as a bound state contribution to the proton-neutron virial
coefficient.

\begin{figure*}[t]
\begin{center}
\includegraphics[scale=0.45,clip=]{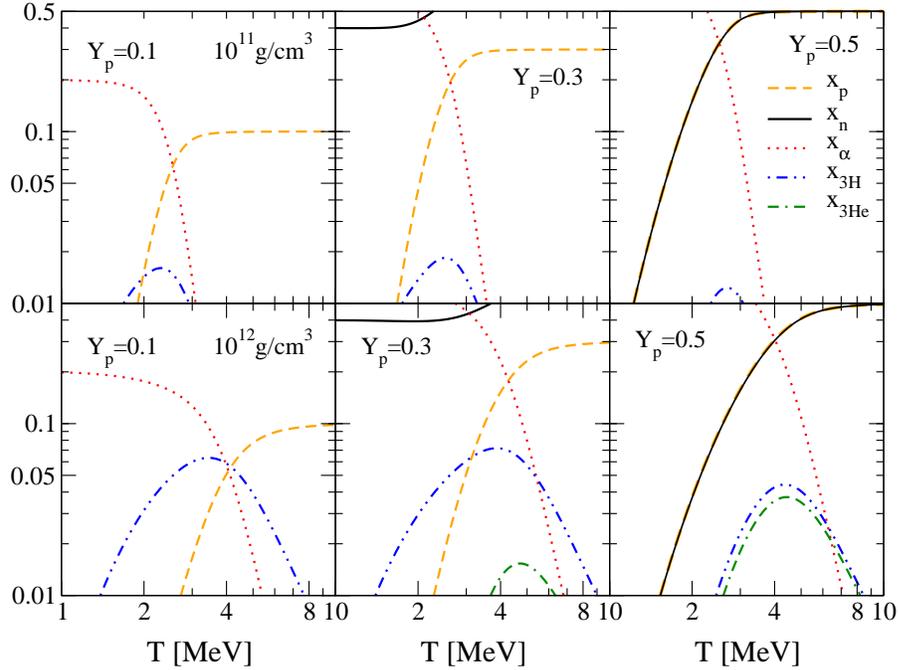}
\caption{Mass fractions of nucleons and $A=3,\,4$ nuclei in chemical
equilibrium as a function of temperature T. The top and bottom rows
correspond to a density of $10^{11}\gcq$ and $10^{12}\gcq$
respectively, and from left to right the proton fractions are $Y_p =
0.1, 0.3$ and $0.5$.} \label{fig:composition}
\end{center}
\end{figure*}

The resulting mass fractions $x_i = A_in_i/n_b$ ($n_i$ is the
density of the $i$ element of mass $A_i$, and $n_b$ is the baryon
density), are shown in Fig.~\ref{fig:composition} for neutrinosphere
densities and temperatures, and various proton fractions $Y_p = (n_p
+ 2n_\alpha + 2n_{\rm{^3He}} + n_{\rm{^3H}})/n_b$, i.e. the ratio of
proton number to baryon number.

The main result of this EoS is the model--independent prediction
that the mass-three fraction can be significant (up to $10 \%$) near
the neutrinosphere. Thus, we present in Sec.~\ref{sec:A3_crs}
predictions for the neutral-current inclusive inelastic
cross-sections on mass-three nuclei, based on microscopic two- and
three-nucleon interactions and meson-exchange currents, including
full final-state interactions via the Lorentz integral transform
(LIT) method.

\subsection{Nucleosynthesis and $\nu$ -- $\alpha$ Interaction}
\label{sub:nucleo}

Due to the high temperature of $\mu$ and $\tau$ neutrinos (and
anti-neutrinos), a substantial amount of them carry more than
$20\mev$, and may dissociate the $^4$He nucleus through inelastic
neutral current reactions. This creates the seed to light element
nucleosynthesis in the surrounding stellar envelope \citep{WO90}. As
a part of this explosive nucleosynthesis, or ``$\nu$
-nucleosynthesis'', a knock out of a nucleon from a $^4$He nucleus
in the helium rich layer, followed by a fusion of the remaining
trinucleus with another $\alpha$ particle, will result in a
$7$--body nucleus. This process is an important source of $^7$Li,
and of $^{11}$B and $^{19}$F through additional $\alpha$ capture
reactions. In fact the production of $^{11}$B in supernovae (also
due to neutrino interaction on $^{12}$C) can be calibrated by
measuring ratio of $^{11}$B to $^{10}$B in meteorites. \citet{WW95}
have suggested that the calculated abundance is too large.
\citet{YO05} have shown that the Li and B production depends highly
on the neutrino temperature, and mainly on the spectrum character of
the high energy tail (due to the high dissociation energy of
$\alpha$). They used this observation to fit the neutrino
temperature to reproduce the measured abundance. However, a correct
description of the process must contain an exact, energy dependent
cross-section for the neutral inelastic $\alpha-\nu$ reaction, which
initiates the process.

The relatively low temperature of $\nu_e$ and $\bar \nu_e$ emitted
from the core suppress the probability for inelastic reactions of
these neutrinos with $^4$He in the supernova scenario. However, the
mixing of the first and the third neutrino mass eigen--states is
enhanced, due to matter effects, in the Oxygen/Carbon layers, just
below the He layer. This significantly changes the spectrum of
electron (anti--)neutrinos reaching the He layer, and yields a
secondary source of energetic electron neutrinos. As indicated by
\citet{YO06}, the resulting charge current inelastic reactions on
$^4$He could double the aforementioned yields. In this reference,
they speculate that a test for neutrino oscillation properties could
be observing the ratio of $^7$Li to $^{11}$B in stars. Yet again,
this could lead to quantitative conclusions only if neutrino--$^4$He
cross--sections are available and accurate.

Elements heavier than iron are produced mainly by a different
nucleosynthesis process -- neutron capture. The so called
$r$--process, in which this capture is rapid, is still missing a
neutron--source in nature. \citet{HA88b} have suggested that
neutrons produced by neutrino spallation of $^4$He in the helium
rich layer could be such a source.

However, the most probable source of $r$-process neutrons is the
neutrino--driven wind blowing from the surface of a newly born
proto-neutron star (PNS). This is a high entropy region, thus all
protons combine with neutrons to form $^4$He, with excess neutrons
left over. This makes the $r$-process picture puzzling, as the free
neutrons that initially accompany the $^4$He, are then converted to
protons by charge current reactions, which then form more $^4$He
\citep{WH92}. Thus it is very hard to maintain a neutron excess
(necessary for the r-process) in the wind. Neutrino cross sections
for $^4$He breakup are part of the detailed balance that determines
what residual neutron abundance will remain in a neutrino-driven
wind.

\section{Weak Interaction Observable -- Beta Decay Rate}
\label{sec:d_r_cal}

In order to give parameter free {\it predictions} of weak processes,
one needs the LECs of the EFT. There is only one unknown LEC, which
is calibrated to fit a weak interaction observable.

A most researched weak interaction observable is the $\beta$ decay.
In this decay, a neutron (which can be free or bound in a nucleus)
spontaneously transforms into a proton, accompanied by electron and
anti--neutrino emission. This instability could be found in many
elements, transforming one element to another. The lightest elements
undergoing this decay are: $n \rightarrow p + e^{-} + \bar{\nu}_e$,
$^3\text{H} \rightarrow ^3\text{He}+ e^{-} + \bar{\nu}_e$ and
$^6\text{He} \rightarrow ^6\text{Li}+ e^{-} + \bar{\nu}_e$.

The $\beta$ decay rate provides a direct test for the weak
interaction model and the axial currents in the nucleus. This aspect
is even more appealing as the $\beta$ transitions are between
bound--states of nuclei, and usually include small momentum
transfer. Thus, the decay is governed by the Gamow--Teller and Fermi
operators, which are proportional to the low momentum transfer
approximation of $E^A_1$ and $C^V_0$ operators. In fact, to leading
order the $\beta$ decay rate is simply given by (see for example
\citep{Wiringa_6He})
\begin{equation}
\Gamma_{fi}^\beta=f_t\frac{G^{(+)2} m_e^5}{2\pi^3}
\big({|\text{F}|^2+g_A^2 |\text{GT}|^2}\big)
\end{equation}
where the Gamow Teller and Fermi reduced matrix elements:
\begin{eqnarray}
\text{F} & \equiv & \bra f \big\| \sum_{i=1}^A \tau_{i,+} \big\| i \ket \\
\text{GT} & \equiv & \bra f \big \| \sum_{i=1}^A \tau_{i,+}
\vec{\sigma}_i
 \big\| i \ket,
\end{eqnarray}
$f_t$ includes the entirety of the kinematics:
\begin{equation}
f_t \equiv \int_1^{\overline{m_i}-\overline{m_f}} d\epsilon \epsilon
\sqrt{\epsilon^2 -1} (\overline{m_i}-\overline{m_f}-\epsilon)^2
F^{(+)}(Z_f,\epsilon)
\end{equation}
here $\epsilon$ is electron energy in units of electron mass,
$\overline{m}$ are masses in the same units, with subscript $i$($f$)
indicating initial (final) state, respectively. Since isospin mixing
is usually small for light nuclei, the Fermi operator can be
evaluated accurately. Thus, by using this formula, one can deduce
the experimental GT to very good accuracy \citep{Beta_A.lt.18}:
\begin{eqnarray}
\text{GT}(n \rightarrow p + e^{-} + \bar{\nu}_e) & = & \sqrt{3}
\cdot (1 \pm 0.003) \\
\text{GT}(^3\text{H} \rightarrow ^3\text{He} + e^{-} + \bar{\nu}_e)
& = & \sqrt{3} \cdot (0.961 \pm 0.005)
\end{eqnarray}
The axial coupling constant, $g_A=1.2670 \pm 0.0004$, is calibrated
using the neutron life--time, where the matrix element is analytic.
For the triton, axial MEC in the nuclei influence the half--life.
Thus, the triton GT is used to calibrate unknown constants in the
MEC model. In the case of MEC based on the EFT introduced in
Chap.~\ref{chap:EFT}, the only unknown is the renormalization LEC
$\hat{d}_r$.

Keeping the philosophy of the EFT* approach, the AV18/UIX
Hamiltonian is used as the nuclear interaction for the calculation
of the trinuclei wave functions. One then calculates the GT
($E^1_A$) using the EFT based MEC, and calibrates $\hat{d}_r$ to
reproduce the half life. This procedure follows \citep{PA03}. One
could use their result for $\hat{d}_r(\Lambda)$. However, as a test
for both the nuclear wave functions and the numerics.

When reproducing, there are additional benchmarks one should note.
First, the calculation can be checked by comparing binding energies
of the trinuclei to those achieved by the FY and the correlated
hyperspherical harmonics methods \citep{Fad_CHH_tri}, with the same
potential model. This appears in Table~\ref{tab:trinuclei}. One can
see a very good comparison between the calculations.
\begin{table}
\begin{center}
\begin{tabular}{|c||c|c|}
\hline
 & \multicolumn{2}{c|} {\bfseries Binding Energy [MeV]} \\ \cline{2-3}
 \raisebox{1.5ex}[0pt]{\bfseries Method}   &     $^3$H      &      {$^3$He}      \\\hline \hline
 EIHH     & $8.471(2)$ & $7.738(2)$     \\
 CHH      & $8.474   $ & $7.742   $     \\
 FY       & $8.470   $ & $7.738   $      \\\hline
 Experimental & $8.482   $ & $7.718   $      \\\hline
\end{tabular}
\end{center}
\caption{Binding energies of $^3$H and $^3$He calculated using
AV18/UIX Hamiltonian model compared to the same calculation done by
using FY equations and expansion on correlated hyperspherical
harmonics (CHH) basis \citep{Fad_CHH_tri}. For the EIHH calculation,
the number in parenthesis indicates the numerical error. Also shown
are the experimental values.} \label{tab:trinuclei}
\end{table}

An additional benchmark is a calculation of the single operator
contribution to the GT amplitude of $^3$H half life (which will be
denoted by $\text{GT}_{1B}$) \citep{pp_fusion} with the same
potential model, using the CHH method. This comparison reads:
$\text{GT}_{1B}(\text{\small{CHH}})=1.596 \Leftrightarrow
\text{GT}_{1B}(\text{\small{EIHH}})=1.598(1)$.

Finally, I reproduce the $\hat{d}_r(\Lambda)$ of \citep{PA03}. My
results are:
\begin{eqnarray} \nonumber
\hat{d}_r= 1.02 \pm 0.02 \pm 0.08 \,\,\,\text{for}\,\,\,\,
\Lambda=500\mev \\ \label{eq:d_r} \hat{d}_r= 1.78 \pm 0.03 \pm 0.08
\,\,\,\text{for}\,\,\,\, \Lambda=600\mev \\ \nonumber \hat{d}_r= 3.8
\pm 0.06 \pm 0.12 \,\,\,\text{for}\,\,\,\, \Lambda=800\mev
\end{eqnarray}
Where the first error is numerical, i.e. convergence of the
calculation, and the second correspond to the experimental
uncertainty in the triton GT strength. This is to be compared with
\citep{PA03},
\begin{eqnarray} \nonumber
\hat{d}_r= 1.00 \pm 0.07 \,\,\,\text{for}\,\,\,\, \Lambda=500\mev
\\ \nonumber \hat{d}_r= 1.78  \pm 0.08
\,\,\,\text{for}\,\,\,\, \Lambda=600\mev \\ \nonumber \hat{d}_r= 3.9
\pm 0.10 \,\,\,\text{for}\,\,\,\, \Lambda=800\mev.
\end{eqnarray}
The minor differences are of the order of those encountered in other
observables. The differences in the error--bars are probably due to
differences in the experimental source.

Summing up, the results of this section give us the needed
information to carry on {\it parameter free predictions} of weak
processes. As a side benefit, I get a verification of the numerical
methods used, and the derived weak currents.
\section{Neutrino Scattering Cross-sections Calculation}

The calculation of the neutral-current inclusive inelastic
cross-sections follows the procedure used to calculate the
photoabsorption process. We solve the bound state problem based on
the Argonne $v_{18}$ nucleon-nucleon potential~\citep{AV18}  and the
Urbana IX three-nucleon interaction~\citep{UIX}. The bound state
binding energies of the trinuclei are given in the previous section,
and those of $^4$He in Table~\ref{tb:4he_gs}.

Neutrino scattering on $A=3$ and $^4$He nuclei only induces
transitions to continuum states, since the nuclei have no bound
excited states. The correct description of the process is achieved
via the LIT method. The resulting Schr\"{o}dinger-like equations are
solved using the effective interaction hyperspherical harmonics
(EIHH) approach of Chap.~\ref{chap:EIHH}.

We use chiral EFT meson-exchange currents (MEC) as presented in
Chap.~\ref{chap:EFT}. The EFT approach is an appropriate
approximation, as $Q \lesssim 60 \mev$ is the typical energy in our
processes of interest, and the cutoff $\Lambda$ is of the order of
the EFT breakdown scale $\Lambda = 400-800 \mev$. In configuration
space, the MEC are obtained from a Fourier transform of propagators
with a cutoff $\Lambda$ (cf. Eq.~(\ref{eq:Fourier})). This leads to
a cutoff dependence, which is renormalized by the cutoff-dependent
counterterm $\hat{d}_r(\Lambda)$, given in Eq.~(\ref{eq:d_r}).
\subsection{Leading Multipole Contributions}
In the supernova scenario one has to consider neutrinos with up to
about $60$ MeV. The main result of this is that it is sufficient to
retain contributions up to $O(q^2)$ in the multipole expansion. This
conclusion is checked on $^4$He, and is accurate to the percentage
level. This is consistent with the discussion of
Sec.~\ref{sec:photo}. The leading contributions to the inelastic
cross-section are due to the axial vector operators
$E^A_1,\,E^A_2,\, M^A_1,\, L^A_2,\, L^A_0 $ and the vector operators
$C^V_1,\, E^V_1,\, L^V_1$. As explained in the previous chapter, the
leading vector operators are all proportional to each other due to
the Siegert theorem, valid in this energy range (see discussion in
Sec.~\ref{sec:photo}). The axial charge operator $C^A_0$ and the
magnetic $M^V_1$ operator are both proportional to the inverse of
the nucleon mass -- thus have no substantial contribution, exactly
as relativistic corrections. In fact, for a percentage level
accuracy, the following long wavelength approximation can be used
\citep{DO76}, for the impulse approximation the spin--orbital parts
of the operators are
\begin{align}
C^V_0(q) & = \frac{1}{\sqrt{4\pi}}
\nonumber \\
C^A_0(q) & = \frac{i}{\sqrt{4\pi}} \vec{\sigma}\cdot\vec{\nabla}
\nonumber \\
L_{0}^A(q) & =  i g_A \frac{qr}{3} [\vec{\sigma} \otimes
{Y}_{1}(\hat{r})]^{(0)} \nonumber \\
E^A_{1M}(q) & =  -i\, g_A
\sqrt\frac{2}{3} [\vec{\sigma} \otimes {Y}_{0}(\hat{r})]^{(1)}_M
\nonumber \\
C^V_{1M}(q) & =  \frac{qr}{3} Y_{1M}(\hat{r})
\nonumber \\
E^V_{1M}(q) & =  -\sqrt{2} \frac{\omega}{q}C^V_{1M}(q)
\nonumber \\
L^V_{1M}(q) & =  - \frac{\omega}{q}C^V_{1M}(q)
\nonumber \\
M^V_{1M}(q) & =-\frac{i}{\sqrt{6\pi}}\frac{q}{2M_N}l_M \nonumber
\end{align}
\begin{align}
M_{1M}^A(q) & =  -g_A \frac{qr}{3} [\vec{\sigma} \otimes
{Y}_{1}(\hat{r})]^{(1)}_M \nonumber \\
E_{2M}^A(q) & =  i
\sqrt{\frac{3}{5}} g_A \frac{q}{3} [\vec{\sigma} \otimes
{Y}_{1}(\hat{r})]^{(2)}_M
\nonumber \\
L_{2M}^A(q) & =  \sqrt{\frac{2}{3}} E_{2M}^A(q) \nonumber
\end{align}

The axial MEC operator, given in Eq.~(\ref{eq:MEC_operator}), is
invariant to an exchange between the particles. Thus, it does not
influence multipoles of opposite symmetry. The MEC will have a
substantial influence only on the GT ($E^1_A$) operator, as it will
be discussed in the following subsections.

\subsection{Inelastic Cross-sections and Energy-transfer}
\label{sub:cross}

The result of the current calculation gives an energy and angle
dependent inclusive cross--section. In order to present the
cross--sections, we use the fact that the neutrino spectra
are approximately thermal. Thus, the calculated cross sections are
averaged over energy and angle, assuming a Fermi-Dirac distribution
for the neutrinos with zero chemical potential, temperature $\tnu$,
and neutrino momentum $k$,
\begin{equation}
f(\tnu,k) = \frac{N}{\tnu^3} \frac{k^2}{e^{k/\tnu}+1} \,,
\end{equation}
where $N^{-1}=2 \sum_{n=1}^{\infty} (-1)^{n+1}/n^3$ is a
normalization factor. The quantities of interest are the
cross-sections and energy transfer cross-sections averaged over
neutrino spectra of temperature $\tnu$,
\begin{align}
\langle \sigma \rangle_\tnu &= \int_{\omega_{\rm th}}^{\infty}
d\omega \int dk_i \: f(\tnu,k_i) \, \frac{d\sigma}{dk_f} \,,
\label{eq:cross} \\[2mm]
\langle \omega \sigma \rangle_\tnu &= \int_{\omega_{\rm
th}}^{\infty} d\omega \int dk_i \: f(\tnu,k_i) \: \omega \,
\frac{d\sigma}{dk_f} \,, \label{eq:etcross}
\end{align}
where $k_{i,f}$ are the initial and final neutrino energy, $\omega =
k_i - k_f$ is the energy transfer, and $\omega_{\rm th}$ denotes the
threshold energy of the breakup reaction. For $^4$He, $^3$He and
$^3$H the threshold energies are $\omega_{\rm
th}=19.95,\,6.23,\,5.5\mev$ respectively.

It is well known that realistic 2--body NN potentials lead to an
under-binding of about $0.5-1$ MeV for the $^3$He and the triton
nuclei and an under-binding of about $3-4$ MeV for $^4$He. The AV18
threshold energies are $16.66\mev$, $5.37 \mev$ and $4.67\mev$ for
$^4$He, $^3$He and triton breakup reaction, respectively. Thus the
AV18 model has a discrepancy, $\Delta \approx 3.3 \mev$ for the
threshold energy of the $^4$He, $\Delta \approx 0.83 \mev$ for
$^3$He break--up, and $\Delta \approx 0.86 \mev$ for $^3$H, with
respect to the experimental inelastic reaction threshold.

Thermal averaging is highly influenced by the threshold energy, thus
very sensitive to this discrepancy. In order to correct for this
difference the response function is shifted to the true threshold,
i.e. $R(\omega)\longrightarrow R(\omega-\Delta)$, when only NN
potentials are used.
\subsection{Neutrino scattering on $^4$He nucleus}

In terms of ``many--body'' nuclear physics, $^4$He is a closed shell
nucleus, i.e. its total angular momentum is zero and its spin
structure is governed by an $S$--wave. Some of the leading operators
are strongly suppressed due to the special structure of the ground
state.

In fact, the Gamow-Teller operator contributes only due to the small
$P$-- and $D$--wave components of the ground state wave function.
This reflects on the axial MEC contribution to the cross--section.
In magnitude, the Gamow--Teller strength is less than $0.25\%$ of
the total cross--section, for all considered nuclear potentials:
AV18, AV8', and AV18/UIX. Due to MEC, the strength grows to $1\% \pm
0.5\%$. The error is due to the $\Lambda$ dependence. In this case,
the part of the error which correspond to the triton half--life is
of no importance. The MEC contribution is calculated only for the
AV18/UIX potential model, and $\hat{d}_r$ is taken from
Eq.~(\ref{eq:d_r}).

The special internal structure of the nucleus affects also the
$M^V_1$ operator, which vanishes for $^4$He. In addition,
$^{4}\mathrm{He}$ is an almost pure zero--isospin state
\citep{NO02}, hence the Fermi operator vanishes.

As a result, the leading contributions to the inelastic
cross-section are due to the axial vector operators $E^A_2, M^A_1,
L^A_2, L^A_0 $ and the vector operators $C^V_1, E^V_1, L^V_1$, which
are all proportional to the momentum transfer $q$.

\begin{figure}
\begin{center}
\rotatebox{270}{ \resizebox{7cm}{!}{
\includegraphics{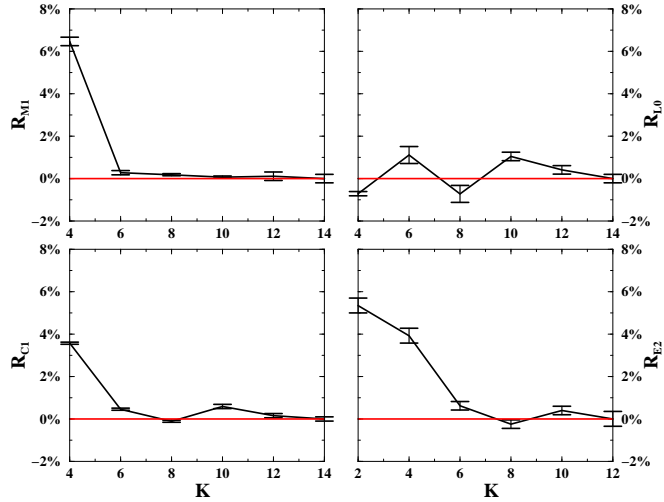} } }
\end{center}
\caption{\label{fig:conv_AV8P} Relative error in the sum-rule of the
leading response functions with respect to the hyper-angular
momentum quantum number $K$ (for $^4$He). Calculated using the AV8'
NN potential model. The error bars reflect the uncertainty in
inverting the LIT.}
\end{figure}

The combination of the EIHH and LIT methods brings to a rapid
convergence in the Response functions. In Fig.~\ref{fig:conv_AV8P},
one can see the relative error in the sum-rule of the main response
functions when calculated with NN potential (in this case Argonne
$v^\prime_8$, but the conclusions hold for AV18 as well), with
respect to the hyper-angular momentum quantum number $K$. It can be
seen that upon convergence the relative error is well below $1\%$.
The error bars presented reflect the error in inverting the LIT.
This convergence pattern holds also for the AV18 potential model.
Bearing in mind that the cross-section, up to kinematical factors,
is the sum of the response functions, this is a measure of the
accuracy in the calculated cross-section.

\begin{figure}
\begin{center}
\rotatebox{0}{ \resizebox{9cm}{!}{
\includegraphics{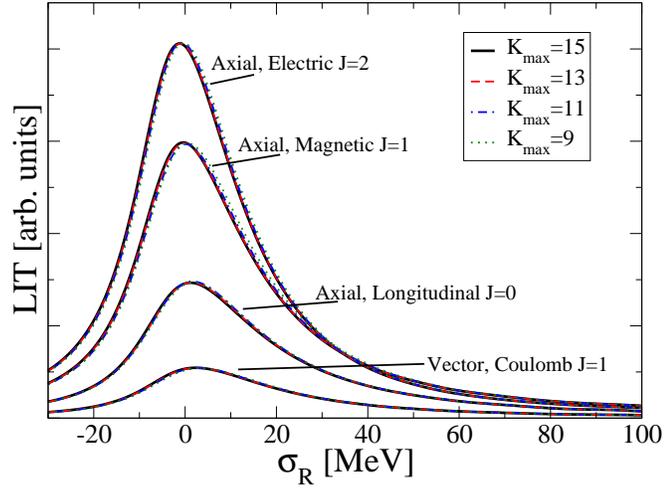}  } }
\end{center}
\caption{\label{fig:conv} Convergence of
${\mathcal{L}}_{\hat{O}_{1}\hat{O}_{2}}/q$ for the leading
operators, as a function of the HH grand angular momenta K (for
$^4$He).}
\end{figure}

The convergence rate is outstanding even when including the 3NF. In
Fig.~\ref{fig:conv} we present for the leading multipoles the
convergence of the LIT as a function of $K$. It can be seen that the
EIHH method results in a rapid convergence of the LIT calculation to
a sub-percentage numerical accuracy. We conclude that the 3NF does
not affect the convergence rate of these operators.

\begin{figure} [t]
\begin{center}
\rotatebox{270}{ \resizebox{6cm}{!}{
\includegraphics{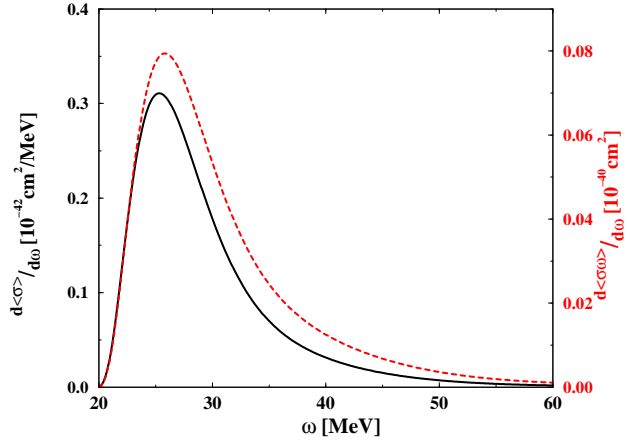}  } }
\end{center}
\caption{\label{fig:crs} Temperature averaged inelastic
cross-sections at temperature $T=10$ MeV for $^4$He. The solid line
is the differential cross-section, $\langle
\frac{d\sigma}{d\omega}\rangle_T=\frac{1}{2} \frac{1}{A} \langle
\frac{d\sigma_\nu}{d\omega}
                   +\frac{d\sigma_{\overline{\nu}}}{d\omega}\rangle_T$,
(left scale). The dashed line is the differential energy transfer
cross-section, $\langle \omega
\frac{d\sigma}{d\omega}\rangle_T=\frac{1}{2} \frac{1}{A} \langle
\omega\frac{d\sigma_\nu}{d\omega}
                   +\omega\frac{d\sigma_{\overline{\nu}}}{d\omega}\rangle_T$,
 (right scale).
}
\end{figure}

The general form of the cross--section is shown in
Fig.~\ref{fig:crs}.
%%%%%%%%%%%%%%%%%%%%%%%%%%%%%%%%%%%%%%%%%%%%%%%%%%%%%%%%%%%%%%%%%%%%
%%%%%%%%%%%%%%%%%%%%%%%%%%%%%%%%%%%%%%%%%%%%%%%%%%%%%%%%%%%%%%%%%%%%%
%%%%%%%%%%%%%%%%%%%%%%%%%%%%%%%%%%%%%%%%%%%%%%%%%%%%%%%%%%%%%%%%%%

\begin{table}
\begin{center}
\begin{tabular}{|c||c|c|c|c|c|}
\hline \hline T  &  \multicolumn{5}{c|} {$\bra \sigma^0_x \ket_T =
\frac{1}{2} \frac{1}{A} \bra \sigma_{\nu_x}^0+
\sigma_{\overline{\nu}_x}^0 \ket_T$ [$10^{-42}cm^{2}$] }  \\
\hline
[MeV]  &  AV8'  & AV18 & AV18+UIX & AV18+UIX+MEC & Woosley (1990)  \\
\hline
 4    &  2.09$\times 10^ {-3}$ & 2.31$\times 10^ {-3}$ & 1.63$\times 10^ {-3}$ & 1.66$\times 10^ {-3}$ & $-$ \\
 6    &  3.84$\times 10^ {-2}$ & 4.30$\times 10^ {-2}$ & 3.17$\times 10^ {-2}$ & 3.20$\times 10^ {-2}$ & 3.87$\times 10^ {-2}$ \\
 8    &  2.25$\times 10^ {-1}$ & 2.52$\times 10^ {-1}$ & 1.91$\times 10^ {-1}$ & 1.92$\times 10^ {-1}$ & 2.14$\times 10^ {-1}$ \\
 10   &  7.85$\times 10^ {-1}$ & 8.81$\times 10^ {-1}$ & 6.77$\times 10^ {-1}$ & 6.82$\times 10^ {-1}$ & 6.87$\times 10^ {-1}$ \\
 12   &  2.05     & 2.29     & 1.79     & 1.80 & 1.63   \\
 14   &  4.45     & 4.53     & 3.91     & 3.93 & $-$  \\
\hline \hline
\end{tabular}
\caption{{\label{tab:pots}} Temperature averaged neutral current
inclusive inelastic cross-section per nucleon as a function of
neutrino temperature. The last column is the calculation in
\citet{WO90}. }
\end{center}
\end{table}
%%%%%%%%%%%%%%%%%%%%%%%%%%%%%%%%%%%%%%%%%%%%%%%%%%%%%%%%%%%%%%%%
In Table~\ref{tab:pots} we present the temperature averaged total
neutral current inelastic cross--section as a function of the
neutrino temperature for the AV8', AV18, and the AV18+UIX nuclear
Hamiltonians and for the AV18+UIX Hamiltonian adding the axial MEC.
From the table it can be seen that the low--energy cross--section is
rather sensitive to details of the nuclear force model. This
sensitivity is visible already when comparing the AV8' and AV18
results, where an effect of about 10\% is discovered.

The 3NF have a bigger influence of about $25\%$. Similar tendency
was also observed for the {\it hep} process \citep{MA01}. The effect
reduces with neutrino temperature, from $35\%$ for $\tnu = 2\mev$,
to $23\%$ for $\tnu=12\mev$.

The results are of the same order of magnitude as previous estimates
\citep{WO90}, also appearing in the Table, though the differences
can reach $25\%$. The current work predicts a stronger temperature
dependence, with substantial increment at high temperatures. This
indicates a different structure of the predicted resonances.

The cross--section is dominated by the axial $E_2$ and $M_1$
multipoles, which for example exhaust about 90\% of the
cross--section at $\tnu=10\mev$.

\begin{table}
\begin{center}
\begin{tabular}{|c||c|c|c|c|}
\hline \hline T &  \multicolumn{4}{c|}{$\langle \sigma \rangle_T$
[$10^{-42}cm^{2}$] }  \\ \hline
 [MeV] &  ($\nu_x$,$\nu^{\,\prime}_x$)             &
          ($\bar{\nu}_x$,$\bar{\nu}^{\,\prime}_x$) &
          ($\nu_e$,e$^-$)                      &
          ($\bar\nu_e$,e$^+$) \\
\hline
 2    & 1.47$\times 10^ {-6}$  & 1.36$\times 10^ {-6}$ & 7.40$\times 10^ {-6}$ & 5.98$\times 10^ {-6}$ \\
 4    & 1.73$\times 10^ {-3}$  & 1.59$\times 10^ {-3}$ & 8.60$\times 10^ {-3}$ & 6.84$\times 10^ {-3}$ \\
 6    & 3.34$\times 10^ {-2}$  & 3.07$\times 10^ {-2}$ & 1.63$\times 10^ {-1}$ & 1.30$\times 10^ {-1}$ \\
 8    & 2.00$\times 10^ {-1}$  & 1.84$\times 10^ {-1}$ & 9.61$\times 10^ {-1}$ & 7.68$\times 10^ {-1}$ \\
 10   & 7.09$\times 10^ {-1}$  & 6.54$\times 10^ {-1}$ &  3.36                 & 2.71    \\
\hline \hline T &  \multicolumn{4}{c|}{$\langle \sigma \omega
\rangle_T$ [$10^{-42}$\mev cm$^{2}$] }  \\ \hline
 [MeV] &  ($\nu_x$,$\nu^{\,\prime}_x$)             &
          ($\bar{\nu}_x$,$\bar{\nu}^{\,\prime}_x$) &
          ($\nu_e$,e$^-$)                      &
          ($\bar\nu_e$,e$^+$) \\
\hline
 2    &  3.49$\times 10^ {-5}$ & 3.23$\times 10^ {-5}$ &  1.76$\times 10^ {-4}$& 1.42$\times 10^ {-4}$ \\
 4    &  4.50$\times 10^ {-2}$ & 4.15$\times 10^ {-2}$ &  2.27$\times 10^ {-1}$& 1.80$\times 10^ {-1}$\\
 6    &  9.26$\times 10^ {-1}$ & 8.56$\times 10^ {-1}$ &  4.56    & 3.70    \\
 8    &  5.85     & 5.43     &  28.4     & 22.9  \\
 10   &  21.7     & 20.2     &  103.8    &  84.4    \\
 \hline \hline
\end{tabular}
\caption{{\label{tab:crsomega_t}} Temperature averaged inclusive
inelastic cross--section (upper part) and energy transfer
cross-section (lower part) per nucleon as a function of temperature.
}
\end{center}
\end{table}

%The small contribution of the MEC can be understood in the following way. The
%symmetry of the vertices in this low--energy approximation, dictate
%a symmetry between the two nucleons interacting via the meson
%exchange. The leading one--body multipoles have negative parity, as
%a result the axial MEC contributions to them is small. In comparison the
%MEC correction to the Gamow-Teller is of the same magnitude as the
%one--body current, however both terms become marginal with
%increasing momentum transfer. Although presented for the neutral
%current, these arguments hold true also for the charged currents
%since the response functions are related by isospin rotation.

In Table \ref{tab:crsomega_t} we present (for AV18+UIX+MEC) the
temperature averaged cross--section and energy transfer as a
function of the neutrino temperature for the various processes. In
both tables it can be seen that the charged current process is
roughly a factor of five more efficient than the neutral current
process.

%Summarizing, we present the first full microscopic study of
%$\nu-\alpha$ reactions, using a state of the art nuclear Hamiltonian
%including MEC. The overall accuracy of our calculation is of the
%order of $5\%$. This error is mainly due to the strong sensitivity
%of the cross--section to the nuclear model, in particular to the
%3NF. The numerical accuracy of our calculations is of the order of
%$1\%$. The contribution of the axial MEC is small. The cutoff
%dependence, and $\hat{d}_r$ uncertainties insert the same scale of
%error.
\subsection{Neutrino scattering on $A=3$ nuclei} \label{sec:A3_crs}

\begin{table} [h]
\begin{center}
\begin{tabular}{|c|cc|cc|} \hline
$\tnu\, \text{[MeV]}$ & \multicolumn{2}{c|}{$^3$H} &
\multicolumn{2}{c|}{$^3$He} \\[1mm] \hline\hline
1  & 1.97$\times 10^{-6}$ & 1.68$\times 10^{-5}$
& 3.49$\times 10^{-6}$ & 2.76$\times 10^{-5}$ \\
2  & 4.62$\times 10^{-4}$ & 4.73$\times 10^{-3}$
& 6.15$\times 10^{-4}$ & 5.94$\times 10^{-3}$ \\
3  & 5.53$\times 10^{-3}$ & 6.38$\times 10^{-2}$
& 6.77$\times 10^{-3}$ & 7.41$\times 10^{-2}$ \\
4  & 2.68$\times 10^{-2}$ & 3.37$\times 10^{-1}$
& 3.14$\times 10^{-2}$ & 3.77$\times 10^{-1}$ \\
5  & 8.48$\times 10^{-2}$ & 1.14
& 9.70$\times 10^{-2}$ & 1.25 \\
6  & 2.09$\times 10^{-1}$ & 2.99
& 2.35$\times 10^{-1}$ & 3.21 \\
7  & 4.38$\times 10^{-1}$ & 6.61
& 4.87$\times 10^{-1}$ & 7.03 \\
8  & 8.20$\times 10^{-1}$ & 13.0
& 9.03$\times 10^{-1}$ & 13.7 \\
9  & 1.41                 & 23.4
& 1.54                 & 24.6 \\
10 & 2.27                 & 39.3 & 2.47                 & 41.2 \\
\hline
\end{tabular}
\caption{Averaged neutrino- and anti-neutrino-$^3$H and -$^3$He
neutral-current inclusive inelastic cross-sections per nucleon
(A=3), $\langle \sigma \rangle_\tnu = \frac{1}{2A} \, \langle \,
\sigma_\nu + \sigma_{\overline{\nu}} \, \rangle_\tnu$ (left
columns), and energy transfer cross-sections, $\langle \omega \sigma
\rangle_\tnu = \frac{1}{2A} \, \langle \, \omega \sigma_\nu + \omega
\sigma_{\overline{\nu}} \, \rangle_\tnu$ (right columns), as a
function of neutrino temperature $\tnu$, in units of $10^{-42} \,
\text{cm}^2$ and $10^{-42} \, \text{MeV}\text{cm}^2$ respectively.}
\label{tab:nu_A3}
\end{center}
\end{table}

\begin{figure}[h]
\begin{center}
\includegraphics[scale=0.3,clip=]{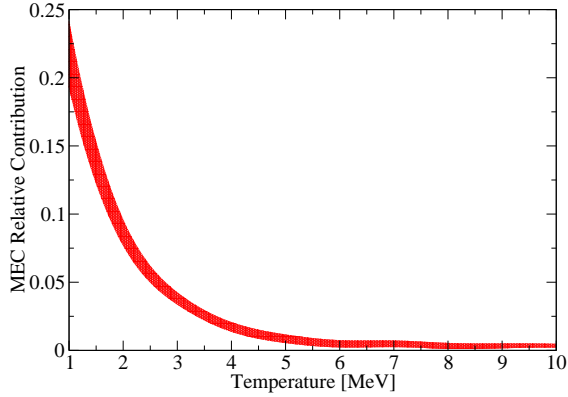}
\caption{Relative contribution of MEC to the energy transfer, i.e.
$\frac{\langle \omega \sigma \rangle_\tnu^{MEC}-\langle \omega
\sigma \rangle_\tnu^{No \,MEC}}{\langle \omega \sigma
\rangle_\tnu^{No \,MEC}}$. The width of the line indicates the error
in the theoretical estimate, due to the cutoff dependance.}
\label{fig:MEC_3H}
\end{center}
\end{figure}
Contrary to $^4$He, the GT is not suppressed for the trinuclei,
which have $J=\frac{1}{2}$ total angular momentum. At low-momentum
transfer, the Gamow-Teller operator dominates for the cross section.
It contributes about 60\% of the cross--section at a temperature of
$1 \mev$. At higher-momentum transfer, higher-order multipoles
(mainly the axial $E_2$ and $M_1$) start to play an important role,
thus the GT part of the cross--section decreases to about $20\%$ at
$\tnu=5 \mev$, and to about $10\%$ at neutrino temperature of
$10\mev$.

As a consequence of the descending importance of GT, the MEC
contribution also decreases with temperature. In
Fig.~\ref{fig:MEC_3H}, one can view the relative contribution of MEC
to the energy transfer, $\frac{\langle \omega \sigma
\rangle_\tnu^{MEC}-\langle \omega \sigma \rangle_\tnu^{No
\,MEC}}{\langle \omega \sigma \rangle_\tnu^{No \,MEC}}$. The width
of the line indicates the error in the theoretical estimate, due to
the cutoff dependance. The cutoff dependence of both cross--sections
is $\sim 2\%$ for $1 \mev$ and $< 1 \%$ for higher temperatures.
This is a rather strict validation of the calculation. The numerical
convergence is very good for the trinuclei $\sim 1\%$.

The 3NF affect the cross--sections by about 15\%, which is less than
the their effect in $^4$He, however still substantial.

While not directly important for the shock revival, the asymmetry
between the scattering of neutrinos and anti-neutrinos increases
with temperature: the difference in the energy transfer grows
gradually from $3 \%$ for a neutrino temperature of $3 \mev$ to $>
50\%$ for $10 \mev$ temperatures.

\subsection{Neutrino energy loss due to inelastic scattering}
\label{dEdx}

\begin{figure}[h]
\begin{center}
\includegraphics[scale=0.45,clip=]{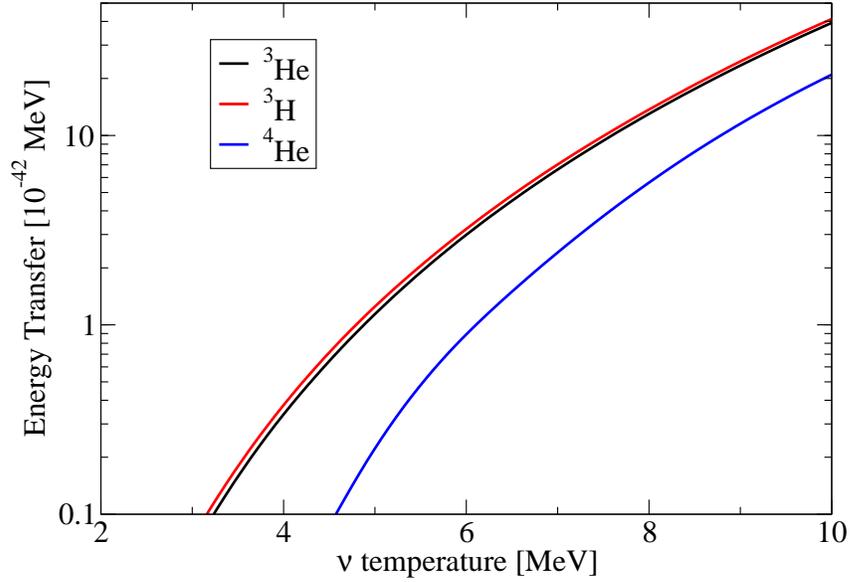}
\caption{ Neutrino total energy transfer cross--section a function
of the neutrino temperature $\tnu$ for $A=3,\,4$ nuclei. }
\label{fig:energy_transfer}
\end{center}
\end{figure}

\begin{figure}[b]
\begin{center}
\includegraphics[scale=0.45,clip=]{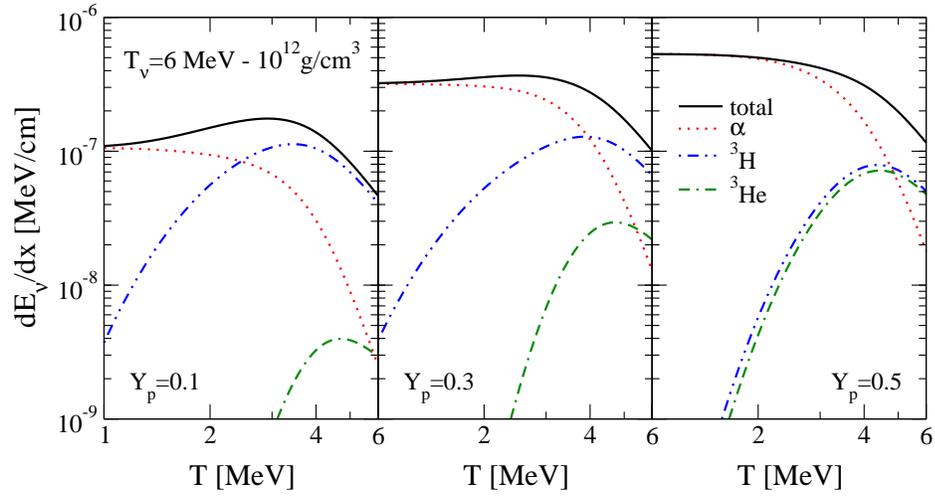}
\caption{ Neutrino energy loss $dE_\nu/dx$ for inelastic excitations
of $A=3,4$ nuclei as a function of the matter temperature $T$ at a
density of $10^{12} \gcq$. We assume that the neutrino energies are
characterized by a Fermi-Dirac distribution with a temperature $\tnu
= 6 \mev$. The contributions from $^3$H, $^3$He and $^4$He nuclei,
and the total neutrino energy loss are shown for proton fractions
$Y_p = 0.1, 0.3$ and $0.5$.} \label{fig:dEdx}
\end{center}
\end{figure}

We can combine the energy transfer cross-sections with the $A=3,4$
mass fractions of Fig.~\ref{fig:composition} to calculate the
neutrino energy loss due to inelastic excitations of $A=3,4$ nuclei.
In Fig.~\ref{fig:energy_transfer} we compare the total energy
transfer cross-sections for $^3$H, $^3$He and $^4$He. The comparison
clearly shows the huge difference between the cross--sections,
originating in the ``open--shell'' character of the trinuclei.

The neutrino of energy $E_\nu$ will decrease due to inelastic
excitations, and heat the matter, at a rate $dE_\nu/dx$ given by
\begin{equation}
\frac{dE_\nu}{dx} = n_b \sum_{i={^3}\hspace*{-0.2mm}{\rm{H}},\,
{^3}\hspace*{-0.2mm}{\rm{He}},\, {^4}\hspace*{-0.2mm}{\rm{He}}} x_i
\, \langle \omega \sigma \rangle_{i,\,\tnu} \,.
\end{equation}
For simplicity, we neglect the energy transfer from nuclei to
neutrinos required by detailed balance. This is strictly correct
only in the limit $T \ll \tnu$.

In Fig.~\ref{fig:dEdx}, the neutrino energy loss due to inelastic
scattering is shown for a density of $10^{12} \gcq$ and neutrino
temperature $\tnu =6 \mev$, as a function of the matter temperature
for various proton fractions. For $T \gtrsim 4 \mev$, the energy
loss is dominated by the contributions from $^3$H nuclei. The total
abundance of $A=3$ nuclei depends only weakly on the proton fraction
(see Fig.~\ref{fig:composition}), which is reflected in the weak
dependence of the neutrino energy loss as a function of proton
fraction. Finally, for lower densities, mass-three nuclei are less
abundant (see Fig.~\ref{fig:composition}), and therefore also their
contributions to the neutrino energy loss.

%% file: summary.tex
\chapter{Conclusions}
In this thesis I have presented the methods and the results of
{\it{ab--initio}} calculations of electro--weak reactions on light
nuclei. The reactions are calculated using realistic NN potential,
Argonne $v_{18}$, and includes the Urbana IX 3NF. The full
interaction is taken into account not only for the ground state, but
also for the continuum states via the LIT method. For the solutions
of the differential equations I use expansions in hyperspherical
harmonics via the EIHH approach.

In the next two sections I will briefly summarize the main results
and implications of the work.

\section{Photoabsorption on $^4$He}
I have presented the first complete calculation of the $^4$He total
photoabsorption cross section using a realistic nuclear force (AV18
NN potential and the UIX-3NF). The results show a rather pronounced
giant dipole peak typical to many body nuclei. This microscopic
calculation of a complex nucleus settles the long living debate
regarding the character of the peak.

The effect of the 3NF reduces the peak height by only 6\%, less than
expected considering its large effect of almost 20\% on the $^4$He
binding energy and its different role in the three-nucleon system.
Beyond the giant dipole resonance 3NF effects become much larger.
With growing energy transfer their importance increases and at pion
threshold one finds a cross-section enhancement of 35\%, about twice
the effect discovered in $^3$H/$^3$He photoabsorption.

The calculated cross--section was integrated with energy-transfer
weights to investigate sum--rules for $^4$He, which can be connected
to important ground state properties. Three well known photonuclear
sum rules for $^4$He are studied, viz the Thomas-Reiche-Kuhn, the
bremsstrahlung and the polarizability SR. Two new equivalences for
the BSR have allowed deduction of information about two-body
properties of the nuclear ground state, like the proton-proton,
neutron-neutron and proton-neutron distances.

A result with some beauty is tying the SR to the configuration
symmetry of the nucleus. In particular, the configuration
tetrahedral symmetry of $^4$He was tested, and found this symmetry
to be slightly broken. An experimental way to access this symmetry
breaking is proposed, via a measurement of the BSR which could be
performed in one of the existing or planned low-intermediate energy
photonuclear facility.

The calculated cross--section can be compared to numerous
experiments estimating the reaction. Close to threshold the
theoretical cross section agrees quite well with experimental data.
In the giant resonance region, where there is no established
experimental cross section, the results are in good agreement with
the data of \citet{Lund} and \citet{Wells}, while a strong
disagreement is found when compared with the data of \citet{Shima}.
In order to understand whether a nuclear force model, which is
constructed in the two- and three-nucleon systems, is sufficient to
explain the four-nucleon photodisintegration, further experimental
investigations are mandatory.

\section{Neutrino Interaction with $^4$He and A=3 nuclei in
Supernova}

The inelastic scatterings of neutrino off $^{4}$He, $^3$H and $^3$He
are calculated microscopically at energies typical for core collapse
supernova environment. These are the first microscopic estimates of
the processes, carried out with realistic nucleon forces. All
breakup channels and full final-state interactions were included via
the LIT method. The contribution of axial meson exchange currents to
the cross sections is taken into account from effective field theory
of nucleons and pions to order ${\cal{O}}(Q^3)$, which reproduce the
triton half--life.

I estimate the overall accuracy of the calculation to be of the
order of $5\%$. This error is mainly due to the strong sensitivity
of the cross--section to the nuclear model, in particular to the
3NF. The numerical accuracy of our calculations is of the order of
$1\%$. The contribution of the axial MEC for $^4$He is small, in the
percentage level. However, for the trinuclei the MEC contribution
could be up to 30\% for low temperature. The cutoff dependence for
both $A=3,\,4$ nuclei, which is indicative to the validity of the
hybrid EFT approach, is negligible, less than a percent for neutrino
temperature above $2\mev$.

Using the virial abundances and the microscopic energy transfer
cross-sections, it is found that mass-three nuclei contribute
significantly to the neutrino energy loss due to inelastic
excitations for $T \gtrsim 4 \mev$, conditions typical to the
neutrinosphere. The predicted energy transfer cross-sections on
mass-three nuclei in this area are approximately one order of
magnitude larger compared to inelastic excitations of $^4$He nuclei.

To fully assess the role of neutrino breakup of $^4$He in
core--collapse, the cross-sections should be included in SN
simulations. The $\nu-\alpha$ interaction is of central importance
also for correct evaluation of the $\nu-$ and $r-$ nucleosynthesis
processes.

With the present work, I have made an important step in the path
towards a more robust and reliable description of neutrino--nuclei
interaction role in core--collapse supernovae.

%% file: app_SM_current.tex
\chapter{Weak Currents In the Standard Model} \label{ap:SM_current}
In this appendix, the general structure of the weak currents in
nuclei are deduced from the standard model.

In the case of neutral neutrino scattering off a point proton,
making use of the Feynman diagrams, with the $Z^0$ boson propagator
approximated as $g_{\mu\nu}/M_Z^2$ one gets:
\begin{eqnarray} \label{eq:proton_nc}
\lefteqn{T_{fi}=-M_Z^{-2}\cdot \bar{u}(k_f) \left[
\frac{-ig}{4cos\theta_W}\gamma^\mu(1-\gamma_5) \right] u(k_i) \cdot}
\\ & & \nonumber
 { \cdot \bar{u}(P_f) \left[ \frac{-ig}{4cos\theta_W}\gamma_\mu
\left( (1-4\sin^2\theta_W) -\gamma_5 \right) \right] u(p_i)=}
\\ & & \nonumber
{= \frac{G}{\sqrt{2}}  \left[ \bar{u}(k_2)
\gamma^\mu(1-\gamma_5)u(k_1) \right ] \cdot \left[ \bar{u}(p_2)
\gamma_\mu \left( \frac{1}{2}(1-\gamma_5) -2\sin^2\theta_W \right)
u(p_1) \right]}
\end{eqnarray}
Where $G \equiv \frac {\sqrt{2} g^2}{8M^2_{Z}cos^2\theta_W}$ is the
Fermi constant. A similar calculation for neutrino neutral
scattering on neutron yields a different result:
\begin{eqnarray} \label{eq:neutron_nc}
{ T_{fi}=\frac{G}{\sqrt{2}}\left[ \bar{u}(k_2)
\gamma^\mu(1-\gamma_5)u(k_1) \right ] \cdot \left[ \bar{u}(p_2)
\gamma_\mu  \left( -\frac{1}{2}(1-\gamma_5) \right) u(p_1) \right]}
\end{eqnarray}
One can combine the results of equations~\ref{eq:neutron_nc} and
\ref{eq:proton_nc}, using the isospin symmetry to get for the
nucleon current:
\begin{eqnarray}
{J^\mu=\bar{u}(p_2)\left[ \frac{\tau_0}{2} \gamma^\mu (1-\gamma_5)
-2\sin^2\theta_W \gamma^\mu \frac{1}{2} (1+\tau_0)\right] u(p_1) }
\end{eqnarray}
For a complex nucleon, one keeps the symmetries of this current to
write
\begin{align}
\mathcal{{J}}^{(0)}_{\mu} & =
(1-2\cdot\sin^2\theta_W)\frac{\tau_0}{2}
{J}^V_\mu+\frac{\tau_0}{2}{J}^A_\mu-2\cdot\sin^2\theta_W\frac{1}{2}
{J}^V_\mu
\\ \nonumber
 & = \frac{\tau_0}{2} \big({J}^V_\mu+{J}^A_\mu \big)
-2\cdot\sin^2\theta_W {J}^{em}_\mu
\end{align}
Where $J_\mu^V$ and $J_\mu^A$ have vector and axial symmetry,
respectively. The second equality takes advantage of the conserved
vector current hypothesis, by which the weak--vector current is an
isospin rotation of the electro--magnetic current.

A similar calculation for the charged current yields:
\begin{equation}
\mathcal{{J}}^{(\pm)}_{\mu}
  = \frac{\tau_{\pm}}{2} \big({J}^V_\mu+{J}^A_\mu \big)
\end{equation}

%% file: app_lepton_current.tex
\chapter{Lepton Current}
\label{app:Lepton_current}

In this appendix the calculation of the lepton current matrix
element is demonstrated. I will show this for the case of a neutrino
scattering, as calculation of this matrix element for $\beta$ decay
or electron capture follow the same line.

The leptonic current is,
\begin{equation} \label{eq:LC}
j_{\mu}(\vec{x})=\bar {\psi}_\nu(\vec{x})
\gamma_\mu(1+\gamma_5)\psi{_{\nu'}}(\vec{x})=l_\mu e^{-i\vec{q}\cdot
\vec{x}}
\end{equation}
Where, for lepton reaction
\begin{equation} \label{eq:nr}
l_{\mu} =\bar u(\nu) \gamma_\mu (1+\gamma_5) u(\nu')
\end{equation}
Here, we keep the incoming particle is always a massless neutrino,
while the outgoing can be either a neutrino or a massive lepton. For
anti-lepton reaction,
\begin{equation} \label{eq:anr}
l_{\mu} =\bar v(-\nu') \gamma_\mu (1+\gamma_5) v(-\nu)
\end{equation}
For completeness, the needed matrix element for $\beta^{-}$ decay
is:
\begin{equation}
l_{\mu} =\bar u(e) \gamma_\mu (1+\gamma_5) v(-\nu)
\end{equation}

The calculation of the cross-section include the expressions
$l_\lambda l^*_{\nu}$. For the evaluation of this expression, the
following facts concerning the $\gamma$ matrices are helpful,
\begin{align}
\nonumber \{ \gamma^\mu,\gamma^\nu \} & =2g^{\mu\nu} \\ \nonumber \{
\gamma_5,\gamma^\mu \} & =0  \\ \nonumber (1+\gamma_5)^2 &
=2(1+\gamma_5) \\ \nonumber \gamma_0\gamma^{\dagger}_\nu\gamma_0 &
=\gamma_\nu \\ \nonumber tr \{ \gamma_{\lambda} \gamma_{\mu}
\gamma_{\nu} \gamma_{\mu'}\} &
=4(g_{\lambda\mu}g_{\nu\mu'}-g_{\lambda\nu}g_{\mu\mu'}+
g_{\lambda\mu'}g_{\mu\nu}) \\ \nonumber tr \{ \gamma_{\lambda}
\gamma_5 \gamma_{\mu} \gamma_{\nu} \gamma_{\mu'}\} & =-4 i
\epsilon_{\lambda \mu \nu \mu'}
\end{align}
where the metric tensor $g^{\mu\nu}=\left( \begin{array}{cccc}
               1  & 0  &  0  & 0 \\
               0  & -1 &  0  & 0 \\
               0  & 0  & -1  & 0 \\
               0  & 0  &  0  & -1
            \end{array} \right)$. \\ For any two spinors
$\psi_1$ and $\psi_2$ and any $4\times4$ matrix $\Gamma$,
\begin{eqnarray}
{\nonumber (\bar{\psi}_1 \Gamma \psi_2)^*= \bar{\psi}_2 \gamma_0
\Gamma \gamma_0 \psi_1}
\end{eqnarray}
thus, for lepton reaction,
\begin{eqnarray}
{l_\lambda l^*_{\nu}
=\bar{u}^{({\alpha}_{2})}(k_2)\gamma_\mu(1+\gamma_5)u^{({\alpha}_{1})}(k_1)
\bar{u}^{({\alpha}_{1})}(k_1)(1-\gamma_5)\gamma_\nu
u^{({\alpha}_{2})}(k_2) }
\end{eqnarray}
To obtain the cross section it is necessary to sum over initial and
average over final lepton helicities.
\begin{eqnarray}
{\overline {l_\lambda l^*_{\nu}} =\frac{1}{2} \sum_{\alpha_1
\alpha_2} \bar{u}^ {({\alpha}_{2})}(k_2) \gamma_\mu(1+\gamma_5)
u^{({\alpha}_{1})}(k_1) \bar{u}^{({\alpha}_{1})}(k_1)
(1-\gamma_5)\gamma_\nu u^{({\alpha}_{2})}(k_2) }
\end{eqnarray}
It is helpful to use the fact that,
\begin{eqnarray}
{\sum_\alpha{{u}^{({\alpha})}_\xi(k) \bar{u}^{({\alpha})}_{\xi'}(k)}
= \left( \frac{k_\mu \gamma^\mu+m}{2E_p} \right) _{\xi \xi'}}
\end{eqnarray}
Thus (the mass term falls due to $\gamma$ matrices algebra),
\begin{eqnarray} {\overline {l_\lambda l^*_{\nu}}
=\frac{\nu^\mu \nu'^{\mu'}}{4 \nu \nu'} tr \{ \gamma_{\lambda}
(1+\gamma_5) \gamma_{\mu} \gamma_{\nu} \gamma_{\mu'}\} }
\end{eqnarray}
finally,
\begin{eqnarray}
{\overline {l_\lambda l^*_{\nu}} =\frac{1}{ \nu \nu'} \{ \nu_\lambda
\nu'_\nu + \nu'_\lambda \nu_\nu  \nu_\mu \nu'_\mu - i
\epsilon_{\lambda \mu \nu \mu'}\nu^\mu  \nu'^{\mu'} \}}
\end{eqnarray}
the calculation for anti-neutrino reaction is similar and gives
\begin{eqnarray}
{\overline {l_\lambda l^*_{\nu}} =\frac{1}{ \nu \nu'} \{ \nu_\lambda
\nu'_\nu + \nu'_\lambda \nu_\nu  \nu_\mu \nu'_\mu + i
\epsilon_{\lambda \mu \nu \mu'}\nu^\mu  \nu'^{\mu'} \}}
\end{eqnarray}
The general neutrino reaction cross-section requires the following
expressions,
\begin{eqnarray} \label{eq:lc}
{\overline {l_0 l^*_{0}} =1+ \hat{\nu} \cdot \hat{\nu'}}
\\ {\overline {l_3 l^*_{3}} =1- \hat{\nu} \cdot \hat{\nu'}
 +2(\hat{\nu} \cdot \hat{q})(\hat{\nu'} \cdot
\hat{q}) } \\ {\overline {l_3 l^*_{0}} =\hat{q} \cdot( \hat{\nu} +
\hat{\nu'}) } \\ {\overline {\frac{1}{2} (\vec{l} \cdot \vec{l^*} -
l_3 l^*_{3})} =1- (\hat{\nu} \cdot \hat{q})(\hat{\nu'} \cdot
\hat{q})} \\ {\overline {(\vec{l} \times \vec{l^*})_3} = S \times 2
i \hat{q} \cdot (\hat{\nu} - \hat{\nu'})}
\end{eqnarray}
where $q^\mu={\nu'}^\mu-\nu^\mu$ is the momentum transfer. $S$ is
$+1$ for neutrino scattering and $\beta^{+}$ decay, and $-1$ for
anti--neutrino scattering and $\beta^{-}$ decay.

%% file: app_spinors.tex
\chapter{Non-relativstic expansions of Dirac spinors expressions}\label{app:spinors}
Dirac spinor of a mass $M$, takes the form:
\begin{equation}
u(p,\sigma)=\sqrt{\frac{E_p+M}{2E_p}}\left(
\begin{array}{c}\chi_\sigma\\
              \frac{\vec{\sigma}\cdot\vec{p}}{E_p+M}\chi_\sigma
\end{array}\right)
\end{equation}
where $E_p=\sqrt{p^2+M^2}$ is the energy of the particle. It is
useful to recall that $\bar{u}=u^\dagger\gamma_0$. This convention
leads to a normalized density, i.e. $u^\dagger u=1$

In this representation,
\begin{equation}
\gamma^0=\left(\begin{array}{cc}1 & 0 \\ 0 & -1\end{array}\right),
\vec{\gamma}=\gamma^0\left(\begin{array}{cc}0 & \vec{\sigma} \\
\vec{\sigma} & 0 \end{array}\right),
\gamma_5=\left(\begin{array}{cc}0 & 1 \\ 1 & 0\end{array}\right),
\end{equation}
where we use $\gamma^0\gamma_5\gamma^0=-\gamma_5$ and
$\gamma^0\gamma_5\vec{\gamma}=-\vec{\sigma}$.

One can now expand the needed matrix elements in the inverse mass:
\begin{align}
\bar{u}(p^\prime,\sigma^\prime)\gamma^0 u(p,\sigma) & =
\chi_\sigma^{\prime\dagger}\left(1-\frac{\vec{q}^2}{8M^2}+i\vec{q}\cdot
\frac{\vec{\sigma}\times\vec{p}}{4M^2}\right)\chi_\sigma \\
\bar{u}(p^\prime,\sigma^\prime)\vec{\gamma} u(p,\sigma) & =
\frac{1}{2M}\chi_\sigma^{\prime\dagger}
(\vec{P}-i\vec{\sigma}\times\vec{q})\chi_\sigma\\
\bar{u}(p^\prime,\sigma^\prime)\gamma_5\gamma^0 u(p,\sigma) & =
-\frac{1}{2M}\chi_\sigma^{\prime\dagger}
(\vec{\sigma}\cdot\vec{P})\chi_\sigma\\
\bar{u}(p^\prime,\sigma^\prime)\gamma_5 u(p,\sigma) & =
\frac{1}{2M}\chi_\sigma^{\prime\dagger}
(\vec{\sigma}\cdot\vec{q})\chi_\sigma\\
\bar{u}(p^\prime,\sigma^\prime)\gamma_5 \vec{\gamma}u(p,\sigma) & =
 -\chi_\sigma^{\prime\dagger}\left(
\vec{\sigma}\left(1-\frac{P^2}{8M^2}\right) +i\frac{\vec{q}\times
\vec{p}}{4M^2}+\frac{\vec{P}(\vec{\sigma}\cdot\vec{P})
-\vec{q}(\vec{\sigma}\cdot\vec{q})}{8M^2}\right)\chi_\sigma
\end{align}
where $\vec{P}=\vec{p}+\vec{p^\prime}=-i( \overrightarrow{\nabla}-\overleftarrow{\nabla}) $ and $\vec{q}=\vec{p}-\vec{p}^\prime$.\\

%% file: app_feynman.tex
\chapter{Feynman Diagrams for $\chi$PT} \label{ap:Feynman}
The Feynman rules are exhausted from the interaction Lagrangian, and
from the currents. This is done by expanding the complete term to
first order in the pion field. The diagrams are then easily
achieved.

It is valuable to remind here the pion propogator
$\frac{i}{k^2-m^2_\pi}$, and the nucleon propogator
$\frac{i}{\not{p}-M}$.

\section{Leading Order Diagrams}
\begin{center}
\begin{figure} \resizebox{10cm}{!} {
\includegraphics{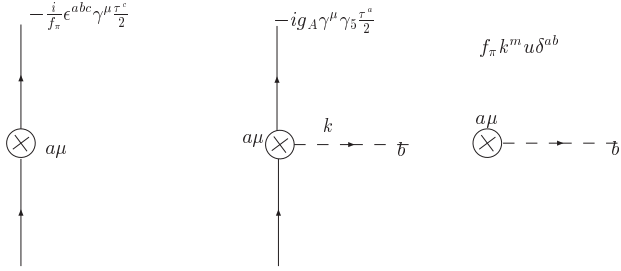}
} \caption{\label{fig:eft_pi_N} Vertices in the axial current.
Crossed circles indicate the attachement to the external probe.}
\end{figure}
\end{center}

The pion-nucleon interaction lagrangian to leading order in the pion
field is given in Eq.~(\ref{eq:L_pi_N_2}). The resulting Feynman
rules appear in Fig.~\ref{fig:pi_N}.
\begin{center}
{\begin{figure} [h]  {
\includegraphics{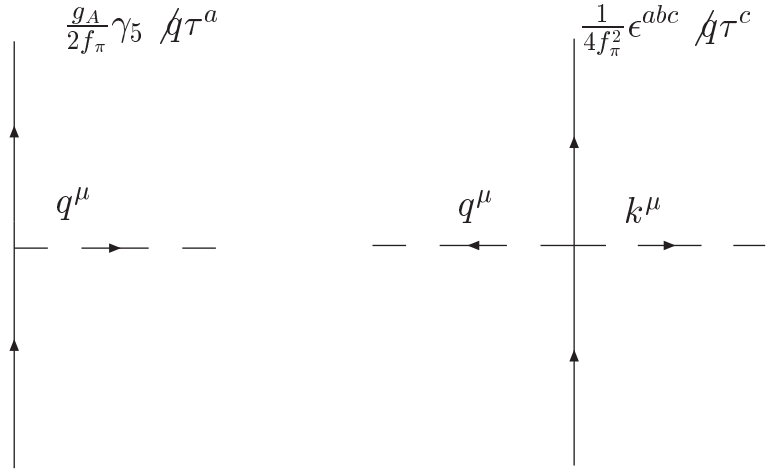}
} \caption{\label{fig:pi_N} The lowest order pion - nucleon Feynman
diagrams. Nucleons are indicated by solid lines, whereas pions are
indicated by dashed lines.}
\end{figure}}
\end{center}
The axial current to the same order is given in
Eqs.~(\ref{eq:AX_pi_pi_2}~-~\ref{eq:AX_pi_N_2}), and appear in
Fig.~\ref{fig:eft_pi_N}.
To lowest order in the pion field, one can get the following vector
vertices from Eqs.~(\ref{eq:V_pi_pi_2}~-~\ref{eq:V_pi_N_2}):\\
$\bullet$ $i\gamma^\mu\frac{\tau^a}{2}$.\\
$\bullet$ $i\frac{g_A}{f_\pi}\epsilon^{abc}\gamma^\mu\gamma_5
\frac{\tau^c}{2}$.\\
$\bullet$ $-\epsilon^{abc}k^\mu$.\\

\section{Next-to Leading Order Diagrams}
The NLO pion-nucleon interaction lagrangian to leading order in the
pion field is given in Eq.~(\ref{eq:L_int_3}). The resulting Feynman
diagram is identical to the diagram drawn in the right part of
Fig.~\ref{fig:pi_N}. The strength of the vertex is:
\begin{equation} \label{eq:L_int_3_strength}
\frac{2i}{f_\pi^2M} \left\{ \hat{c}_4 \epsilon^{abc}
\frac{\tau_c}{2} k_\mu q_\nu \sigma^{\mu\nu} - \hat{c}_3 k_\mu q^\mu
\delta^{ab} \right\}
\end{equation}
The NLO contribution to currents within the nuclei appear in
Eqs.~(\ref{eq:AX_pi_N_2}~-~\ref{eq:V_pi_N_2}). The resulting vector
current includes two pions, thus of higher order. The resulting
axial vector diagram is identical to the one which appears in the
center of Fig.~\ref{fig:eft_pi_N}. The additional strength is:
\begin{equation}
\frac{2}{f_\pi M} \left\{ -\hat{c}_4 \epsilon^{abc} \frac{\tau_c}{2}
q_\nu \sigma^{\mu\nu} + \hat{c}_3 q^\mu \delta^{ab} \right\}
\end{equation}

The NLO Lagrangian includes in addition contact terms, as given in
Eq.~(\ref{eq:contact}). The form of the vertex is a contact term
with an outgoing pion line, and the strength of the resulting vertex
is $-\frac{g_A D_1}{f_\pi} \gamma_5 \not{q} \tau_a$.

The contact term contribution to the vector currents is zero in the
non--relativistic limit. However, the contribution to the axial
current diagram does not vanish $D_1 \gamma^\mu \gamma_5
\frac{\tau_a}{2}$ with the diagram in Fig.~\ref{fig:eft_2b}.
\begin{center}
{\begin{figure}  {
\includegraphics{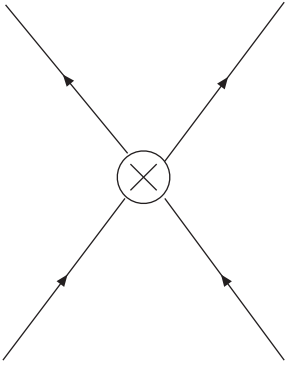}
} \caption{\label{fig:eft_2b} Contact axial vertex in NLO.}
\end{figure}}
\end{center}

\section{Example: Single Nucleon Operators For Neutrino Scattering}
One can use the Feynman diagrams of the previous section to draw
neutrino scattering on a single nucleon:
\begin{center}
{\begin{figure} [h] \resizebox{10cm}{!} {
\includegraphics{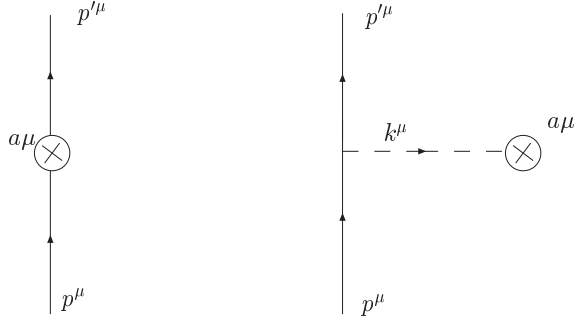}
} \caption{\label{fig:eft_1b} Leading order 1-body axial vector
current.}
\end{figure}}
\end{center}
The resulting strength is identical to that achieved by standard
nuclear physics approach (SNPA):
\begin{equation}
M^{a\mu}(1)=ig_A\bar{u}(p')\gamma_5\left\{\gamma^\mu-\not\mspace{-3mu}{k}
\frac{k^\mu}{k^2-m_\pi^2}\right\}\frac{\tau^a}{2}u(p)=
i\bar{u}(p')\left\{F_A\gamma^\mu\gamma_5 +{F_P}\gamma_5
k^\mu\right\} \frac{\tau^a}{2}u(p)
\end{equation}
The second equality uses the known SNPA axial form factor
$F_A=-g_A$, and pion form factor $F_P=-\frac{2Mg_A}{k^2-m^2_\pi}$.
By non--relativistic expansion of the nucleon spinor, as explained
in Appendix~\ref{app:spinors}, one gets the single nucleon
scattering operators of Eqs.~(\ref{eq:V0}~-~\ref{eq:A}).

\section{Example: Axial MEC Operators For Neutrino Scattering}
\begin{center}
\begin{figure} \resizebox{10cm}{!} {
\includegraphics{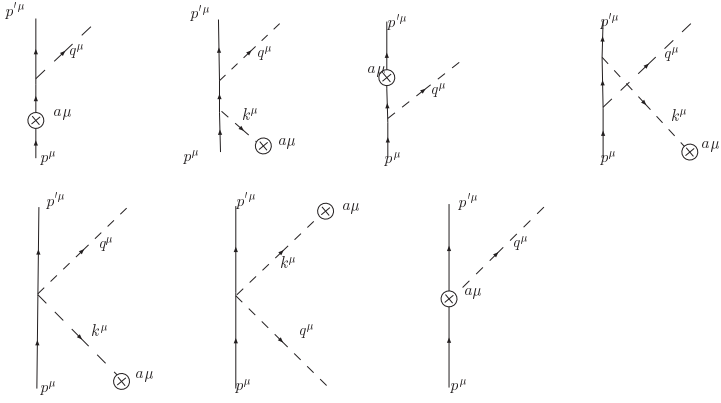}
} \caption{\label{fig:eft_pipro} Pion-production on a nucleon.}
\end{figure}
\end{center}

In order to continue to axial currents that involve two nucleons
with a pion exchange, we will find the amplitude for pion
production. This is done in \citet{AN02}, and the resulting Feynman
diagrams are given in Figure~\ref{fig:eft_pipro}. This is however
wrong, as the first three diagrams (starting upper left side)
include pion exchange between nucleons without any interaction of
this pion with the external probe. The correct pion production
amplitude is thus,
\begin{eqnarray}
2f_\pi M^{ab\mu}(\pi)=\bar{u}(p')\left\{ -g_A^2 \not\mspace{-3mu}{q}
\gamma_5\frac{1}{(\not\mspace{-2.5mu}{p}^\prime+\not\mspace{-2.5mu}{q}-M)}
\gamma_5 \not\mspace{-3mu}{k}\frac{k^\mu}{k^2-m^2_\pi} \frac{\tau^b
\tau^a}{2} \right. \\ \nonumber \left.
+i\epsilon^{abc}\frac{\tau_c}{2}\left[(\not\mspace{-3mu}{k}-
\not\mspace{-3mu}{q})
\frac{k^\mu}{k^2-m^2_\pi}-2\gamma^\mu\right]\right\}u(p)
\end{eqnarray}
The two nucleon axial current is achieved through the exchange of a
pion, as observed in Figure~\ref{fig:eft_NN}. The amplitude is:
\begin{equation} {\label{eq:exchange}}
M^{a\mu}(2)=M^{ab\mu}(\pi)\frac{i}{q^2-m^2_\pi}\frac{g_A}{f_\pi}
\bar{u}(p^{\prime}_2)\not\mspace{-3mu}{q}\gamma_5\frac{\tau^b}{2}u(p_2)
+momenta\mspace{5mu}{permutations}
\end{equation}
Taking the non-relativstic limit ($O(1/M)$) and the soft pion limit
($q,k \rightarrow 0$), we get for the pion production,
\begin{equation}
2f_\pi M^{ab\mu}(\pi) \approx
-i\bar{u}(p^\prime)\epsilon^{abc}\tau^c\gamma^\mu u(p)
\end{equation}
hence,
\begin{equation} {\label{eq:ex_cu}}
\vec{M}^{a} \approx
\frac{ig_A}{8Mf_\pi^2}(\tau^{(1)}\times\tau^{(2)})^a(i\vec{P}
_1+\vec\sigma_1\times\vec{q}+\vec{\sigma}_1\times{\vec{k}_2})\frac{\vec{\sigma}_2
\cdot\vec{k}_2}{\vec{k}_2^2+m_\pi^2}+(1\leftrightarrow 2),
\end{equation}
\begin{equation}
{M}^{a0} \approx \frac{g_A}{4f_\pi^2}(\tau^{(1)}\times\tau^{(2)})^a
\frac{\vec{\sigma}
\cdot\vec{k}_2}{\vec{k}_2^2+m_\pi^2}+(1\leftrightarrow 2).
\end{equation}
where $\vec{P}_1=\vec{p}+\vec{p}^\prime$,
$\vec{k}_i=\vec{p}_i-\vec{p}
^\prime_i$, and $\vec{q}$ is the neutrino's momentum.\\
\begin{center}
{\begin{figure} \resizebox{5cm}{!} {
\includegraphics{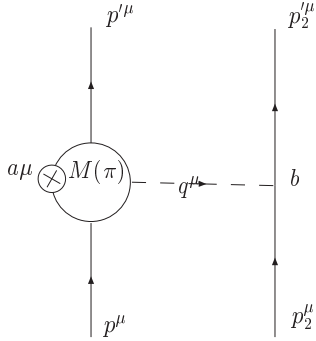}
} \caption{\label{fig:eft_NN} Pion exchange contribution to the
scattering amplitude .}
\end{figure}}
\end{center}
An additional contribution to the pion production amplitude comes
from NLO (calculated using the diagrams in
Figure~\ref{fig:eft_pi_nlo}):
\begin{eqnarray}
\lefteqn{ M^{ab\mu}_{NLO}=-\frac{2\hat{c}_4}{f_\pi M}\epsilon^{abc}
\left\{g^\mu_\lambda-\frac{k^\mu k_\lambda}{k^2-m^2_\pi}\right\}
q_\nu\bar{u}(p^\prime)\sigma^{\lambda\nu}\frac{\tau^c}{2}u(p)+}
\nonumber \\ && {+\frac{2\hat{c}_3}{f_\pi M}\epsilon^{abc}
\left\{q^\mu-\frac{k^\mu}{k^2-m^2_\pi}k\cdot q\right\} \delta^{ab}
\bar{u}(p^\prime)u(p)},
\end{eqnarray}
using Eq.~(\ref{eq:exchange}) we can now get the NLO contribution to
the pion exchange currents:
\begin{eqnarray}
\lefteqn{M^{a\mu}_{NLO}(2)=i\frac{g_A}{f_\pi^2}\left\{-\frac{2\hat{c}_4}{M}\epsilon^
{abc}\left[g^\mu_\lambda-\frac{k^\mu
k_\lambda}{k^2-m^2_\pi}\right]\times\right.} \nonumber \\ && {\left.
\left(\bar{u}(p_1^{\prime})\sigma^{\lambda\nu}\frac{\tau^c}{2}u(p_1)
\frac{q_\nu
q_\sigma}{q^2-m_\pi^2}\bar{u}(p_2^{\prime})\gamma^\sigma\gamma_5
\frac{\tau^b}{2}u(p_2)\right)\right.} \nonumber \\ &&
{+\left.\frac{2\hat{c}_3}{M}\left[q^\mu-\frac{k^\mu}{k^2-m^2_\pi}k\cdot
q\right] \frac{q_\sigma}{q^2-m_\pi^2}
\left(\bar{u}(p_1^{\prime})u(p_1)\bar{u}(p_2^{\prime})\gamma^\sigma\gamma_5
\frac{\tau^a}{2}u(p_2)\right)\right\}}.
\end{eqnarray}
\begin{center}
{\begin{figure} \resizebox{10cm}{!} {
\includegraphics{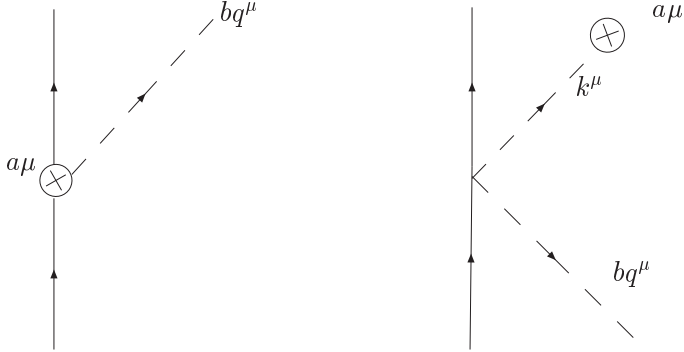}
} \caption{\label{fig:eft_pi_nlo} Pion production contribution from
NLO lagrangian.}
\end{figure}}
\end{center}
The non-relativistic reduction leads to no contribtion to the axial
charge density and to the following expression for the axial
current:
\begin{eqnarray}
{\vec{M}^a_{NLO}(2)=\frac{ig_A}{2Mf_\pi^2}\left\{\hat{c}_4
\left(\vec{\tau}^{(1)}\times\vec{\tau}^{(2)}\right)\vec{\sigma}_1\times
\vec{k}_2+2\hat{c}_3\vec{\tau}^{(2)a}\vec{k}_2\right\}
\frac{\vec{\sigma}_2\cdot\vec{k}_2}{\vec{k}_2^2+m_\pi^2}}.
\end{eqnarray}
If we combine this with Eq.~(\ref{eq:ex_cu}), we get
\begin{eqnarray}
\lefteqn{ \vec{M}^{a} \approx \frac{ig_A}{2Mf_\pi^2}\left\{
\frac{i}{4}(\vec{\tau}^{(1)}\times\vec{\tau}^{(2)})^a \vec{P}_1
+2\hat{c}_3\vec{\tau}^{(2)a}\vec{k}_2+\right.}\nonumber \\ &&
{\left.
+\left(\frac{1}{4}+\hat{c}_4\right)(\vec{\tau}^{(1)}\times\vec{\tau}^{(2)})^a
\vec{\sigma}_1\times{\vec{k}_2}+\right.}\nonumber \\ && {\left.
+\frac{1}{4}(\vec{\tau}^{(1)}\times\vec{\tau}^{(2)})^a(\vec\sigma_1\times\vec{q})\right\}
\frac{\vec{\sigma}_2\cdot\vec{k}_2}{\vec{k}_2^2+m_\pi^2}+(1\leftrightarrow
2)},
\end{eqnarray}
The contribution of the contact Lagrangian vanishes for the axial
charge operator in the non--relativistic limit. Its contribution to
the MEC reads:
\begin{equation}
M^{a\mu}_4\approx
iD_1\left(\vec{\sigma}_1\frac{\vec{\tau}^{(1)a}}{2}+
\vec{\sigma}_2\frac{\vec{\tau}^{(2)a}}{2}\right).
\end{equation}
In order to get the scattering operators, one has to transform to
configuration space. After doing thus, one gets the final form of
the MEC operators, appearing in Eq.~(\ref{eq:MEC_operator}).

%% file: app_2BME.tex
\chapter{Reduced Matrix Elements of different opearators} \label{ap:2BME}
In this chapter, I give a list of the needed matrix elements when
calculating in the HH expansion, as implemented in the program. In
the following $\vec\Sigma$ is a spin spherical tensor operator of
order $1$.
\subsubsection{ME1}
\begin{eqnarray}
\lefteqn { \langle K (l s) j \|j_J(qx)Y_J(\hat{x})\| K' (l' s') j'
\rangle=} \nonumber \\ & & {=\delta_{s,s'}(-)^{J+j'+l+s}
\sqrt{(2j+1)(2j'+1)} \sixj{l}{j'}{J}{s}{l'}{j}}\times \nonumber \\ &
& {\langle K l \|j_J(qx)Y_J(\hat{x})\|K' l'\rangle}
\end{eqnarray}
using the fact that,
\begin{equation} \label{eq:YL}
{\langle l\|Y_J(\hat{x})\|l'\rangle=
(-)^{l}\sqrt{\frac{(2l+1)(2J+1)(2l+1)}{4\pi}}
\threej{l}{0}{J}{0}{l'}{0}}
\end{equation}
We finally get,
\begin{eqnarray}
\lefteqn { \langle K (l s) j \|j_J(qx)Y_J(\hat{x})\| K' (l' s') j'
\rangle=} \nonumber \\ & & {=\delta_{s,s'}(-)^{J+j'+s'}
\sqrt{\frac{(2l+1)(2j+1)(2J+1)(2l'+1)(2j'+1)}{4\pi}}}\times
\nonumber \\ & & {\sixj{l}{j'}{J}{s}{l'}{j}
\threej{l}{0}{J}{0}{l}{0} \langle K l |j_J(qx)|K' l'\rangle}
\end{eqnarray}
\subsubsection{ME2}
We use the fact that $\vec{\Sigma} \cdot \vec{Y}_{JLM}(\hat{x})=
\left[\Sigma^{(1)}\otimes Y_L\right]^{(J)}_M$,
\begin{eqnarray}
\lefteqn {\langle K (l s) j\|j_J(qx)\vec{\Sigma}\cdot
\vec{Y}_{JLM}(\hat{x})\|K' (l' s') j' \rangle=
\sqrt{(2j+1)(2J+1)(2j'+1)}} \nonumber \\ & & {\times
\ninej{l'}{s'}{j'}{L}{1}{J}{l}{s}{j} \langle K l
\|j_J(qx)Y_{L}(\hat{x})\|K' l' \rangle \langle s \|\vec{\Sigma}\| s'
\rangle}
\end{eqnarray}
putting now Eq.~(\ref{eq:YL}), we get
\begin{eqnarray}
\lefteqn {\langle K (l s) j\|j_J(qx)\vec{\Sigma}\cdot
\vec{Y}_{JLM}(\hat{x})\|K' (l' s') j' \rangle=
(-)^{l}\ninej{l'}{s'}{j'}{L}{1}{J}{l}{s}{j}} \nonumber
\\ & &
{\times\sqrt{\frac{(2j+1)(2l+1)(2J+1)(2L+1)(2j'+1)(2l'+1)} {4\pi}}}
\nonumber \\ & & {\times\threej{l}{0}{L}{0}{l}{0} \langle s
\|\vec{\Sigma}\| s' \rangle \langle K l \|j_J(qx)\|K' l'\rangle}
\end{eqnarray}
\subsubsection{ME3}
\begin{eqnarray}
\lefteqn {\langle K l \|\vec Y_{JL}(\hat{x})\cdot \vec {\nabla}\| K'
l' \rangle=(-)^{J+1}\sqrt{\frac{(2l+1)(2J+1)(2L+1) (2l'+1)}{4\pi}}}
\nonumber \\ & & {\times [ \sixj {J}{l'-1}{l'}{L}{l}{1}
\frac{\threej{l}{0}{L}{0}{l'-1}{0}}{\threej{l'}{0}{1}{0}{l'-1}{0}}
\frac{l'}{2l'+1}\langle K l |\left( \frac{d}{dr}+ \frac{l'+1}{r}
\right)|K' l' \rangle} \nonumber \\ & & {+\sixj
{J}{l'+1}{l'}{L}{l}{1}
\frac{\threej{l}{0}{L}{0}{l'+1}{0}}{\threej{l'+1}{0}{1}{0}{l'}{0}}
\frac{l'+1}{2l'+1}\langle K l |\left( \frac{d}{dr}- \frac{l'}{r}
\right)|K' l' \rangle ]}
\end{eqnarray}
\subsubsection{ME4}
\begin{eqnarray}
\lefteqn {\langle K (l s) j \|\vec Y_{JL}(\hat{x})\cdot\vec
{\nabla}\| K' (l' s') j' \rangle=} \nonumber \\ & & {=\delta_{s,s'}
(-)^{J+j+l+s}\sqrt{(2j+1)(2j'+1)} \sixj{l}{j'}{J}{s}{l'}{j} \langle
K l \|\vec Y_{JL}(\hat{x})\cdot\vec {\nabla}\| K' l'  \rangle= }
\nonumber
\\ & & {=\delta_{s,s'}(-)^{j+l+s+1}
\sqrt{\frac{(2j+1)(2l+1)(2J+1)(2L+1)(2j'+1)(2l'+1)}{4\pi}}
\sixj{l}{j'}{J}{s}{l'}{j}} \nonumber \\ & & {\times [ \sixj
{J}{l'-1}{l'}{L}{l}{1}
\frac{\threej{l}{0}{L}{0}{l'-1}{0}}{\threej{l'}{0}{1}{0}{l'-1}{0}}
\frac{l'}{2l'+1}\langle K l |\left( \frac{d}{dr}+ \frac{l'+1}{r}
\right)|K' l' \rangle} \nonumber \\ & & {+\sixj
{J}{l'+1}{l'}{L}{l}{1}
\frac{\threej{l}{0}{L}{0}{l'+1}{0}}{\threej{l'+1}{0}{1}{0}{l'}{0}}
\frac{l'+1}{2l'+1}\langle K l |\left( \frac{d}{dr}- \frac{l'}{r}
\right)|K' l' \rangle ]}
\end{eqnarray}
\subsubsection{ME5}
We note that
\begin {equation}
{Y_{LM}(\hat{x})\vec {\nabla}\cdot\vec {\Sigma}=\sum_J (-)^{L-J}
\sqrt {\frac{2J+1}{2L+1}} \left[ \left[ Y_L\otimes \vec {\nabla}
\right] ^J \otimes \vec{\Sigma} \right]_M^L}
\end{equation}
Thus
\begin{eqnarray}
\lefteqn{\langle K (l s) j \| Y_{L}(\hat{x})\vec {\nabla}\cdot \vec
{\Sigma} \|K' (l' s') j' \rangle= \sum_J (-)^{L-J} \sqrt
{(2j+1)(2J+1)(2j'+1)}} \nonumber \\ & &{ \times \langle s
\|\vec{\Sigma}\| s' \rangle \ninej{l'}{s'}{j'}{J}{1}{L}{l}{s}{j}
\langle K l \|\vec Y_{JL}(\hat{x})\cdot\nabla\| K' l' \rangle}
\end{eqnarray}
\subsubsection{Radial Integrals} The radial matrix element to be
calculated is,
\begin{eqnarray}
\lefteqn {\langle(K_{N-1} l_N) K_N | \hat{O}(L
q\rho \sin \theta_N) | (K_{N-1} l'_N) K'_N \rangle=} \nonumber \\
& & {={\mathcal{N}}_n^{(\alpha \beta)} {\mathcal{N}}_{n'}^{(\alpha'
\beta')} \int_0^{\frac{\pi}{2}}d\theta_N \sin^{2}\theta_N
\cos^{3N-4}\theta_N} \nonumber \\ & & {\times \sin^{l_N}\theta_N
\cos^{K_{N-1}} \theta_N P^{(\alpha \beta)}_n(\cos2\theta_N) }
\nonumber \\ & & {\times \hat{O}(Lq\rho \sin \theta_N)
\sin^{l'_N}\theta_N \cos^{K_{N-1}} \theta_N P^{(\alpha'
\beta')}_{n'}(\cos2\theta_N)}
\end{eqnarray}
where $N=A-1$, $n=\frac{K'_N-K_{N-1}-l'_N}{2}$,
$\alpha=l_N+\frac{1}{2}$, and $\beta=K_{N-1}+\frac{3N-5}{2}$.
\begin{eqnarray}
{I_0 =\langle(K_{A-1} l_N) K_N | j_J(L q\rho
\sin \theta_N) | (K_{A-1} l'_N) K'_N \rangle} \\
{I_1 =\langle(K_{A-1} l_N) K_N | j_J(L q\rho \sin{\theta_N})
\frac{1}{Lq\rho \sin{\theta_N}}|
 (K_{A-1} l'_N) K'_N \rangle} \\
{I_2 =\langle(K_{A-1} l_N) K_N | j_J(L q\rho \sin{\theta_N})
\frac{d}{d(Lq\rho \sin{\theta_N})}| (K_{A-1} l'_N) K'_N \rangle}
\end{eqnarray}
using the fact that
\begin{eqnarray}
\lefteqn {\frac{d}{d\sin\theta_N} \left(\sin^{l_N}\theta_N
\cos^{K_{N-1}} \theta_N P^{(\alpha \beta)}_n(\cos2\theta_N)\right) =
} \nonumber \\ & & {=\sin^{l_N}\theta_N \cos^{K_{N-1}} \theta_N
P^{(\alpha \beta)}_n(\cos2\theta_N)} \nonumber \\ & & {\times \left[
\frac{l_N}{\sin{\theta_N}}-\frac{K_{N-1}\sin{\theta_N}}{\cos^2{\theta_N}}
+\frac{2n(n+\beta)}{(2n+\alpha+\beta)\sin \theta_N
\cos^2\theta_N}-2n\frac{\sin \theta_N}{\cos^2\theta_N}\right]}
\nonumber
\\ & & {-\frac{2(n+\alpha)(n+\beta)}{2n+\alpha+\beta} \sin^{l_N-1}\theta_N
\cos^{K_{N-1}-2} \theta_N P^{(\alpha \beta)}_{n-1}(\cos2\theta_N)}
\end{eqnarray}
We can now calculate numerically the integrals above.

\section{2-Body Matrix Elements}
The $2$-body matrix elements needed for the evaluation of the axial
MEC include several operators, whose reduced matrix elements are not
completely trivial.

A much needed equalities are the following reduced matrix elements,
important also for the isospin part, of pauli matrices:
\begin{eqnarray} {\label{eq:SME1}}
{ \langle (\frac{1}{2}\frac{1}{2}) S'  ||
\vec{\sigma}_1+\vec{\sigma}_2 ||
  (\frac{1}{2}\frac{1}{2}) S   \rangle = \delta_{S,S'} \sqrt{S(S+1)(2S+1)}} \\
{\label{eq:SME2}} { \langle (\frac{1}{2}\frac{1}{2}) 0  ||
\vec{\sigma}_1-\vec{\sigma}_2 ||
  (\frac{1}{2}\frac{1}{2}) 1   \rangle =
-\langle (\frac{1}{2}\frac{1}{2}) 1  ||
\vec{\sigma}_1-\vec{\sigma}_2 ||
  (\frac{1}{2}\frac{1}{2}) 0   \rangle =-2\sqrt{3}} \\
{\label{eq:SME3}} { \langle (\frac{1}{2}\frac{1}{2}) 0  ||
\vec{\sigma}_1 \times \vec{\sigma}_2 ||
  (\frac{1}{2}\frac{1}{2}) 1   \rangle =
\langle (\frac{1}{2}\frac{1}{2}) 1  || \vec{\sigma}_1 \times
\vec{\sigma}_2 ||
  (\frac{1}{2}\frac{1}{2}) 0   \rangle =-2\sqrt{3}}i.
\end{eqnarray}
These are used both for the spin and isospin operators.

By making use of the known spherical harmonics coupling:
\begin{eqnarray}
{[Y^{(l_1)} \otimes Y^{(l_2)}]^{(l)}_m=
(-)^{l_1-l_2}\sqrt{\frac{[l_1][l_2]}{4\pi}} \left(\begin{array}{ccc}
l_1 & l_2 & l
                       \\  0  &  0  & 0 \end{array}\right)
Y^{(l)}_{m}}
\end{eqnarray}
(here $[l]=2l+1$) we can achieve the following formula,
\begin{eqnarray}
\lefteqn {\langle (L'  S') J' || [ Y_1 \otimes (\vec{\sigma}_1 \odot
\vec{\sigma}_2)^{(1)}]^{(0)}Y_L
 || (L S) J \rangle=}  \nonumber \\ &&
{=(-)^{J+J'+L+L'+S+S'}\frac{\sqrt{[L'][J'][L][L][J]}}{4\pi} \langle
S'  || \vec{\sigma}_1 \odot \vec{\sigma}_2 ||   S \rangle }\nonumber
\\ && {\sum_{L''}[L''] \left(\begin{array}{ccc} L'' & L & L
                       \\   0   & 0  & 0   \end{array}\right)
\left\{\begin{array}{ccc} L'' & L  & L
                       \\ J & S' & J' \end{array}\right\}
\left(\begin{array}{ccc}  L'' & 1  & L'
                       \\   0   & 0  & 0   \end{array}\right)
\left\{\begin{array}{ccc} L'' & 1  & L'
                       \\ S' & J' & S \end{array}\right\}}
                       \nonumber
\end{eqnarray}
Thus,
\begin{eqnarray}
{\vec{\mathcal{O}}_{\odot}^a\cdot\vec{Y}_{JLM}=
(\vec{\tau}^{(1)}\odot\vec{\tau}^{(2)})^a [\vec{\Sigma}_\odot\otimes
Y_L]^{(J)}_M}
\end{eqnarray}
\begin{eqnarray}
{\vec{\mathcal{T}}_{\odot}^a\cdot\vec{Y}_{JLM}=
(\vec{\tau}^{(1)}\odot\vec{\tau}^{(2)})^a \left\{
(-)^L\sqrt{4\pi[L]} \left(\begin{array}{ccc}    1   &  L & J
                       \\   0   & 0  & 0   \end{array}\right)
[\vec{\Sigma}_\odot\otimes Y_1]^{(0)}Y_{JM}
-\frac{1}{3}[\vec{\Sigma}_\odot\otimes Y_L]^{(J)}_M\right\}}
\nonumber \\ && {}
\end{eqnarray}
where $\vec{\Sigma}_\odot=\vec{\Sigma}_1 \odot \vec{\Sigma}_2$, for $\odot=\times,+,-$.\\

%% file: Acknowledgments.tex
    \chapter*{\Huge\sc Acknowledgements \\ \vskip 10pt \hrule}
        \addcontentsline{toc}{chapter}{Acknowledgements}
Writing this section is always a pleasure, not only because I write
it at the end of the work, but also since I treasure the help given
to me by many people in the period of the PhD.

First of all, I would like to thank Nir Barnea, who has been more
than a supervisor, who put his trust in me and gave me the
opportunity to research with him. This gratitude stretches to the
long coffee-breaks, and the even longer arguments.

I am more than grateful to Prof. G. Orlandini and Prof. W.
Leidemann, for all the physics I have learned from them, and for
giving me the feeling of a colleague rather than a student. I would
also like to thank each and every one of the other co-authors of the
papers which constitute this work: Sonia Bacca, Evan O'connor, Achim
Schwenk, and Chuck Horowitz.

Third, this work would not have been such joy without the people of
the Physics department, with whom every day is a new experience.
Among others Yuval Birnboim deserves some kind of a prise for
tolerating my changing moods; Shmulik Balberg and Eli Livne for many
fruitful discussions; TT, Ami, Shimon, who made these times
memorable.

Most important are of course all my family: my parents, my sisters
and my grandparents. For the constant support, reassurance and
encouragement, which made this difficult session more tolerable.

Last, and definitely not least, I would like to thank my wife Shani,
for helping through difficult times; for taking on every burden; and
of course for convincing me to take Daffy...